\documentclass[aps,pre,showpacs,twocolumn,superscriptaddress,nofootinbib]{revtex4}
\pdfoutput=1
\usepackage{graphicx}
\usepackage{booktabs}
\usepackage{array}
\usepackage{bm}
\usepackage{times}
\usepackage{balance}
\usepackage{epic,eepic}
\usepackage{graphics}
\usepackage{epsfig}
\usepackage{pifont}
\usepackage{subfigure}
\usepackage{bm}
\usepackage{natbib}
\usepackage{graphicx}
\usepackage{soul}

\usepackage{color}
\usepackage{amssymb}
\usepackage{hyperref}

\usepackage{enumerate}
\usepackage{amsthm,amsfonts,amsmath}
\usepackage{graphics,epsf,psfrag,pifont,dsfont,bbold}
\usepackage{siunitx}
\usepackage{xspace}
\usepackage{accents}

\def \vv  {{\bm v}}
\def  \xx  {{\bm x}}

\def \taus {\tau_{\rm s}}

\def \Rey  {\mbox{Re}}
\def \Wi   {\mbox{Wi}}

\def \Teta {T_{\eta}}

 \def \St  {{\rm St}}

\def \Poincare  {Poincar\'{e}}

\newcommand{\ddt}[1]{\frac{d #1}{dt}}

\newsavebox{\astrutbox}
\sbox{\astrutbox}{\rule[-5pt]{0pt}{20pt}}

\newcommand{\rem}[1]{}
\DeclareMathAlphabet{\mathbi}{OML}{cmm}{b}{it}

\newcommand{\bel}{\begin{equation}\label}
\newcommand{\ee}{\end{equation}}
\newcommand{\beq}{\begin{eqnarray}\label} 
\newcommand{\eeq}{\end{eqnarray}} 
\newcommand{\bc}{\begin{center}} 
\newcommand{\ec}{\end{center}} 
\newcommand{\ben}{\begin{enumerate}}
\newcommand{\een}{\end{enumerate}}
\newcommand{\bit}{\begin{itemize}}
\newcommand{\eit}{\end{itemize}}

\graphicspath{{matlab/}}

\begin{document}

\title{Melting of a nonequilibrium vortex crystal in a fluid film with polymers: elastic versus fluid turbulence}
\author{Anupam Gupta}
\email{anupam1509@gmil.com}
\affiliation{Department of Physics and INFN, University of ``Tor Vergata'', Via della Ricerca Scientifica 1, 00133 Rome, Italy.}
\affiliation{Laboratoire de G\'enie Chimique, Universit\'e de Toulouse, INPT-UPS, 31030, Toulouse, France.}
\author{Rahul Pandit}
\email{rahul@physics.iisc.ernet.in}
\affiliation{Centre for Condensed Matter Theory, Department of Physics, Indian
Institute of Science, Bangalore 560012, India.}
\altaffiliation[also at~]{Jawaharlal Nehru Centre For Advanced
Scientific Research, Jakkur, Bangalore, India.}


\begin{abstract}

We perform a direct numerical simulation (DNS) of the forced, incompressible
two-dimensional Navier-Stokes equation coupled with the FENE-P equations for
the polymer-conformation tensor. The forcing is such that, without polymers and
at low Reynolds numbers $\Rey$, the film attains a steady state that is a square
lattice of vortices and anti-vortices. We find that, as we increase the
Weissenberg number ${\Wi}$, a sequence of nonequilibrium phase
transitions transforms this lattice, first to spatially distorted, but
temporally steady, crystals and then to a sequence of crystals that oscillate
in time, periodically, at low ${\Wi}$, and quasiperiodically, for
slightly larger ${\Wi}$.  Finally, the system becomes disordered and
displays spatiotemporal chaos and elastic turbulence. We then obtain the
nonequilibrium phase diagram for this system, in the ${\Wi} - \Omega$ plane, where $\Omega \propto \Rey$,
and show that (a) the boundary between the crystalline and turbulent
phases has a complicated, fractal-type character and (b) the Okubo-Weiss
parameter $\Lambda$ provides us with a natural measure for characterizing the
phases and transitions in this diagram.

\end{abstract}

\maketitle{}

\section{Introduction}\label{sec:intro}

The equilibrium melting transition, from a spatially periodic crystal to a
homogeneous liquid, has been studied
extensively~\cite{ramakrishnan1979first,haymet1987theory,chaikin2000principles,oxtoby1991liquids,singh1991density,perlekar2010turbulence}.
Nonequilibrium analogs of this transition have been explored in, e.g., the
melting of colloidal crystals by shear~\cite{das2002melting} and the
turbulence-induced melting of a periodic array of vortices and anti-vortices in
a forced, two-dimensional (2D), fluid film~\cite{perlekar2010turbulence}. We
elucidate, via extensive direct numerical simulations (DNSs), the
nonequilibrium melting of a periodic array of vortices and anti-vortices in a
forced, 2D, fluid film with polymer additives; we refer to this periodic array
as a nonequilibrium, 2D crystal; this is also called a cellular
flow~\cite{gotoh1984instability,gotoh1987instability} in some studies.

We show that such a crystal can be melted either (a) by increasing the Reynolds numbers $\Rey$\footnote{
Strictly speaking, we should use the Grashof number $\rm{Gr}$, the nondimensionalized force~\cite{doering1995applied}, because we control the forcing, which leads eventually to the Reynolds number $\rm{Re}$ of the statistically steady of our fluid system. However, it has been conventional to use $\rm{Re}=n\Omega$ in studies of vortex crystals or cellular flows~\cite{perlekar2010turbulence,gotoh1984instability}; therefore, we use $\rm{Re}$ too.} or (b) by increasing
the Weissenberg number ${\Wi}$. Case (a) can be thought of as a
nonequilibrium melting transition, which is driven by fluid
turbulence~\cite{perlekar2010turbulence,ouellette2007curvature,ouellette2008dynamic}
and in which the disordered phase is a turbulent fluid that shows dissipation
reduction, because of the polymer
additives~\cite{perlekar2010direct,perlekar2006manifestations,gupta2015two}.
Case (b) is also a nonequilibrium melting transition; in this case the vortex
crystal melts into a disordered state, which is a polymeric fluid that shows
elastic turbulence~\cite{groisman2000elastic,groisman2001efficient} or
rheochaos~\cite{ganapathy2006intermittency,majumdar2011universality}. We show
that this system of driven vortices and anti-vortices provides an excellent
laboratory for studying the crossover from the dissipation-reduction to the
elastic-turbulence regimes in a fluid film.

Case (a), namely, turbulence-induced melting, has been explored in
experiments~\cite{ouellette2007curvature,ouellette2008dynamic}, which use a
spatially periodic force to impose a nonequilibrium vortex crystal on a fluid
film at low $\Rey$; this crystal melts, at high $\Rey$, into a disordered,
nonequilibrium, liquid-type phase. This melting problem has been studied by
using linear-stability
analysis~\cite{gotoh1984instability,gotoh1987instability}, for the first
instability of the vortex crystal, and direct numerical simulations
(DNS)~\cite{perlekar2010turbulence,braun1997bifurcations,molenaar2007attractor}.
These DNSs have explored the nonequilibrium phase diagram of this driven,
dissipative system by using techniques from nonlinear
dynamics~\cite{perlekar2010turbulence,braun1997bifurcations,molenaar2007attractor}
and generalizations of \textit{order parameters} that are used to characterize
equilibrium
crystals~\cite{ramakrishnan1979first,haymet1987theory,chaikin2000principles,perlekar2010turbulence}. 

To the best of our knowledge, case (b), namely, elastic-turbulence-induced
melting of a vortex crystal, has not been explored in experiments, even though
there have been several experimental studies of elastic
turbulence~\cite{groisman2000elastic,groisman2001efficient} and the related
phenomenon of
rheochaos~\cite{ganapathy2006intermittency,majumdar2011universality}. Direct
numerical simulations of 2D fluid films with polymer additives have begun to
explore the~\cite{berti2008two,berti2010elastic} elastic-turbulence-induced
melting of spatially periodic flows; the Kolmogorov-flow pattern investigated
in Ref.~\cite{berti2008two,berti2010elastic} is a one-dimensional crystal. 

The nonequilibrium vortex crystal, whose melting we study, is a 2D crystal. To
describe the ordering in this crystal and its subsequent melting, either by
elastic or fluid turbulence, we follow the treatment of
Ref.~\cite{perlekar2010turbulence}, which has investigated this problem in the
absence of polymers.  In particular, we solve the incompressible, 2D,
Navier-Stokes (NS) equation, for the fluid film, coupled to the FENE-P
equation, for the polymer-conformation tensor $\mathcal{C}$; this provides a
good description of our system if we can neglect the film thickness and 
the Marangoni effect, and if the Mach number is low~\cite{perlekar2010turbulence,chomaz1990soap,chomaz2001dynamics,perlekar2009statistically}.
The force we use yields, at low $\Rey$, a spatially periodic array of vortices
and antivortices that is stationary in time; we call this a vortex crystal.

As we increase either (a) the forcing amplitude and, thereby
$\Rey$ or (b) the polymer-relaxation time $\tau_P$ and, hence, $\Wi$,
this crystal loses stability and gives way to a variety of nonequilibrium
states. We show that time series of the energy of the fluid, their power
spectra, and \Poincare $~$maps help us to understand the spatiotemporal evolution
of this system and its sequence of nonequilibrium transitions (cf.,
Refs.~\cite{perlekar2010turbulence,braun1997bifurcations,molenaar2007attractor}
for turbulence-induced melting in a fluid film without polymers).

Our exploration of these transitions, which is inspired by the
density-functional theory of the crystallization of a
liquid~\cite{ramakrishnan1979first,chaikin2000principles,oxtoby1991liquids,singh1991density,perlekar2010turbulence},
is based on order parameters for our vortex crystal.  The crystal density is
{\it periodic} in space; the density-functional theory uses its Fourier
coefficients as order parameters. 


We follow Ref.~\cite{perlekar2010turbulence} and use the Okubo-Weiss field
$\Lambda\equiv \det(A)$ in the vortex-crystal instead of the crystal density
$\rho({\bf r})$, in the density-functional theory mentioned above; $A$ has
elements $A_{ij} \equiv \partial_i u_j$, with $u_j$ the $j^{th}$ component of
the velocity.  The Okubo-Weiss
criterion~\cite{okubo1970horizontal,weiss1991dynamics}, first developed to
characterize  the local topology of inviscid, incompressible, 2D, fluid flow,
is also useful if viscosity and air-drag-induced friction are
present~\cite{perlekar2009statistically,rivera2001universal,elhmaidi1993elementary};
this criterion states that $\Lambda > 0$, where the flow is vortical, and
$\Lambda < 0$, where it is extensional or
strain-dominated~\cite{okubo1970horizontal,weiss1991dynamics}.  In a
nonequilibrium vortex crystal, $\Lambda({\bf r})$ has the Fourier expansion
\begin{equation}
  \Lambda({\bf r}) = \sum_{\bf k} 
  {\hat{\Lambda}}_{\bf k} 
  \exp(\imath {\bf k} \cdot {\bf r}) ,
\end{equation}
where ${\bf k}$ are the reciprocal-lattice vectors 
and the ${\hat{\Lambda}}_{\bf k}$ are the natural order parameters
for the vortex crystal. In our 2D vortex crystal,
the counterpart of the structure factor $S({\bf k})$, for a 
crystal~\cite{ramakrishnan1979first,chaikin2000principles,oxtoby1991liquids,singh1991density,perlekar2010turbulence}, is the 2D spectrum
\begin{equation}
  E_{\Lambda}({\bf k})\equiv 
  \langle {\hat{\Lambda}}_{\bf k}
  {\hat{\Lambda}}_{\bf -k} \rangle,
\end{equation}
where the angular brackets denote the average over the nonequilibrium state of
our system (and not a Gibbsian thermal average); this state can (a) be
independent of time, (b) exhibit periodic or quasiperiodic oscillations, or (c)
be spatiotemporally chaotic and turbulent, but statistically steady.  We define
\begin{equation}
  G({\bf r}) = \langle 
  \overline{\Lambda({\bf x + r}) 
    \Lambda({\bf x})} \rangle ,\label{eq:auto_cf}
\end{equation} 
an autocorrelation function whose spatial Fourier transform is related to
$E_{\Lambda}({\bf k})$. This characterizes the spatial correlations in our
system (the overbar in Eq.~\ref{eq:auto_cf} indicates the average over the origin ${\bf x}$). As we show below, the near isotropy of the turbulent phase implies
that, to a very good approximation, $G$ depends only on $r \equiv \mid {\bf r}
\mid$ here; and it characterizes the short-range vortical order in our 2D fluid
film, exactly as the density correlation function $g(r)$ does in an isotropic
liquid at
equilibrium~\cite{ramakrishnan1979first,chaikin2000principles,oxtoby1991liquids,singh1991density,perlekar2010turbulence}.
We also follow the spatiotemporal evolution of the polymer-conformation tensor
${\cal C}$ and relate it to the spatiotemporal evolution of $\Lambda$.

We use the order parameters, defined above, to obtain many interesting results
for the elastic-turbulence-induced melting of such a 2D vortex crystal in a
fluid film with polymer additives.  We first describe our principal results
qualitatively.  We show that, as $\Rey$ or ${\Wi}$ increase, the steady,
ordered, vortex crystal proceeds to the disordered, turbulent state via a
sequence of nonequilibrium transitions between different nonequilibrium phases,
which we list in Table~\ref{tablech5:para} (column four): SX is the original,
{\it temporally steady}, crystal, with a square unit cell that is imposed by
the force; this is followed by phases of type SXA, which are {\it temporally
steady} crystals that are distorted relative to SX; we then find distorted
crystals that displays periodic (OPXA) or quasiperiodic (OQPXA) oscillations in
time; finally, the system exhibits {\it spatiotemporal chaos and turbulence}
(SCT). We conjecture that OPXA and OQPXA are actually several (perhaps an
infinity) different nonequilibrium spatiotemporal crystals, which are periodic
in space and time; given the spatial and temporal resolutions of our
calculation, we can identify only some of these. In addition to the order
parameters and correlation functions mentioned above, we also examine the
spatiotemporal evolution of the polymer-conformation tensor $\mathcal{C}$ in
these phases. Our work leads to a rich, nonequilibrium phase diagram for our
system. This has not been studied hitherto. 

The remainder of this paper comprises the following Sections. Section 2, which
is devoted to the equations that we use to model 2D fluid films with polymer
additives and the numerical methods we employ. Section 3 contains our results.
Section 4 is devoted to a discussion of our results and to conclusions.

\section{Model and Numerical Methods}\label{sec:model}

To study a 2D fluid film with polymer additives, we write the 2D,
incompressible, NS and FENE-P equations in terms of the streamfunction $\psi$
and the vorticity $\omega = \nabla \times \bf u(\bf x, t)$, where ${\bf
u}\equiv(-\partial_y \psi, \partial_x \psi)$  is the fluid velocity at the
point $\bf x$ and time $t$, as follows (we use the nondimensional form
suggested in Ref.~\cite{platt1991investigation} for 2D fluid films without
polymer additives):
\begin{eqnarray}
D_t{\bf \omega} &=& \nabla^2 {\bf \omega}/{\Omega}+
  \frac{\beta}{\Omega \Wi} \nabla \times \nabla.[f(r_P){\cal C}] 
        - \alpha \omega   + F_{\omega} ;    
                \label{eqch5:ns_vor}\\
\nabla^2 {\bf \psi}  &=& {\bf \omega}; 
            \label{eqch5:poisson}\\
D_t{\cal C}&=& {\cal C}. (\nabla {\bf u}) + 
                {(\nabla {\bf u})^T}.{\cal C} - 
 \frac{{f(r_P){\cal C} }- {\cal I}}{\Wi};
                                                   \label{eqch5:FENE}
\end{eqnarray}
here, $D_t \equiv \partial_t + {\bf u}\cdot\nabla $, the uniform solvent
density $\rho=1$, $\alpha$ is the non-dimensionalized, air-drag-induced
friction coefficient, 
$\beta = \mu / \nu$, the Weissenberg number ${\Wi} = \tau_P (F_{amp}/(n k\nu))$, 
$\nu$ is the kinematic viscosity, $\mu$ the zero-shear viscosity
parameter for the solute (FENE-P), $\tau_P$ the polymer relaxation time, and
$F_{\omega} \equiv -n^3 [\cos(n x) + \cos(n y)]/\Omega$ is the
non-dimensionalized, spatially periodic force, whose injection wave vector is
related to $n$, $\alpha=n\nu\alpha^\prime k/F_{amp}$, and
$\Omega=nF_{amp}/(\nu^2 k^3)=n{\Rey}$, where $F_{amp}$ is the forcing amplitude, and
$\alpha^\prime$ is the coefficient of friction, ${\Rey}$ is the Reynolds number. We non-dimensionalize lengths
by a factor $k/n$, with $k$ a wave number or inverse
length~\cite{platt1991investigation}. The $x$ and $y$ components of the
velocity are $u_1\equiv u$ and $u_2 \equiv v$, respectively. The superscript
$T$ denotes a transpose, ${\cal C}_{\alpha\beta}\equiv
{\langle{R_\alpha}{R_\beta}\rangle}$ are the elements of the
polymer-conformation tensor (angular brackets indicate the average over polymer
configurations), the identity tensor ${\cal I}$ has the elements
$\delta_{\alpha \beta}$, $f(r_P)\equiv{(L^2 -2)/(L^2
- r_P^2)}$ is the FENE-P potential that ensures finite extensibility, and $ r_P
  \equiv \sqrt{{\rm Tr}(\cal C)}$ and $ L $ are, respectively, the length and
the maximal possible extension of the polymers.

Our direct numerical simulation (DNS) of
Eqs.~(\ref{eqch5:ns_vor})-(\ref{eqch5:FENE}) uses the MPI code that we have
developed to study fluid turbulence with polymer additives in fluid
films~\cite{gupta2015two}. We use periodic boundary conditions, a square
simulation domain of side ${\mathbb L}=2\pi$, and $N^2$ collocation points;
these boundary conditions are well suited for studying vortex crystals with
square unit cells. We use a fourth-order, Runge-Kutta scheme, with time step
$\delta t$, for time marching and an explicit, fourth-order,
central-finite-difference scheme in space and the Kurganov-Tadmor (KT)
shock-capturing scheme~\cite{kurganov2000new} for the advection term in
Eq.~(\ref{eqch5:FENE}), because this scheme (see Eq. (7) of
Ref.~\cite{perlekar2010direct}) successfully resolves steep gradients in ${\cal
C}_{\alpha\beta}$ and, thereby, minimizes dispersion errors. We solve the
Poisson equation~(\ref{eqch5:poisson}) in Fourier space by using a
pseudospectral method and the FFTW library~\cite{frigo1999fftw}. We choose
$\delta t$ to be small enough to prevent $r_P$ from becoming larger than $L$
(if $r_P > L$, there is a numerical instability).  To preserve the
symmetric-positive-definite (SPD) nature of $\cal C$ at all times we adapt to
2D the Cholesky-decomposition scheme of
Refs.~\cite{perlekar2010direct,perlekar2006manifestations,gupta2015two,vaithianathan2003numerical}.
In particular, we define ${\cal J} \equiv f(r_P) {\cal C}$, so
Eq.~(\ref{eqch5:FENE}) becomes 
\begin{equation}  
D_t{\cal J} = {\cal J}. (\nabla {\bf u}) 
+ ({\nabla \bf u})^T .{\cal J} -s({\cal J} - {\cal I})+ q {\cal J},
\label{eqch5:conj} 
\end{equation} 
where $s=(L^2 -2+ j^2)/(\tau_P L^2)$, $q=[d/(L^2 -2)-(L^2 -2+
j^2)(j^2 -2)/(\tau_P L^2(L^2 -2))]$, $j^2\equiv Tr({\cal J})$,
and $d = Tr[ {\cal J}. (\nabla{\bf u}) + (\nabla{\bf u})^T .{\cal
J}].$ Note that ${\cal C}$, and hence ${\cal J}$, are SPD
matrices, so ${\cal J}= {\cal LL}^T$, where ${\cal L}$ is a
lower-triangular matrix with elements $\ell_{ij}$, such that
$\ell_{ij}=0$ for $j>i$. We can now use Eq.(\ref{eqch5:conj}) to
obtain, for $1\le i \le2$ and $ \Gamma_{ij}=\partial_i u_j,$  
\begin{eqnarray}
\nonumber
{D_t \ell_{11}}  &=& \Gamma_{11} \ell_{11} + \Gamma_{21} \ell_{21}
+ \frac{1}{2}\Big[(q-s)\ell_{11}+ \frac{s}{\ell_{11}}\Big], \\    
\nonumber    
{D_t \ell_{21}}  &=& \Gamma_{12} \ell_{11} + 
\Gamma_{21} \frac {\ell_{22}^2}{\ell_{11}} + 
\Gamma_{22} \ell_{21}\\
\nonumber
&&{}+ \frac{1}{2}\Big[(q-s)\ell_{21} - s \frac{\ell_{21}}{\ell^2_{11}} \Big], \\    
\nonumber    
{ D_t \ell_{22}}  &=& - \Gamma_{21} \frac {\ell_{21} \ell_{22}}{\ell_{11}} 
+ \Gamma_{22} \ell_{22} \\
&&{}+ \frac{1}{2}\Big[(q-s)\ell_{22} - \frac {s}{\ell_{22}}  \left( 1 + \frac {\ell^2_{21}}{\ell^2_{11}} \right) 
\Big]. 
\label{eqch5:ellij}
\end{eqnarray}
Equation(\ref{eqch5:ellij}) preserves the SPD property of  $\cal C$ if
$\ell_{ii} > 0$; we enforce this, as in
Refs.~\cite{perlekar2010direct,perlekar2006manifestations,gupta2015two}, by
considering the evolution of $\ln(\ell_{ii})$ instead of $\ell_{ii}$.

In most of our studies, we use a time step $\delta t = 10^{-3}$ to $10^{-4}$
for $N=128$; its precise value depends on the value of the polymer-relaxation
time $\tau_P$; for $\tau_P \simeq 1$, $\delta t = 10^{-3}$ and $\delta t =
10^{-4}$ otherwise ($\tau_P>1$). Our results do not change significantly if we
use $N=256$ or $N=1024$ (we have checked this in representative cases).  We
obtain long time series for several variables (see below) to make sure that the
temporal evolution of our system is captured accurately; most of our runs are
at least as long as $5\times10^6 \delta t$. We monitor the time-evolution of
the total kinetic energy $E(t)\equiv \overline{{\bf u}^2}$ and calculate the
stream function $\psi$, vorticity $\omega$, polymer-conformation tensor
$\mathcal{C}$, the Okubo-Weiss parameter $\Lambda$, and the kinetic energy
spectrum $E(k)\equiv\sum_{k-1/2< k' \le k+1/2} k'^2 \langle |\hat {\bf
\psi}({\bf k'},t)|^2 \rangle _t$, where $\langle \rangle_t$ indicates the time
average over the statistically steady state. From these we obtain
$E_{\Lambda}({\bf k})$ and $G({\bf r})$, by averaging over $40$ configurations
of $\Lambda({\bf r})$, which are separated from each other by $10^5 \delta t$;
to eliminate the effects of initial transients, we remove data from the first
$10^6$ time steps before we collect data for these averages.  From $\psi$ we
obtain the velocity and thence the ${\bf k} = (1,0)$ component of the Fourier
transform $\hat{v}$ of the $y$ component $v$ of ${\bf u}$. We use the time
series of $E(t)$ to obtain its temporal Fourier transform $E(f)$ and from that
the power spectrum $\mid E(f) \mid$, which we use to differentiate between
periodic, quasiperiodic, and chaotic temporal evolutions. We also use
\Poincare-type sections, namely, plots of $\Im\hat{v}_{(1,0)}$ versus
$\Re\hat{v}_{(1,0)}$ at successive times (for the Kolmogorov flow see
Ref.~\cite{platt1991investigation}).

The equation of motion for a small, rigid particle (henceforth, an inertial
particle) in an incompressible flow~\cite{maxey1983equation,gatignol1983faxen}
assumes the following simple form~\cite{bec2006acceleration,boffetta2004large},
if the particle is much heavier than the advecting fluid:
\begin{eqnarray}
\ddt{\xx} = \vv(t); 
\ddt{\vv} = - \frac{\vv(t) -{\bf u}[\xx(t),t]}{\taus};
            \label{maxey_riley}
\end{eqnarray}
here, $\vv$ and $\xx$ are, respectively, the velocity and position of an
inertial particle, $\taus = (2 a^2)/(9\nu \rho_{\rm f})$ is the Stokes or
response time of the particle, whose ratio with
the Kolmogorov dissipation time $\Teta$ is the Stokes number $\St =
\taus/\Teta$. We assume that the radius of a particle $a
\ll\eta$, where $\eta$ is the dissipation scale of the carrier fluid, and that
the particle density is so low that we can neglect interactions between
particles. We also assume, as in Ref.~\cite{de2012control}, that the polymer
additives affect the inertial-particle velocity $\vv(t)$ only by virtue of
their effect on the Eulerian velocity ${\bf u}[\xx(t),t]$ of the fluid at the
position of the inertial particle. We show below that the spatial distribution
of such inertial particles in the flow can also be used to distinguish the
nonequilibrium crystalline and turbulent states in our system.

\section{Results}\label{sec:results}

We now present the results of our DNS. We begin with an overview, in the first
subsection. In the next subsection, we present the nonequilibrium phase diagram
that we obtain from our calculations.  We then present the results of our
numerical simulations for $n=4$ and the ranges of parameters given in
Table~\ref{tablech5:para}. In the third and fourth subsections, respectively,
we present our results for $\Omega=1$ and $\Omega=22$. In the fifth subsection,
we report our results for the order parameters, spatial correlation functions,
and kinetic-energy spectra that we have defined above; and the last subsection
contains our results for the spatial patterns of polymers and inertial
particles.

\subsection{Overview}

We demonstrate below how our crystal of vortices melts, as we increase $\Wi$
for a fixed, low value of $\Omega$, through a sequence of transitions. At
somewhat higher values of $\Omega$, at which the vortex crystal has already
reached a melted state because of fluid turbulence, we find that this turbulent
state moves first to a frozen state, as we increase $\Wi$; this frozen state
melts again as $\Wi$ increases some more. The air-drag-induced friction
delays these transitions; we have checked this in some representative cases.
However, to make contact with earlier studies of turbulence-induced melting of
a vortex
crystal~\cite{perlekar2010turbulence,braun1997bifurcations,molenaar2007attractor}
without polymers, we do not include air-drag-induced friction here; our
conclusions are not altered qualitatively by this.

The steady-state solution~\cite{platt1991investigation,perlekar2010turbulence},
of Eq.~(\ref{eqch5:ns_vor}) is $\omega_{s,n}=-n[\cos(nx) + \cos(ny)]$, if
$\Wi = 0$ and $\Omega < \Omega_{s,n}$; here the subscript $s$ stands for
steady state. We examine the destabilization of this solution, by increasing
$\Wi$, while we hold $\Omega$ fixed, for one representative value of $n$,
namely, $n=4$ for which $\Omega_{s,4} \simeq 5.657$ (see
Ref.~\cite{perlekar2010turbulence}). The initial vorticity is taken to be
$\omega=\omega_{s,n} + 10^{-4}\sum_{m_1=0,m_2=0}^{2,2}[\sin(m_1x + m_2y) +
\cos(m_1x + m_2y)] m_2^2/\sqrt{(m_1^2+m_2^2)}$; and the system then evolves
under the force $F_{\omega}$. For a fixed value of $\Omega$, we increase
$\Wi$ in steps of $0.1$; and we carry out simulations for $\Omega = 1,2$, and
from then to $\Omega = 26$ in steps of $2$. To track the nonequilibrium
transitions in our system, say from SXA to SCT, we have also carried out runs
in which we increase $\Wi$ in steps of $\simeq 0.01$. We have validated our
numerical scheme by comparing our results with those of
Refs.~\cite{platt1991investigation,perlekar2010turbulence}. Reference
~\cite{platt1991investigation} studies a Kolmogorov flow $F_{\omega} = n \cos(n
y)$; and Ref.~\cite{perlekar2010turbulence} uses an external force of the form
$F_{\omega} \equiv -n^3 [\cos(n x) + \cos(ny)]/\Omega$, which is exactly the
same as the one we employ.

\begin{table}[ht]
  \begin{center}
    \begin{tabular}{@{\extracolsep{\fill}} c c c c c}
      \hline \hline 
      $ Run $ & $\Omega$ & $\Wi$ & $\rm{Order}$ \\ 
      \hline \hline 
      ${\tt R1-1}$ & $1$ & $\Wi \leq 1.9$ & $\rm{SX}$ \\
      ${\tt R1-2}$ & $1$ & $ 2 \leq \Wi \leq 9 $  & $\rm{OPXA}$ \\
      ${\tt R1-3}$ & $1$ & $\Wi \geq 10$  & $\rm{SCT}$ \\
      \hline \hline 
      ${\tt R22-1}$ & $22$ & $\Wi = 0 $ & $\rm{SCT}$ \\
      ${\tt R22-2}$ & $22$ & $0<\Wi<0.45$ & $\rm{OPXA}$ \\
      ${\tt R22-3}$ & $22$ & $\Wi = 0.5 $ & $\rm{SXA}$ \\
      ${\tt R22-4}$ & $22$ & $0.55<\Wi<0.8$ & $\rm{OPXA}$ \\
      ${\tt R22-5}$ & $22$ & $\Wi\geq 1$ & $\rm{SCT}$ \\ \hline
      \hline
    \end{tabular}
  \end{center}

\caption{Table indicating the number of the Run (e.g., R1-1), the value of 
$\Omega$, and the type of order. SX : original, {\it steady}
square crystal (the pattern depends on $n$); SXA : {\it temporally steady} 
crystals slightly distorted compared to SX; OPXA : crystal that is distorted 
slightly relative to SX and with temporal oscillations; OQPXA : OPXA but with 
{\it quasiperiodic temporal oscillations}; SCT : disordered phase with 
{\it spatiotemporal chaos and turbulence}. We have carried out $\simeq 400$ such Runs.}
\label{tablech5:para}
\end{table} 

\begin{figure*}
\includegraphics[width=0.47\linewidth]{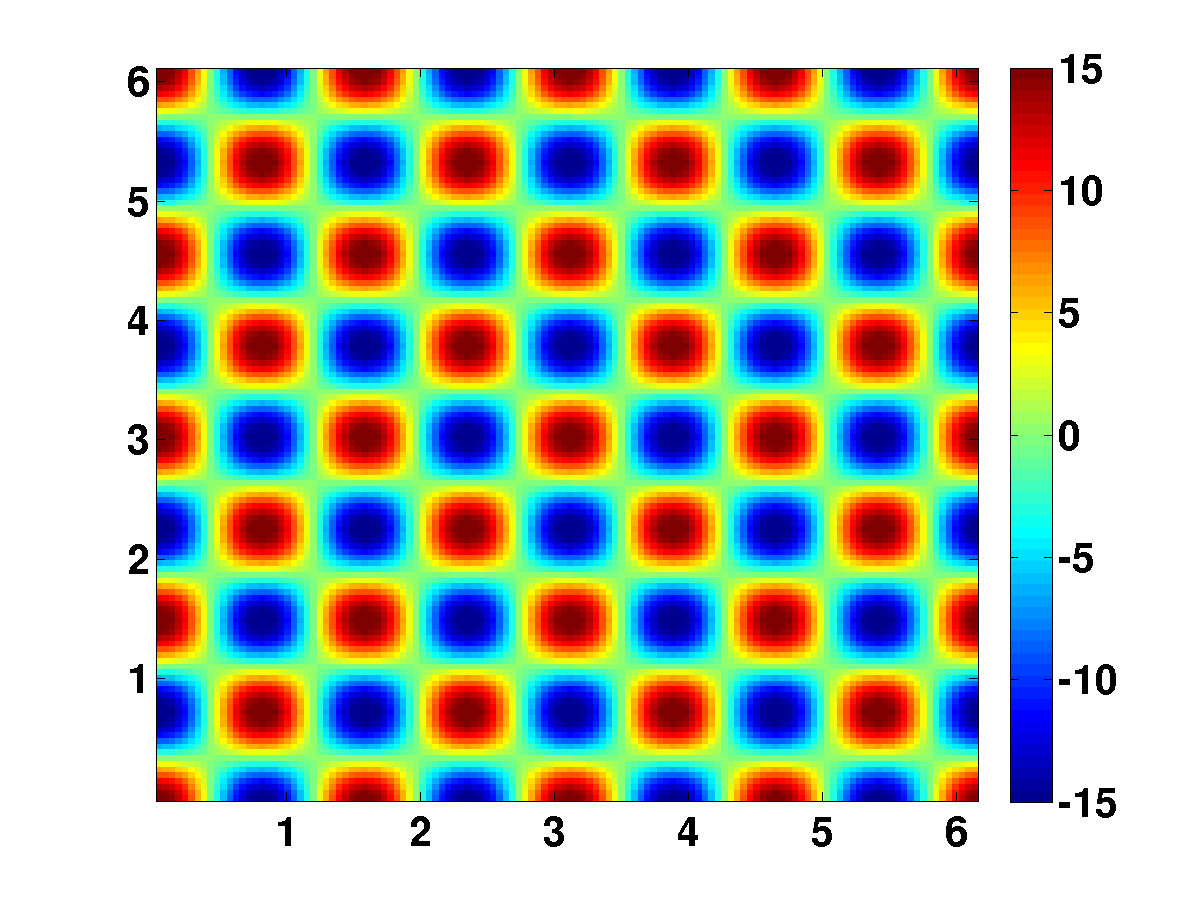}
\put(-65,148){\color{white}{ {\bf (a)} } }
\includegraphics[width=0.47\linewidth]{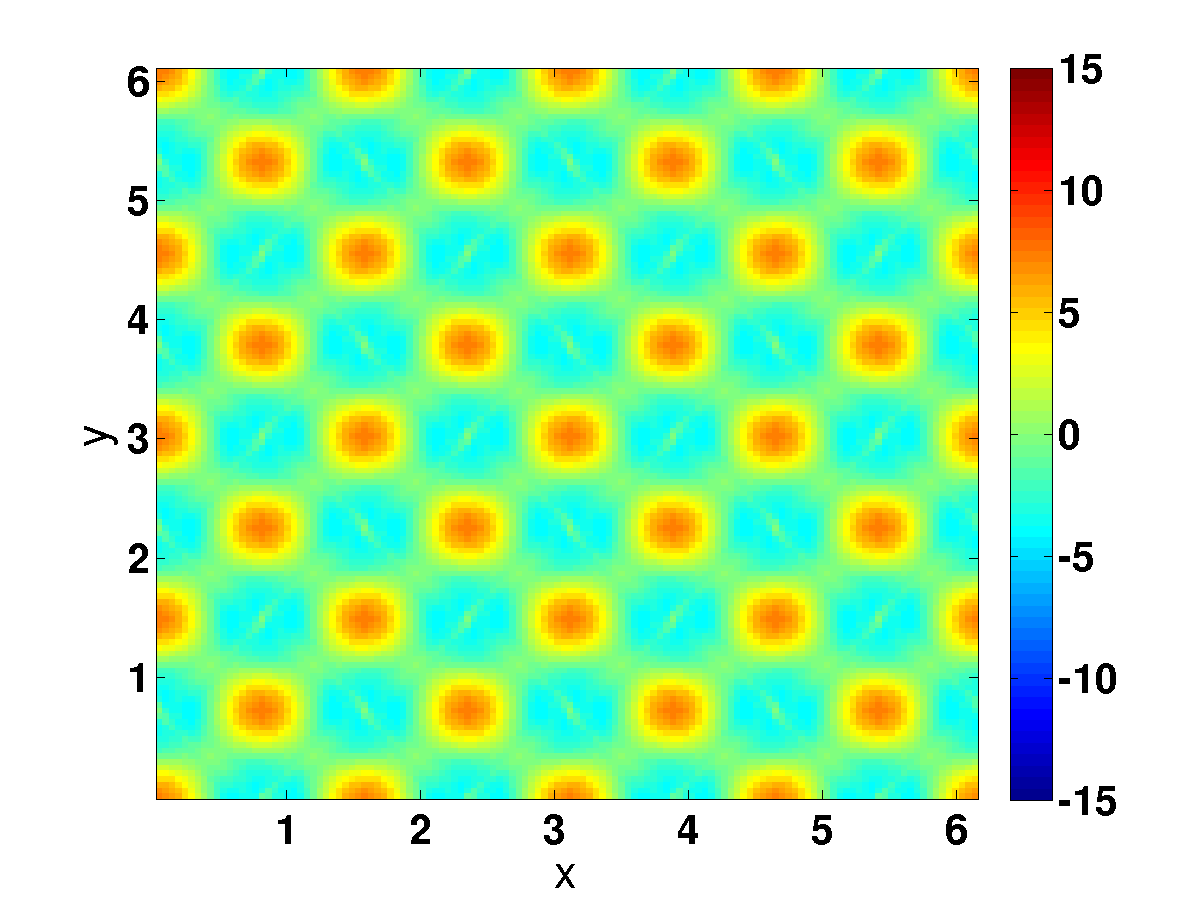}
\put(-65,148){\color{black}{ {\bf (b)} } }
  \caption{(Color online) Pseudocolor plots of the Okubo-Weiss parameter
$\Lambda$ for (a) $\Omega = 1$ and $\Wi=0 $ and (b) $\Omega =
20$ and $\Wi=1 $; these show vortex-crystal states. Given our color bar
in (a), vortical regions are red and strain-dominated regions are 
dark blue.}
  \label{figch5:inlam}
\end{figure*}

\begin{figure*}
\includegraphics[width=0.47\linewidth]{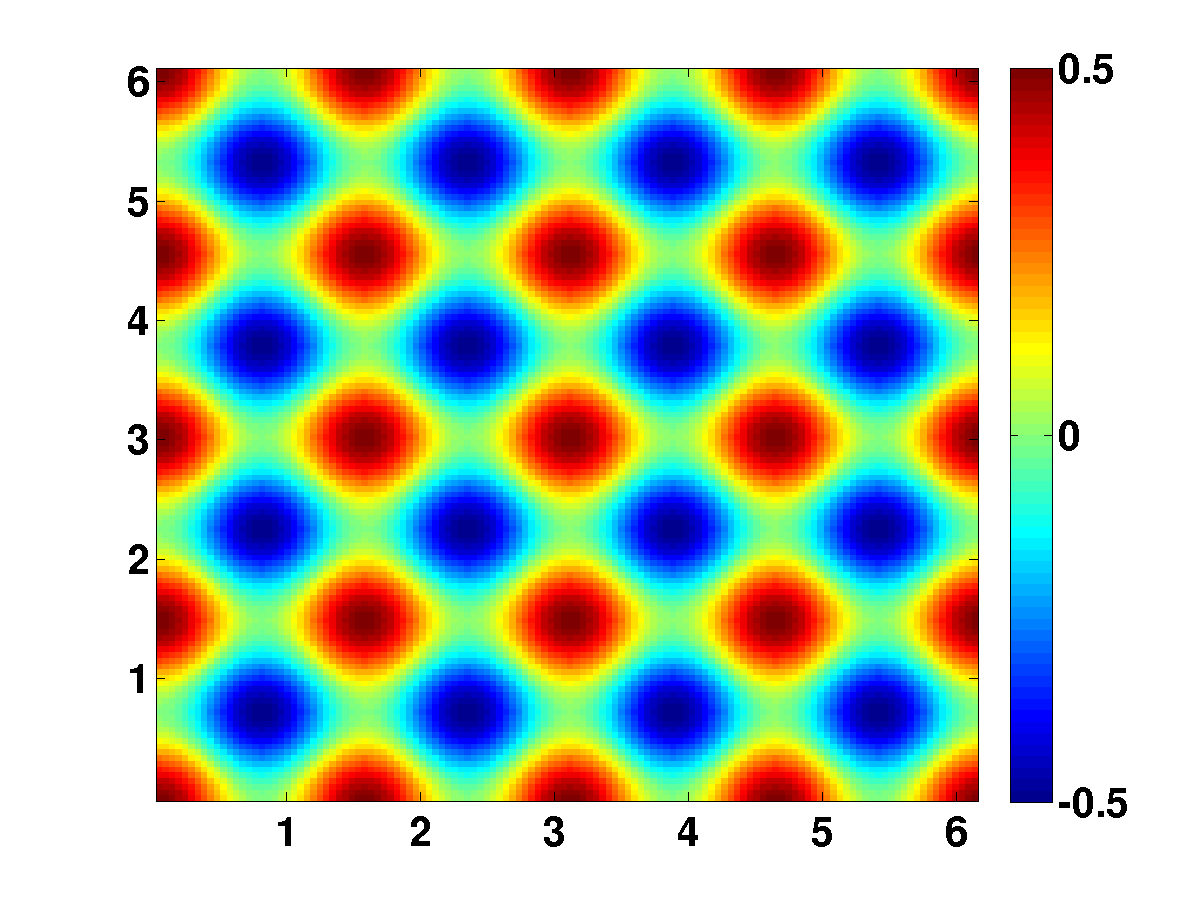}
\put(-65,148){\color{white}{ {\bf (a)} } }
\includegraphics[width=0.47\linewidth]{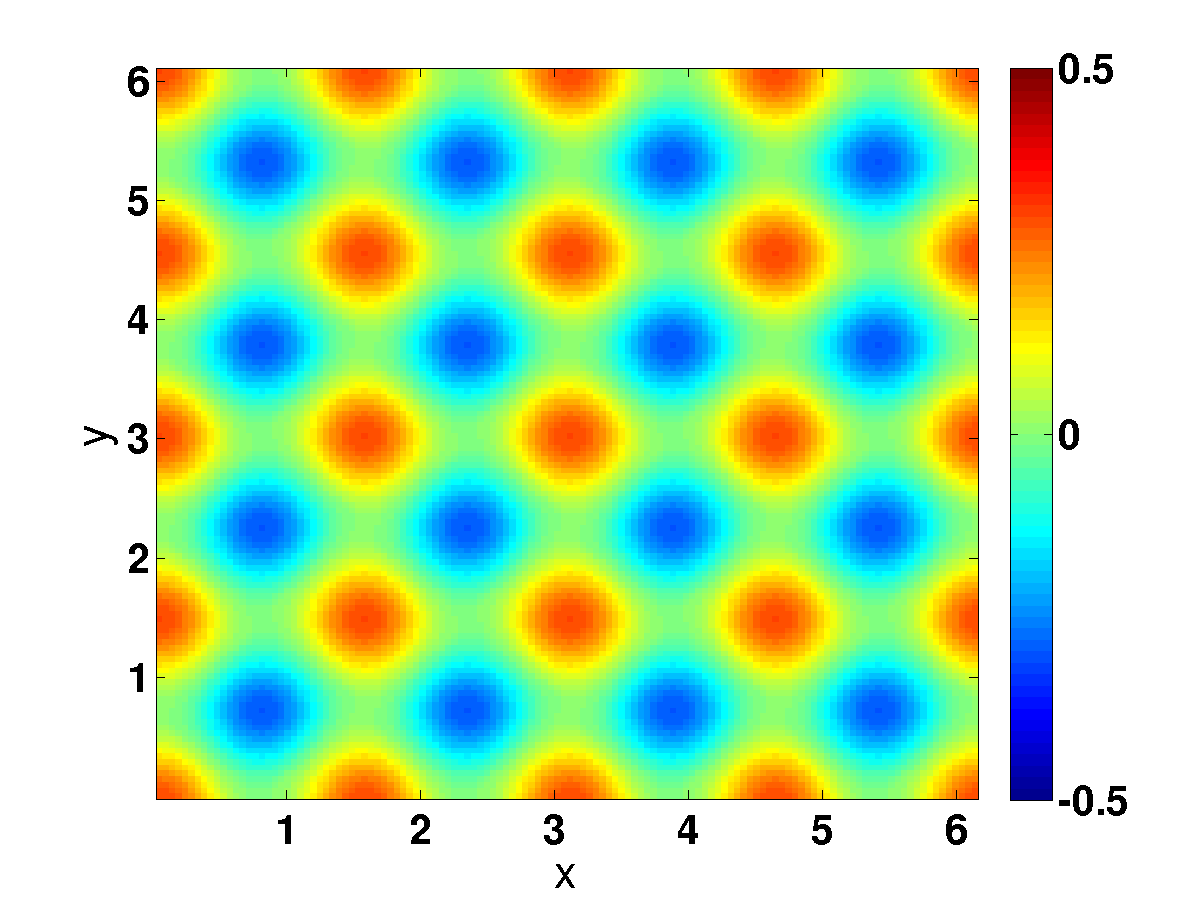}
\put(-65,148){\color{black}{ {\bf (b)} } }
\caption{(Color online) Pseudocolor plots of the stream function
$\psi$, illustrating vortex-crystal states, 
for (a) $\Omega = 1$ and $\Wi=0 $ and (b)  $\Omega = 20$ and
$\Wi=1$.}
  \label{figch5:inpsi}
\end{figure*}

For $\Wi=0$ and $\Omega < \Omega_{s,n}$, the $\Lambda$ field
shows alternating vortical and extensional regions, which are
also referred to as centers and saddles; these are arranged in a
two-dimensional square lattice, which we illustrate via
pseudocolor plots of $\Lambda$, for $n=4$, in
Fig.~\ref{figch5:inlam} (a).  For $\Omega =20$, which is well
beyond $\Omega_{s,4} \simeq 5.657$, the fluid becomes turbulent;
but, on the addition of polymers, $\Wi$ increases, and the
fluid again forms a 2D vortex lattice, which we illustrate via
pseudocolor plots of $\Lambda$, for $n=4$, in
Fig.~\ref{figch5:inlam} (b). In Figs.~\ref{figch5:inpsi} (a) and (b) 
we show the corresponding pseudocolor plots of the streamfunction $\psi$.

\subsection{Nonequilibrium Phase Diagram}

\begin{figure*}
\begin{center}
  \includegraphics[width=0.43\linewidth]{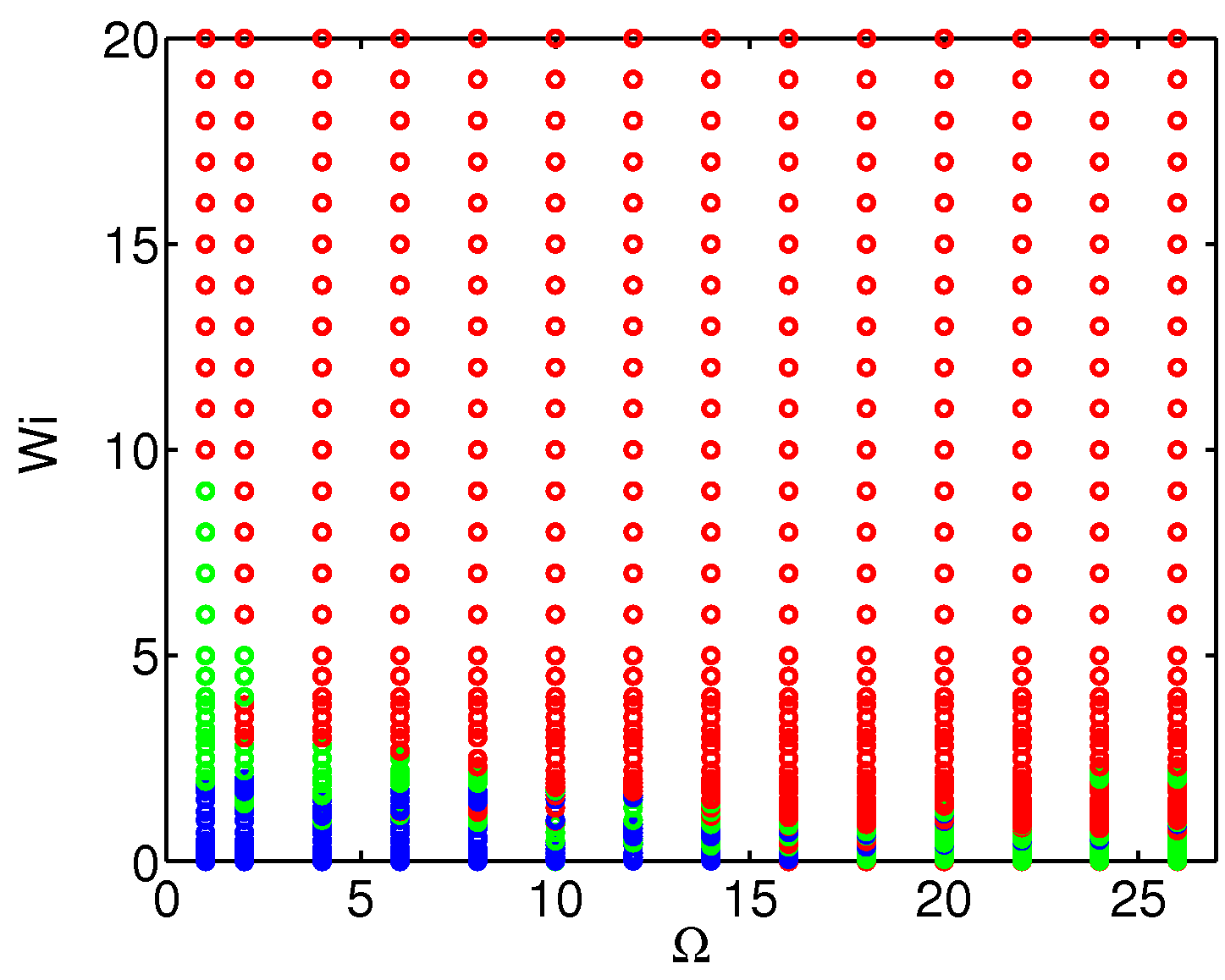}
\put(-40,148){\color{black}{ {\bf (a)} } }
\hspace{1.5cm}
  \includegraphics[width=0.43\linewidth]{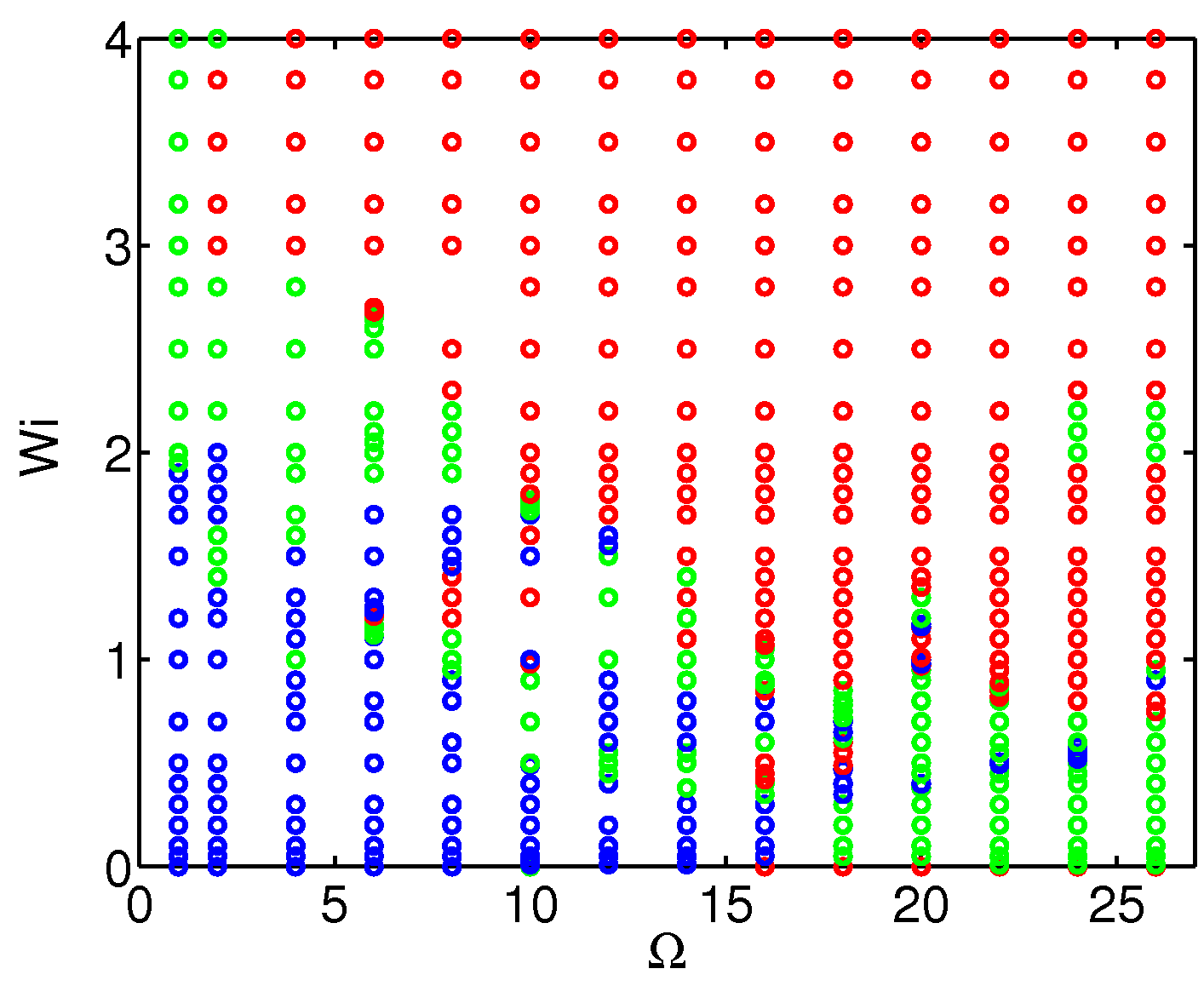}
\put(-40,148){\color{black}{ {\bf (b)} } }
  \caption{\label{figch5:phase}(Color online) (a) Phase diagram of our 2D system with
     polymer additives in the $\Wi$-$\Omega$ plane, where blue  
     circles represent the regions SX and SXA, green circles 
     represent the regions OPXA and OQPXA, and the red circles
     represents the region SCT; (b) an enlarged version of a region
     of the phase diagram in (a).}
\end{center}
\end{figure*}

We have carried out a \textit{very large number} ($\simeq 400$) of DNS studies
of our system for \textit{a wide range of parameter values} in the
$(\Wi$-$\Omega)$ plane; only a very small fraction of these parameter values
are listed in Table~\ref{tablech5:para}.  For each one of these parameter sets
we calculate the quantities, such as $E(t)$, which we have defined above; and
from these calculations we obtain the rich, nonequilibrium phase diagram of
Fig.~\ref{figch5:phase} (a), in which blue circles represent the {\it steady
crystal} SXA, green circles represent either {\it temporally periodic} (OPXA)
or {\it quasiperiodic} (OQPXA) states, and the red circles represent the state
SCT that displays {\it spatiotemporal chaos and turbulence}.  The most striking
feature of this nonequilibrium phase diagram is that the boundaries between the
different states are very complicated, with a fractal-type interleaving of the
states SXA, OPXA/OQPXA, and SCT; this is especially clear in
Fig.~\ref{figch5:phase}(b), which shows a detailed view of the phase diagram in
the vicinity of the these boundaries.  Earlier studies of elastic turbulence
seem to have missed the complicated nature of these boundaries because they
have not been able to examine these transition in as much detail as we have
done.

Fractal-type boundaries between different nonequilibrium states have been
obtained in nonequilibrium phase diagrams for some other extended dynamical
systems. Examples of such boundaries can be found in the transition to
turbulence in pipe flow~\cite{schneider2007turbulence}, the transitions between
different spiral-wave patterns in mathematical models for cardiac
tissue~\cite{shajahan2007spiral,shajahan2009spiral}, and the transition from
no-dynamo to dynamo regimes in a shell model for magnetohydrodynamic (MHD)
turbulence~\cite{sahoo2010dynamo}. In many dynamical systems, a fractal basin
boundary can separate  the basin of attraction of a fixed point or a limit
cycle from the domain of attraction of a strange attractor; thus, a small
change in the initial condition can lead either to the simple dynamical
evolutions, associated with fixed points or limit cycles, or to chaos, because
of a strange attractor. The 2D Navier-Stokes (NS) and FENE-P equations that we
study here comprise a $4 N^2$-dimensional dynamical systems, where $N^2$ is the
number of grid points; the factor of $4$ appears here because we have the 2D
vorticity and the three independent components of the $\cal L$ matrix at each
grid point.  It is not feasible to find the complete basin boundaries for such
a high-dimensional dynamical system (we use $N \sim 10^2$ so $4 N^2 \sim 4
\times 10^4$); however, we can reasonably assume that complicated, fractal-type
boundaries separate the basins of attraction of the SXA, OPXA/OQPXA, and SCT
states. Note that we do not change the initial condition; instead, we change
our dynamical system by changing the parameters $F_{amp}$ and $\tau_P$  and,
hence, ${\Omega}$ and ${\Wi}$. This change affects the long-time
evolution of our 2D NS and FENE-P system as sensitively as does a change in the
initial conditions because the fractal basin boundary itself changes with these
parameters. Reference~\cite{schneider2007turbulence} has found such a
sensitive dependence on model parameters in the transition to turbulence and the
edge of chaos in pipe flow (see their Fig. 3, which shows that the border, in
parameter space, between the laminar and turbulent regions is very complicated,
in much the same way as the phase boundary is in our study). In
Refs.~\cite{shajahan2007spiral,shajahan2009spiral}, it has been shown that the
spatiotemporal evolution of spiral waves of electrical activation depends
sensitively on inhomogeneities in partial-differential-equation models for
ventricular tissue; changes in the positions of these inhomogeneities in models
for cardiac tissue correspond in our system to changes in parameters such as
$\Omega$ and $\Wi$. A shell-model study of the dynamo transition in MHD has
also found such a fractal-type boundary, separating dynamo and no-dynamo
regimes, in the plane of the magnetic Reynolds number and the magnetic Prandtl
number~\cite{sahoo2010dynamo}. 

In the next two subsections we describe in detail the results of our
calculations for two representative values of $\Omega$, namely, $1$ and $22$,
and a range of values of $\Wi$ (Table~\ref{tablech5:para}).  We have carried
out many such calculations ($\simeq 400$) to obtain the phase diagram of
Fig.~\ref{figch5:phase}. 

\subsection{The case $\Omega=1$}

\begin{figure*}[]
    \includegraphics[width=0.323\linewidth]{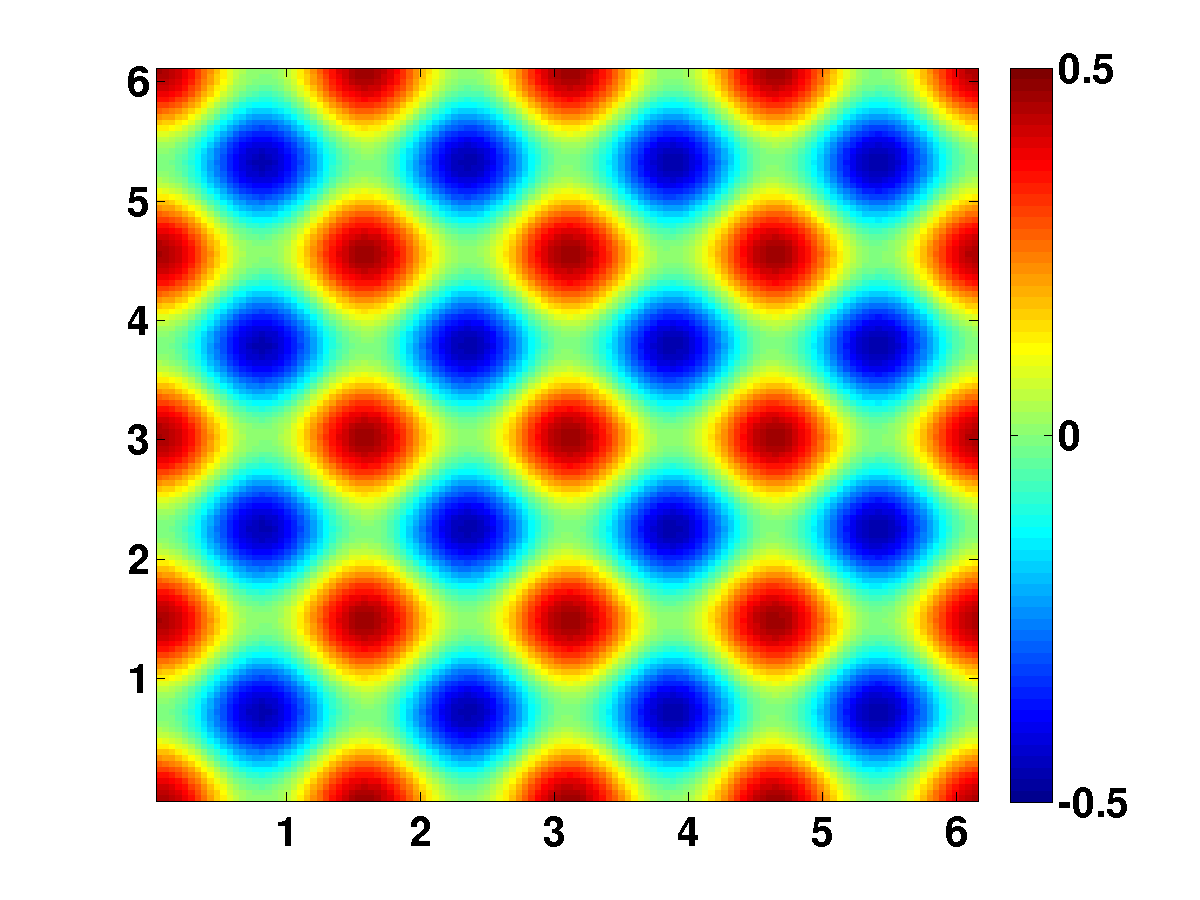}
  \put(-53,95){\color{white}{ {\bf (a)} } }
     \includegraphics[width=0.323\linewidth]{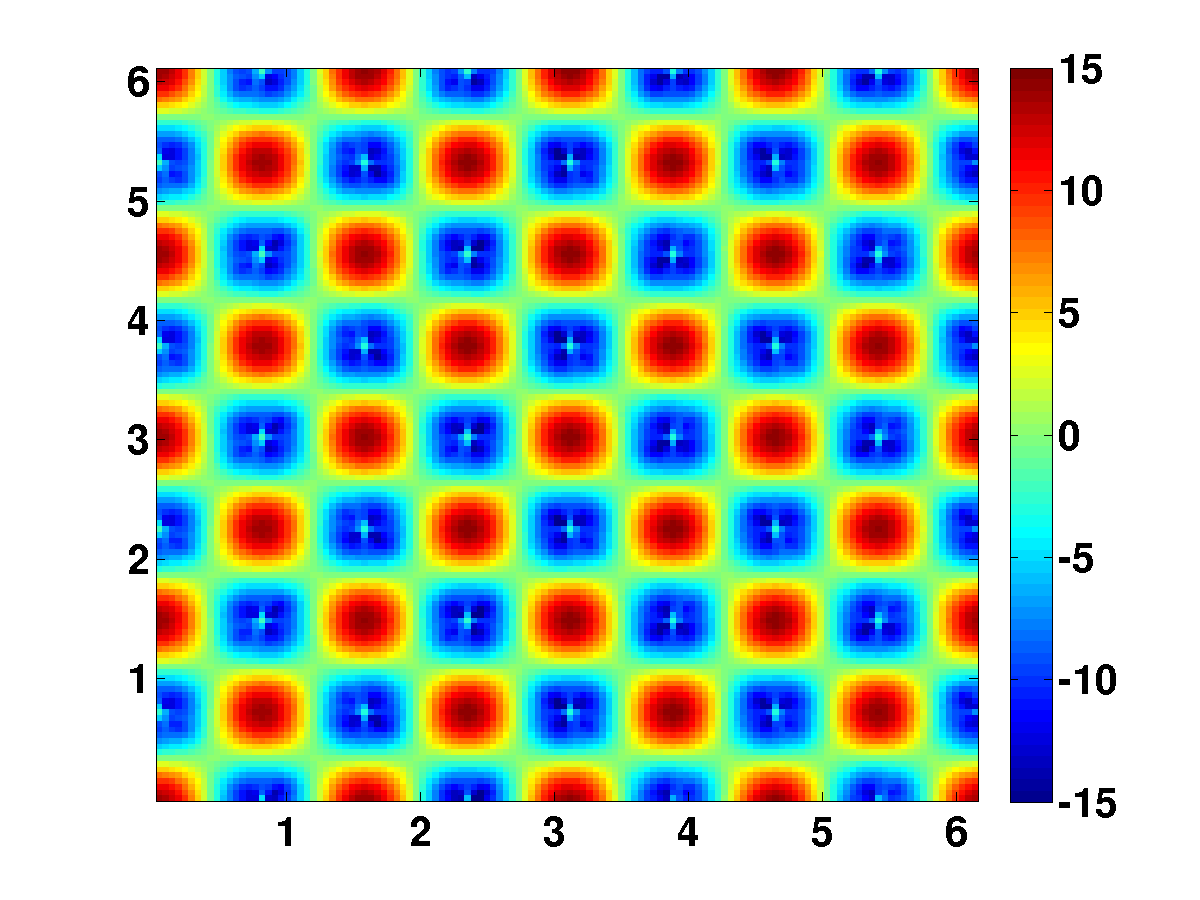}
  \put(-53,95){\color{white}{ {\bf (b)} } }
    \includegraphics[width=0.323\linewidth]{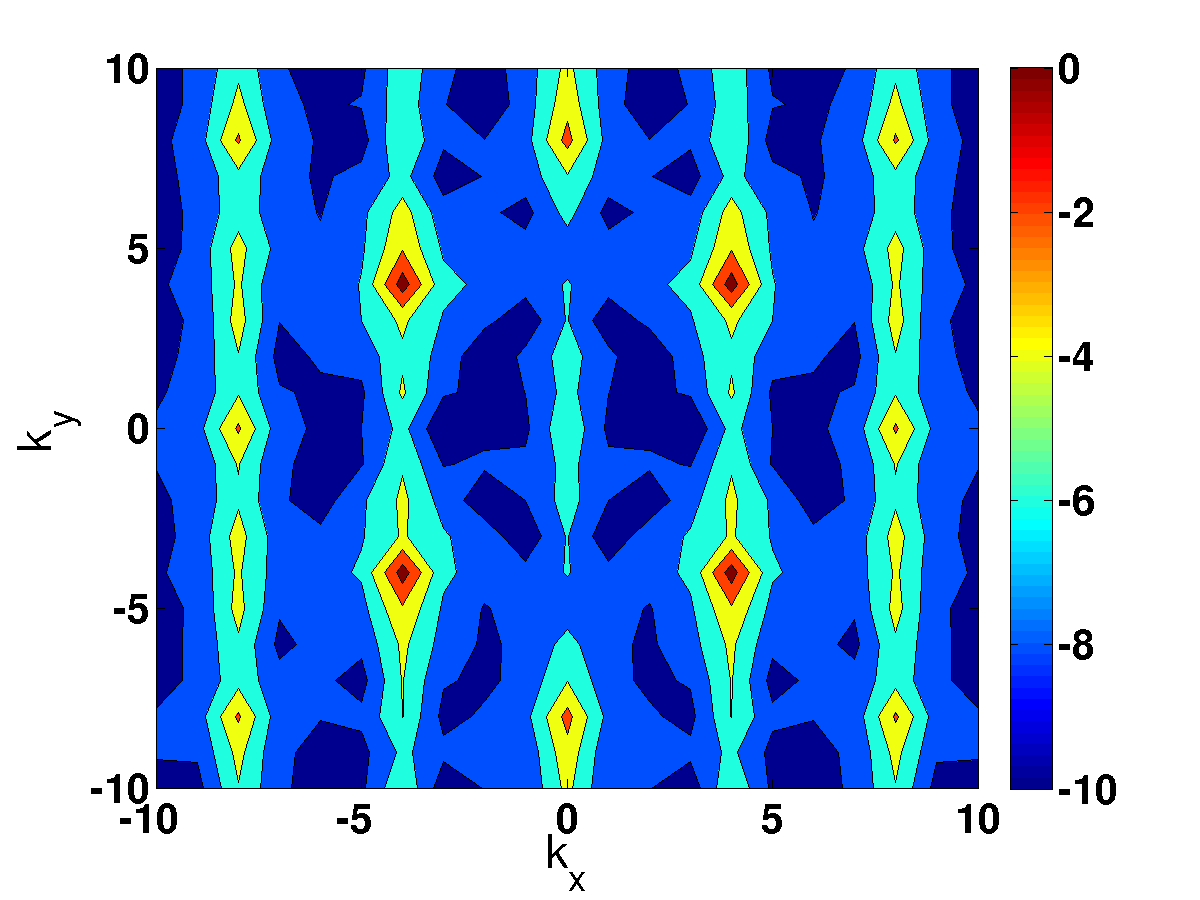}
  \put(-53,95){\color{white}{ {\bf (c)} } }
  \caption{\label{figch5:R42psilamk}(Color online) Pseudocolor plots
    for $\Omega=1$ and $\Wi=1$ : (a) the streamfunction $\psi$ and (b) the
    Okubo-Weiss parameter $\Lambda$; (c) a filled contour plot of the $E_{\Lambda}$, 
    reciprocal-space spectrum; this shows peaks at the
    forcing wave vectors.}
\end{figure*}

If $\Omega = 1$, which lies in region $\Omega <\Omega_{s,n=4}$, the
steady-state solution is $\omega_{s,n=4}$, in the absence of polymers. When we
add FENE-P polymers to the 2D Navier-Stokes fluid and increase the Weissenberg 
number $\Wi$, the steady-state solution $\omega_{s,n=4}$ becomes
unstable. For the range of values of $\Wi$ in our runs ${\tt R1-1}$
(Table~\ref{tablech5:para}), we obtain a steady vortex crystal, which is shown
by the pseudocolor plots of $\psi$ and  $\Lambda$ in
Figs.~\ref{figch5:R42psilamk}(a) and (b), respectively, for the illustrative
value $\Wi = 1$; for this set of parameters, Fig.~\ref{figch5:R42psilamk}(c)
shows the reciprocal-space spectrum $E_{\Lambda}$, which has clear, dominant
peaks at the forcing wave vectors.

\begin{figure*}[]
  \includegraphics[width=0.32\linewidth]{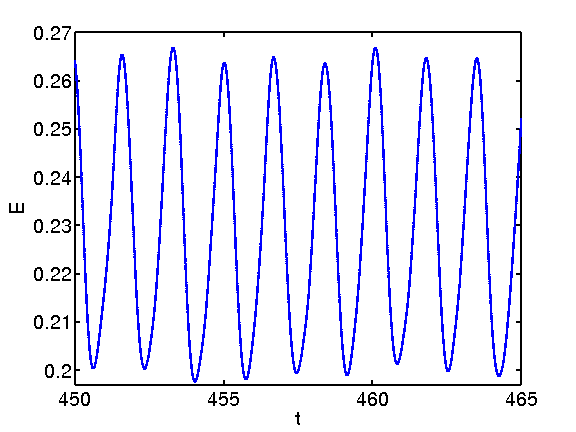}
  \put(-50,95){\color{black}{ {\bf (a)} } }
  \includegraphics[width=0.32\linewidth]{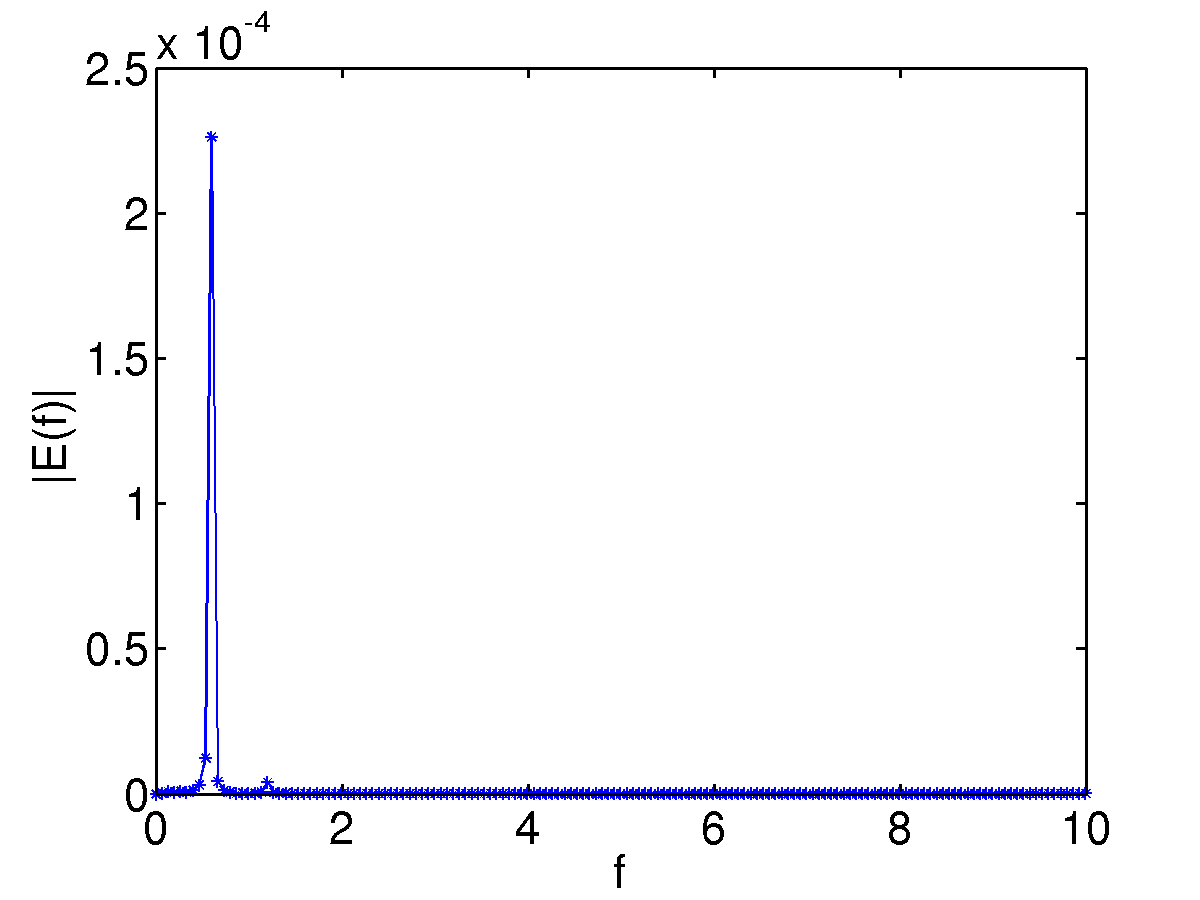}
  \put(-53,95){\color{black}{ {\bf (b)} } }
  \includegraphics[width=0.32\linewidth]{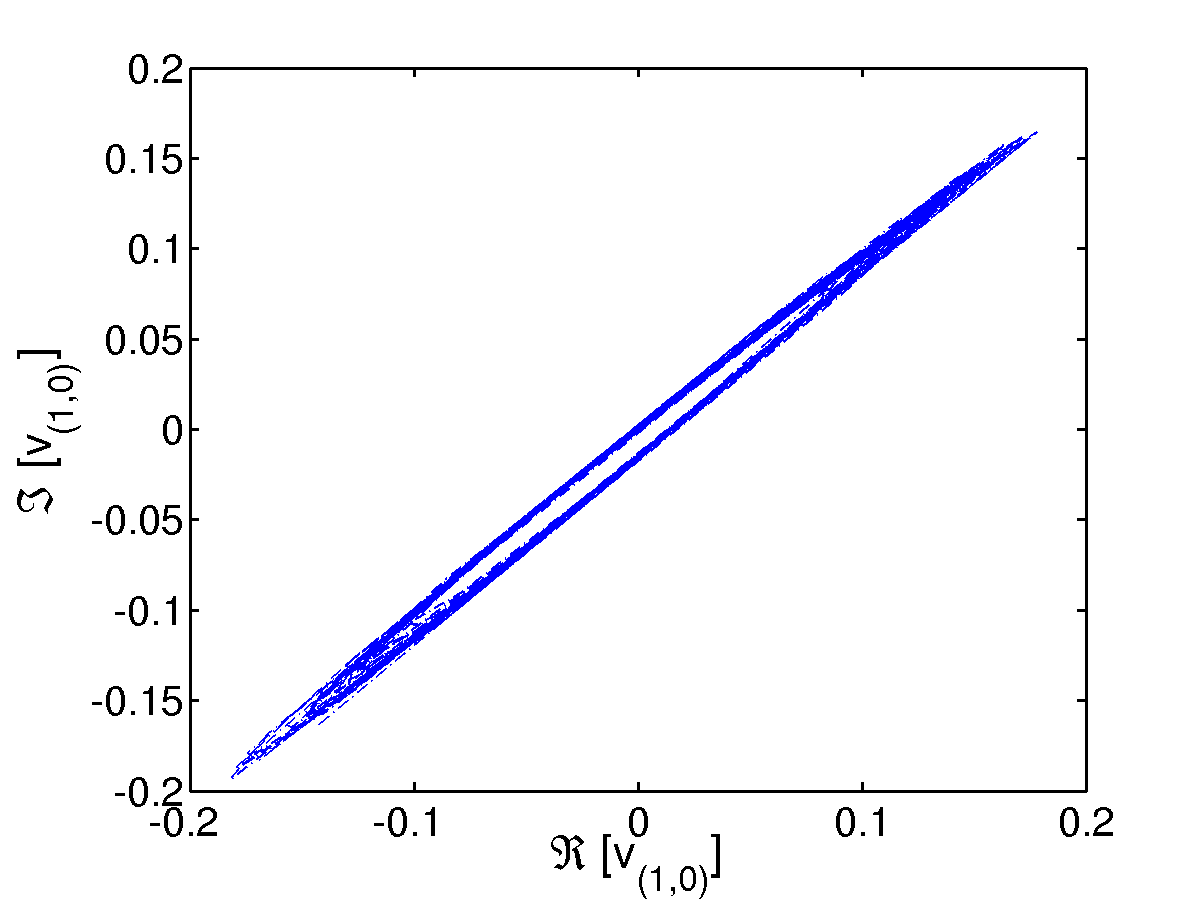}
  \put(-53,95){\color{black}{ {\bf (c)} } }\\
  \includegraphics[width=0.323\linewidth]{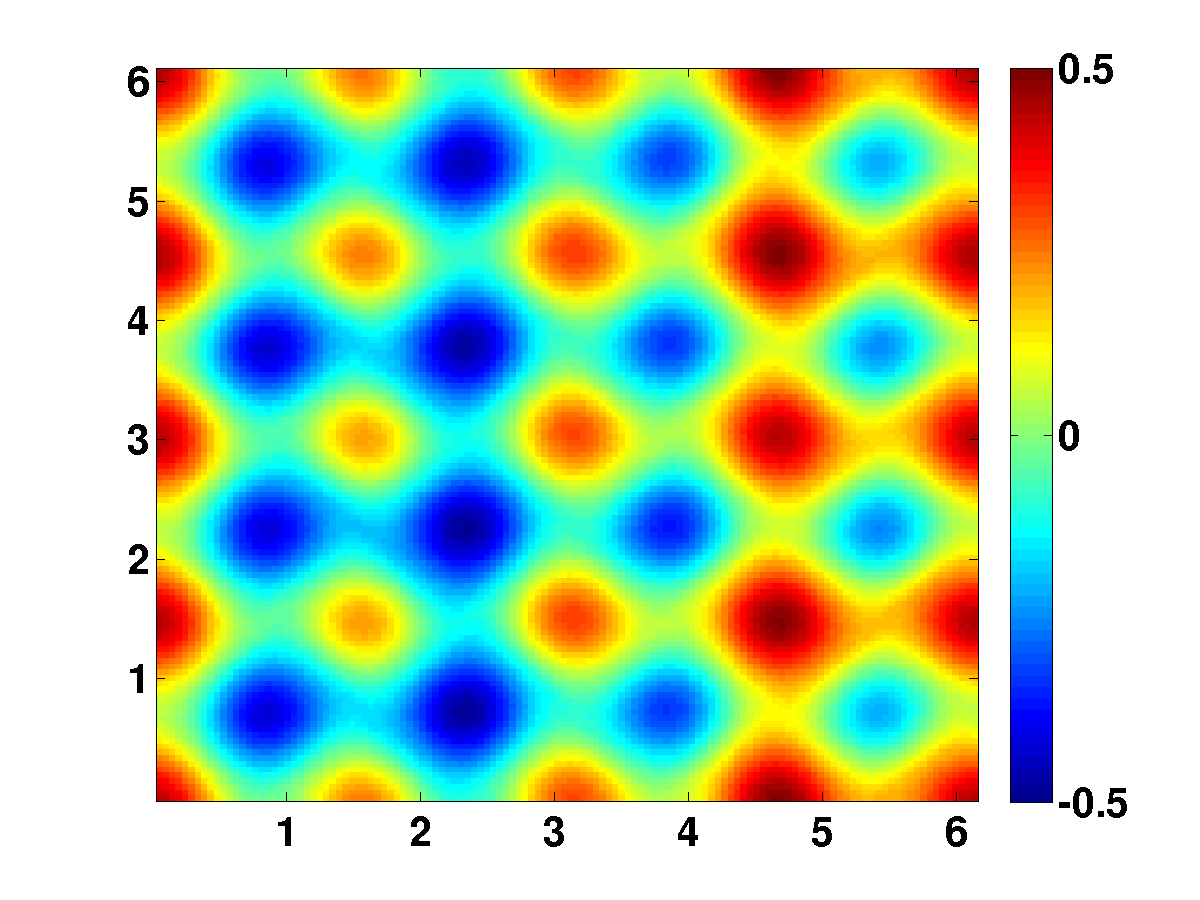}
  \put(-53,95){\color{black}{ {\bf (d)} } }
  \includegraphics[width=0.323\linewidth]{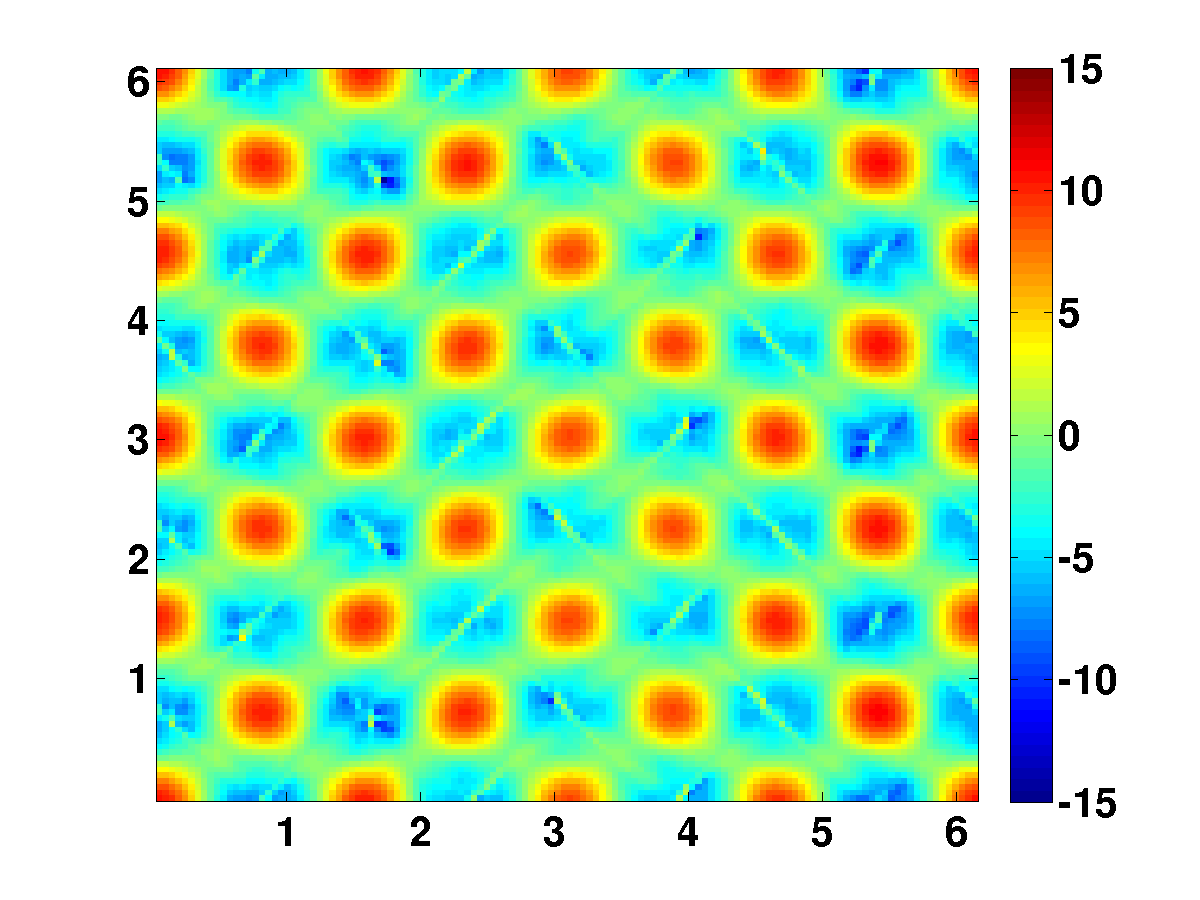}
  \put(-53,95){\color{black}{ {\bf (e)} } }
  \includegraphics[width=0.323\linewidth]{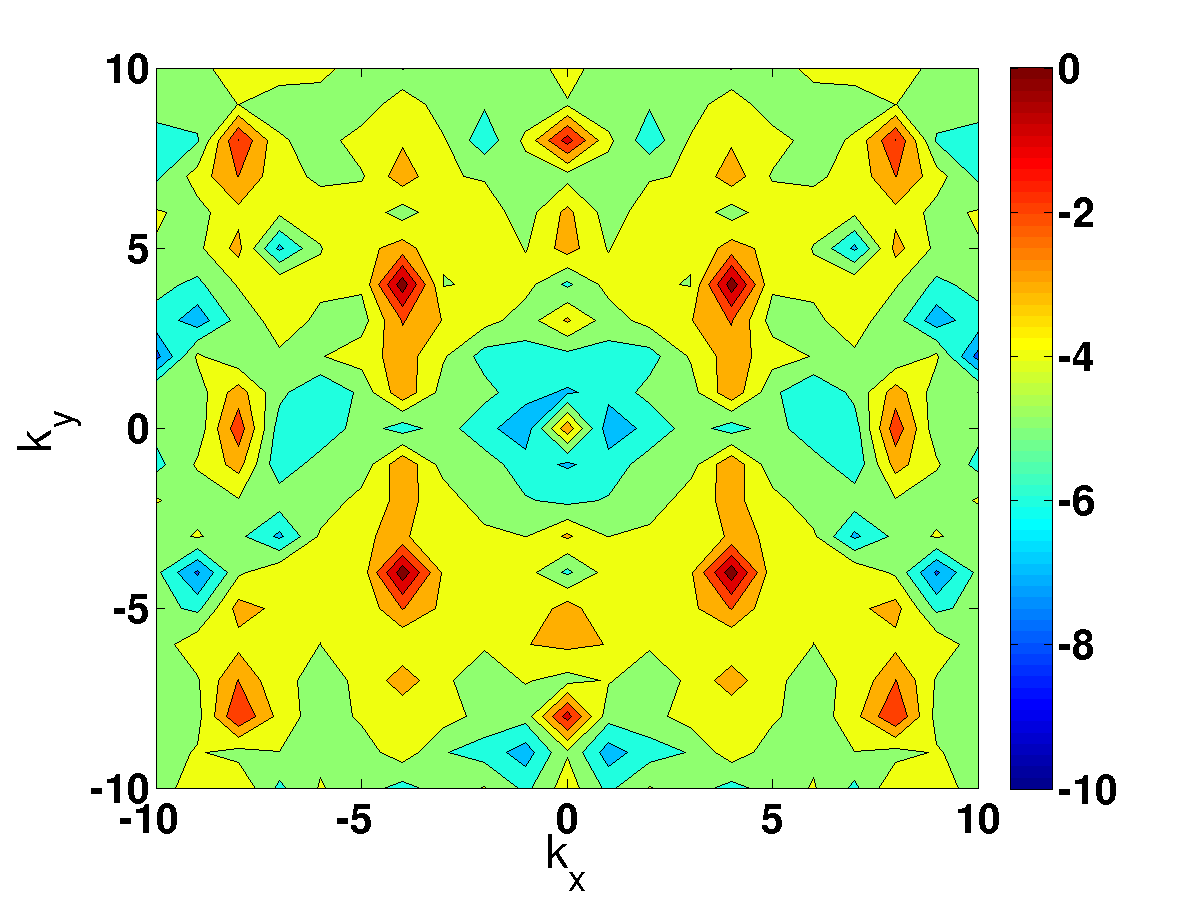}
  \put(-53,99){\color{black}{ {\bf (f)} } }
  \caption{\label{figch5:R43psilamk}(Color online) Plots, for
    $\Omega=1$ and $\Wi=3$ : (a) the time evolution of the energy 
    $E(t)$, (b) its power spectrum $|E(f)|$ versus the frequency $f$, 
    and (c) the \Poincare-type section in the plane 
    $(\Re[\hat{v}(1,0)],\Im[\hat{v}(1,0)])$.
    Pseudocolor plots of (d) the streamfunction
    $\psi$, (e) the Okubo-Weiss parameter $\Lambda$,
    and (f) a filled contour plot of the
    reciprocal-space spectrum $E_{\Lambda}$; the amplitudes of 
    the principal peaks in (f) are lower than the amplitudes of their
    counterparts,  at the forcing wave vectors, in 
    Fig.~\ref{figch5:R42psilamk}(c).}
\end{figure*}

As we increase $\Wi$ beyond $1.9$, new dynamical regimes
appear  in the range of values for our runs $\tt R1-2$. In
particular, we observe, for $\Omega = 1$ and $\Wi=3$, that the
time series of $E(t)$ is periodic (Fig.~\ref{figch5:R43psilamk}
(a)) and its power spectrum $|E(f)|$
(Fig.~\ref{figch5:R43psilamk}(b)) shows one dominant peak, with
hardly any sign of higher harmonics. Thus, the \Poincare-type map
in the $(\Re[\hat{v}(1,0)],\Im[\hat{v}(1,0)])$ plane
(Fig.~\ref{figch5:R43psilamk}(c)) displays a simple attractor.
In Figs.~\ref{figch5:R43psilamk}(d) and (e), we show, for these
parameter values, pseudocolor plots of $\psi$ and $\Lambda$,
respectively; there is spatial undulation in the former 
and the latter is deformed relative to SX (Fig.~\ref{figch5:inlam}(a)).
In the reciprocal-space spectrum
$E_{\Lambda}$ (Figure~\ref{figch5:R43psilamk}(f)), we do not see
any major peaks, other than the ones at the forcing wave vectors,
because the deformation of the original crystal is weak; the
amplitudes of these dominants peaks are smaller than those of
their counterparts, at the forcing wave vectors, in
Fig.\ref{figch5:R42psilamk}(c).

\begin{figure*}[]
  \includegraphics[width=0.32\linewidth]{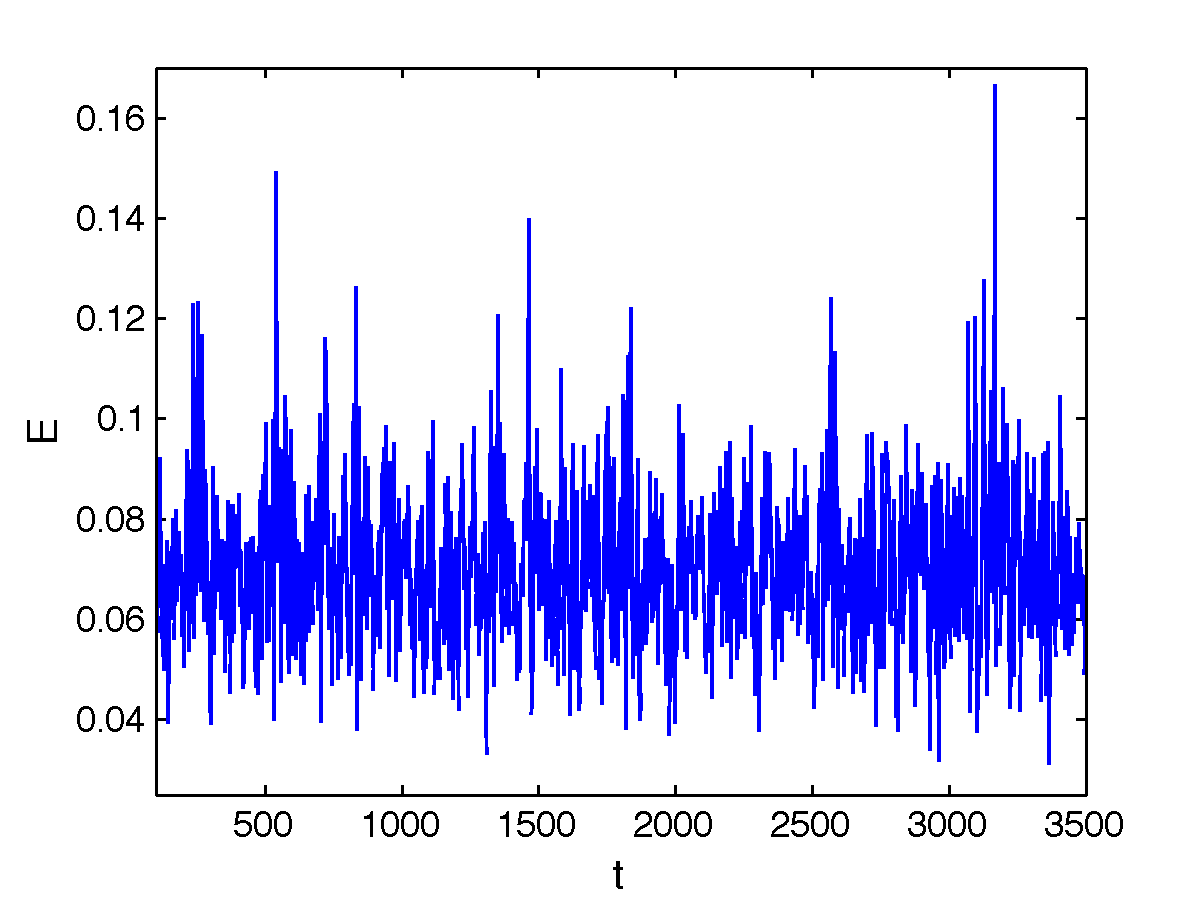}
  \put(-53,95){\color{black}{ {\bf (a)} } }
  \includegraphics[width=0.32\linewidth]{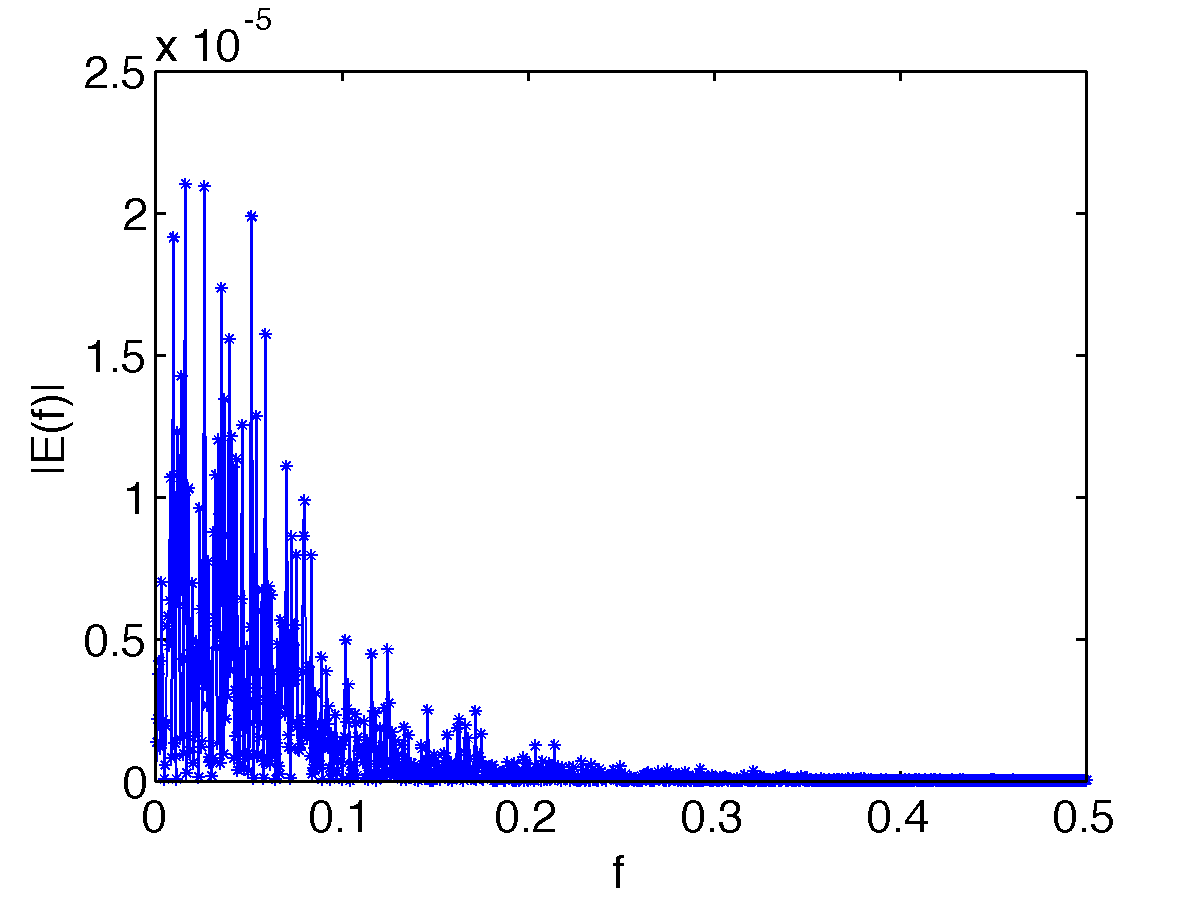}
  \put(-53,95){\color{black}{ {\bf (b)} } }
  \includegraphics[width=0.32\linewidth]{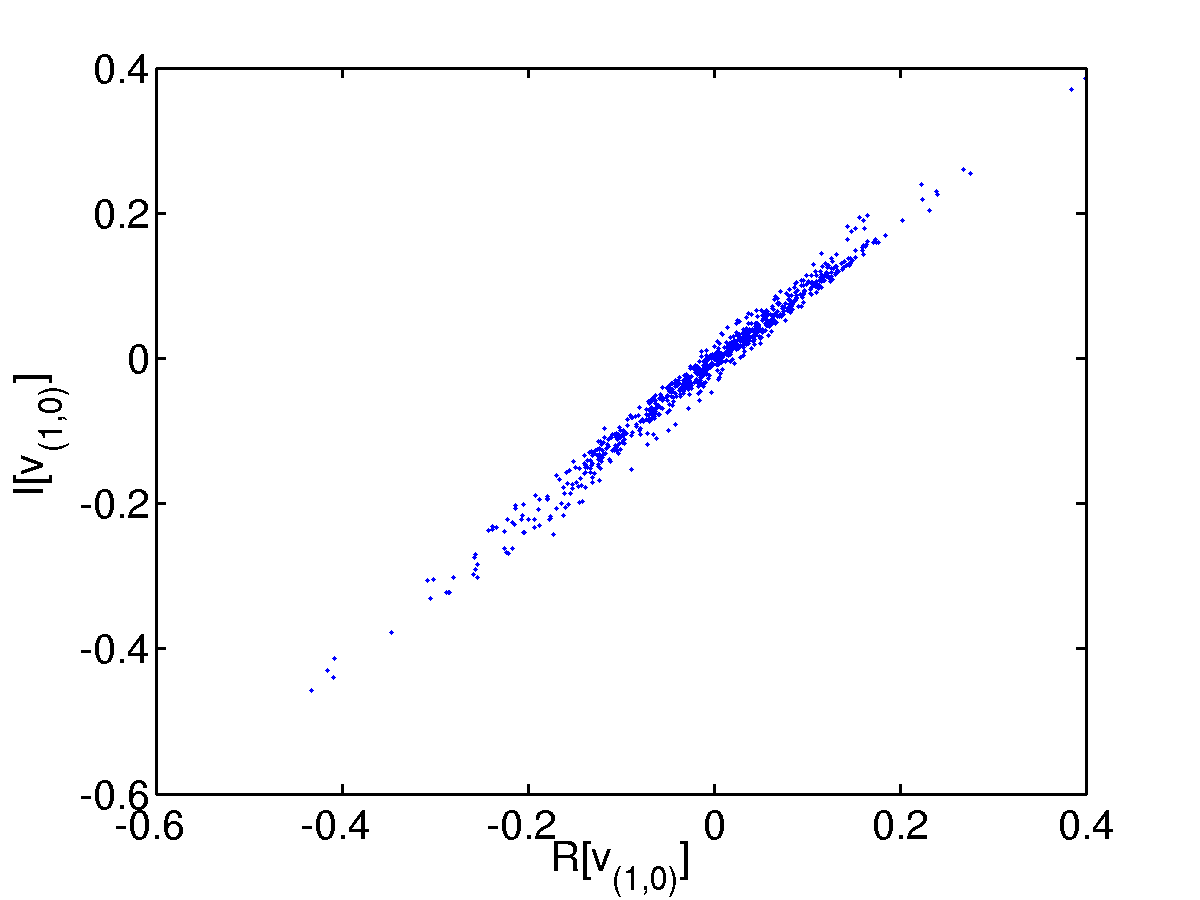}
  \put(-73,95){\color{black}{ {\bf (c)} } }\\
  \includegraphics[width=0.323\linewidth]{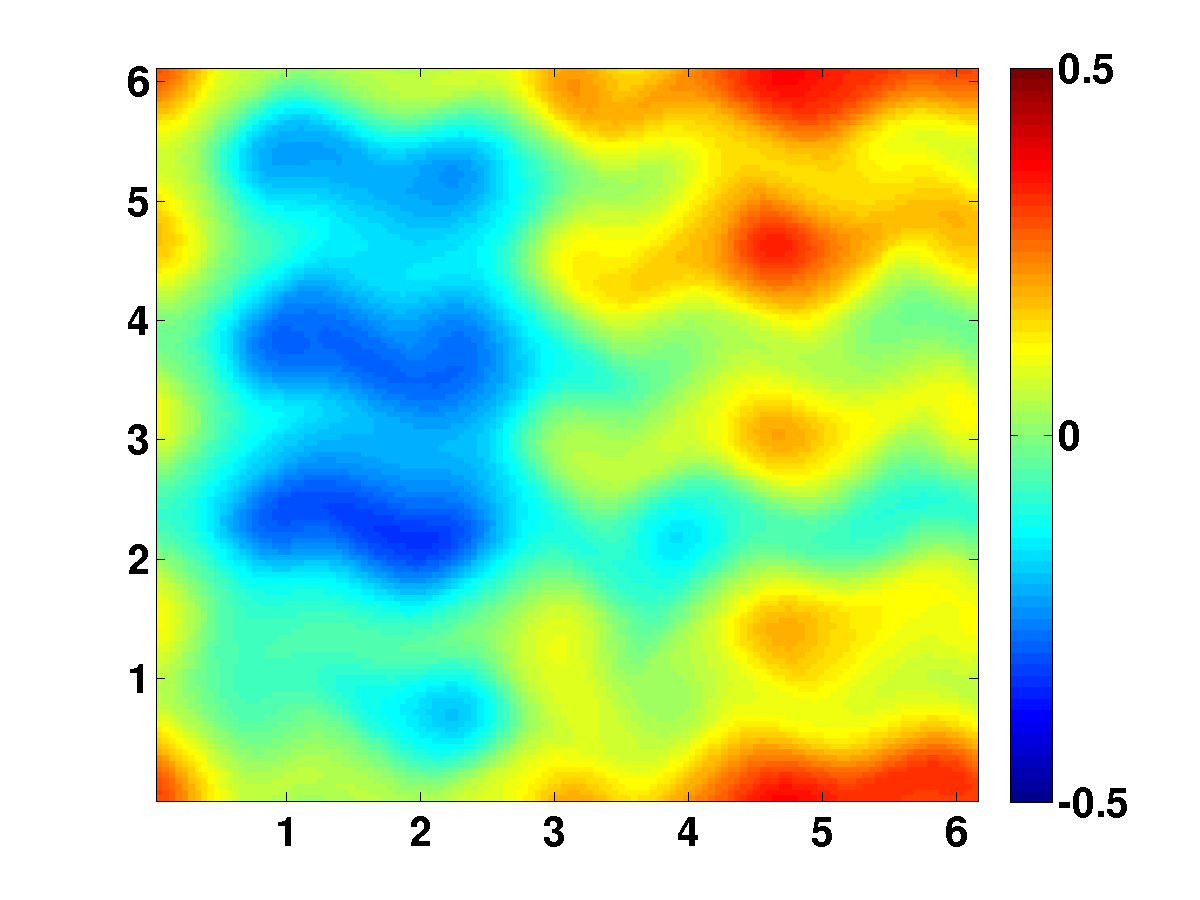}
  \put(-53,95){\color{black}{ {\bf (d)} } }
  \includegraphics[width=0.323\linewidth]{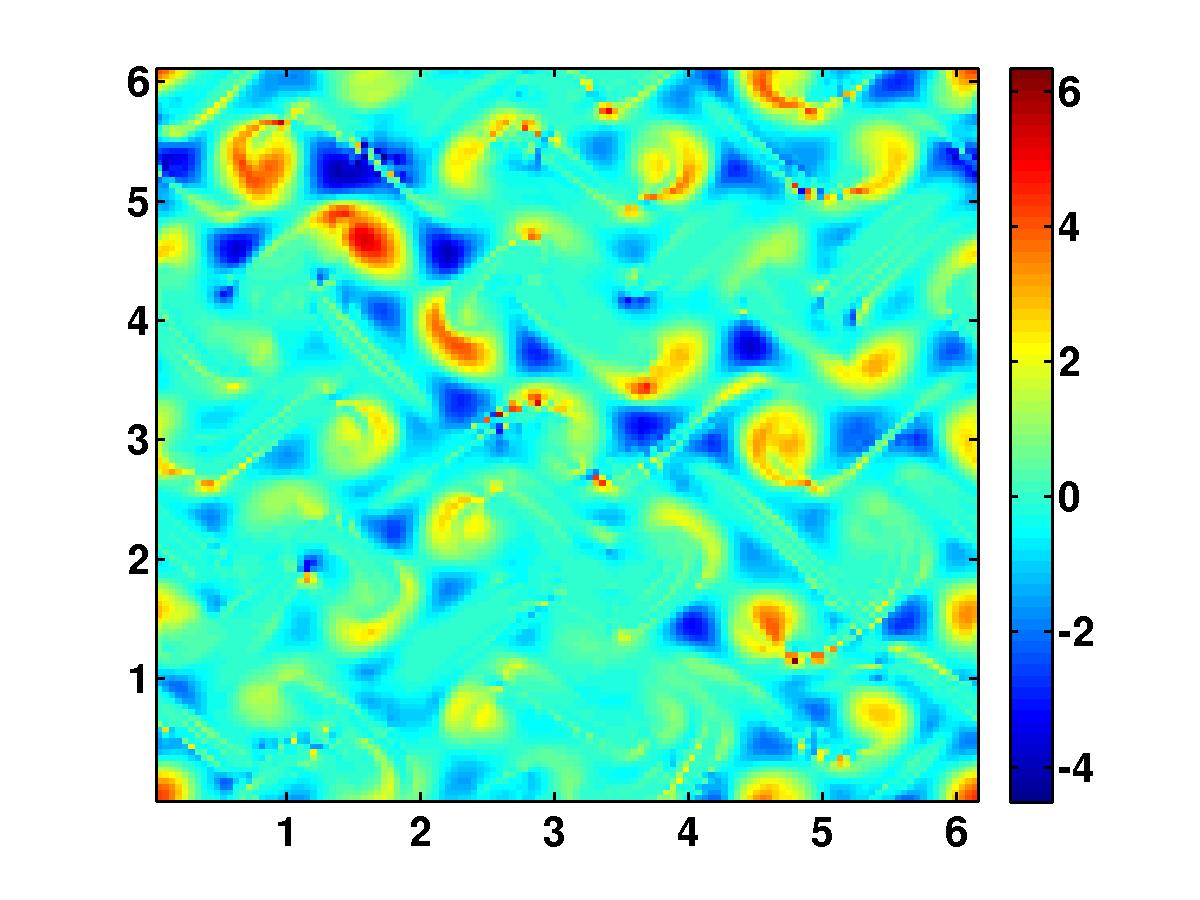}
  \put(-53,95){\color{black}{ {\bf (e)} } }
  \includegraphics[width=0.323\linewidth]{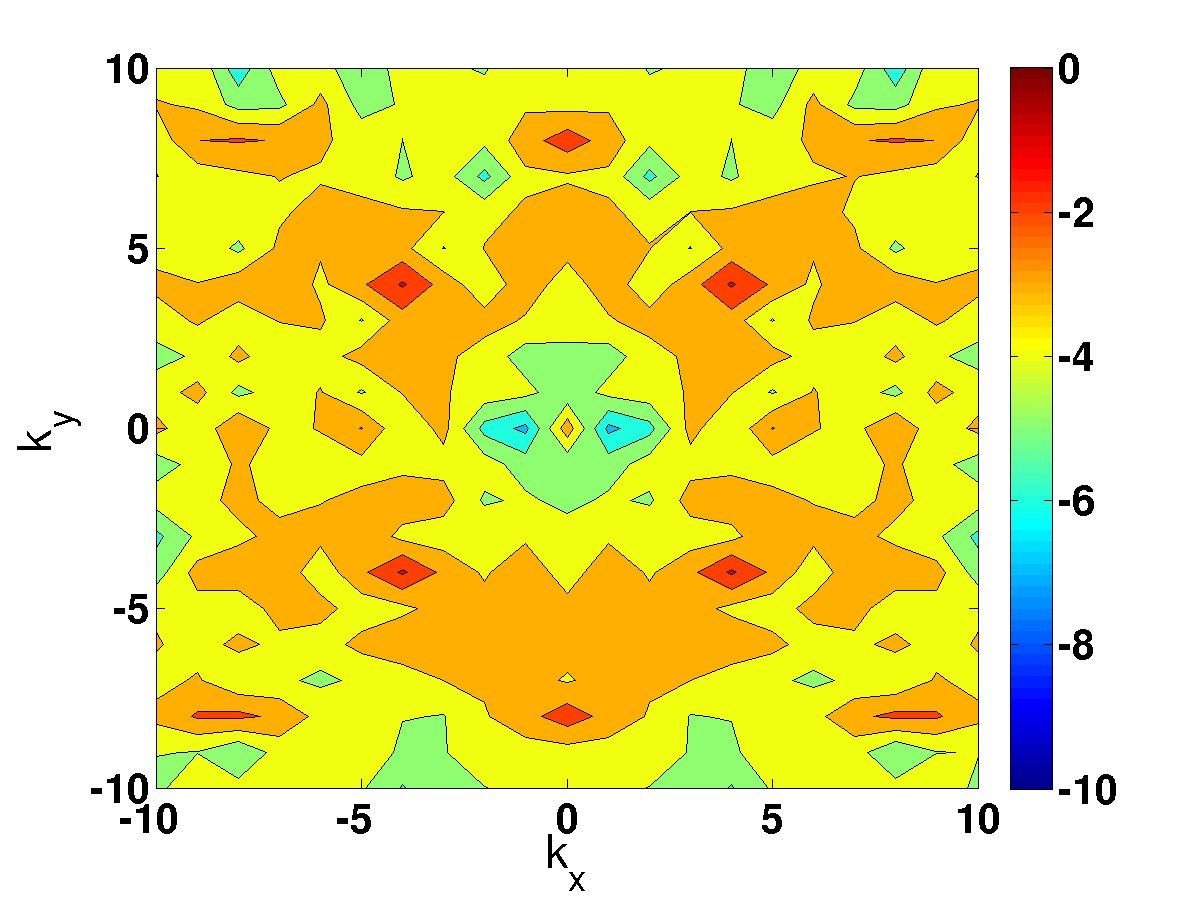}
  \put(-53,95){\color{black}{ {\bf (f)} } }
  \caption{(Color online) Plots for $\Omega=1$ and $\Wi=20$ : (a) the time 
   evolution of the energy $E(t)$, (b) its power spectrum $|E(f)|$ 
   versus the frequency $f$, and (c) the \Poincare-type section in
    the plane $(\Re[\hat{v}(1,0)],\Im[\hat{v}(1,0)])$.
    Pseudocolor plots of (d) the streamfunction
    $\psi$ and (e) the Okubo-Weiss parameter $\Lambda$; and (f) a filled contour plot of the
    reciprocal-space spectrum $E_{\Lambda}$, which shows that 
    various Fourier modes are present in addition to those at 
    the forcing wave vectors.}
  \label{figch5:R44psilamk}
\end{figure*}

An additional increase of $\Wi$ (runs ${\tt R1-3}$ with $\Wi > 10$) leads
to chaotic temporal evolution and a disordered set of vortices and antivortices
in space. Thus, we have states with spatiotemporal chaos and turbulence; for
our nonequilibrium system, turbulent states with spatiotemporal chaos are the
analogs of the disordered liquid state, which appears on the melting of an
equilibrium
crystal~\cite{ramakrishnan1979first,haymet1987theory,chaikin2000principles,oxtoby1991liquids,singh1991density,perlekar2010turbulence}.
We illustrate a state with spatiotemporal chaos for $\Omega =1$ and $\Wi =
20$ by the plots in Fig.~\ref{figch5:R44psilamk}.  In particular,
Fig.~\ref{figch5:R44psilamk}(a) shows the chaotic energy time series $E(t)$,
whose power spectrum $|E(f)|$ (Fig~\ref{figch5:R44psilamk}(b)) shows a broad
background, which indicates temporal chaos. The scattered points in the
\Poincare-type section in the $(\Re[\hat{v}(1,0)],\Im[\hat{v}(1,0)])$ plane
confirm the presence of chaos. The disorder of this state is shown in the
pseudocolor plots of $\psi$ and $\Lambda$ in Figs.~\ref{figch5:R47psilamk}(d)
and (e), respectively; and the associated reciprocal-space spectrum
$E_\Lambda({\bf k})$ (Fig.~\ref{figch5:R47psilamk}(f)) shows many new Fourier
modes in addition to the residual peaks at the forcing wave vectors.

Thus, we see that the onset of elastic turbulence can lead to
vortex-crystal melting, through a very rich sequence of
transitions, even when $\Omega$ is itself too small to melt this
nonequilibrium crystal.  The spatial autocorrelation function
$G({\bf r})$ and the evolution of the order parameters $\langle
{\hat \Lambda_{\bf k}} \rangle$ with $\Wi$ are given in the
fifth subsection.

\subsection{The case $\Omega=22$}

\begin{figure*}[]
  \includegraphics[width=0.32\linewidth]{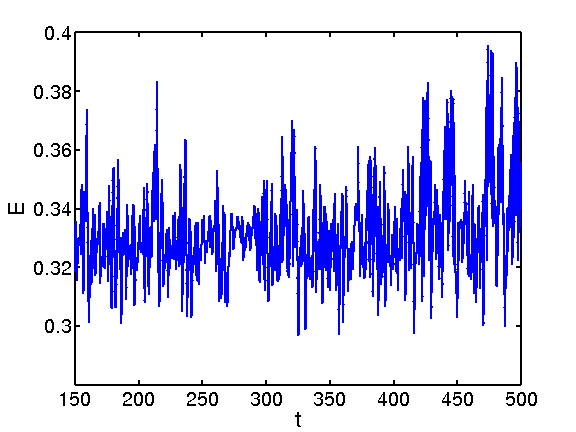}
  \put(-53,95){\color{black}{ {\bf (a)} } }
  \includegraphics[width=0.32\linewidth]{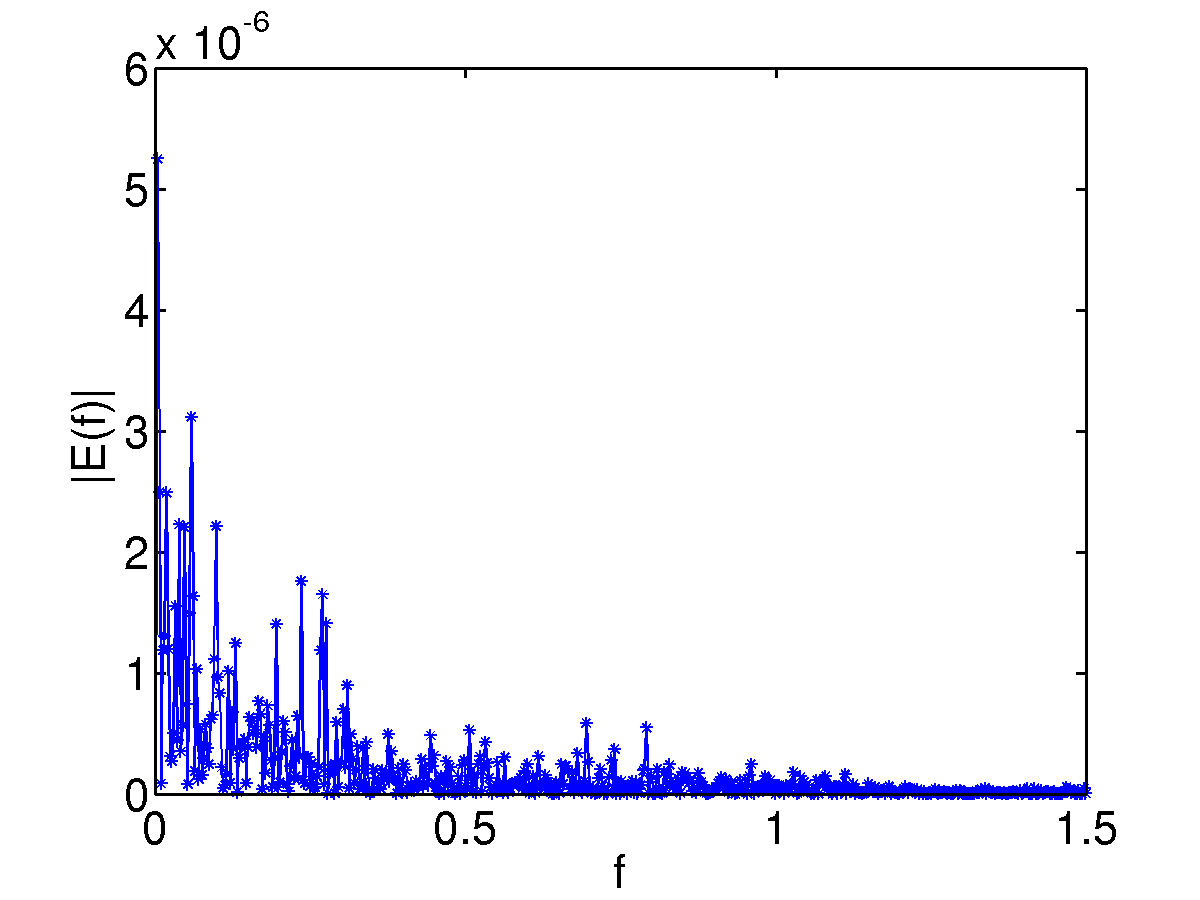}
  \put(-53,95){\color{black}{ {\bf (b)} } }
  \includegraphics[width=0.32\linewidth]{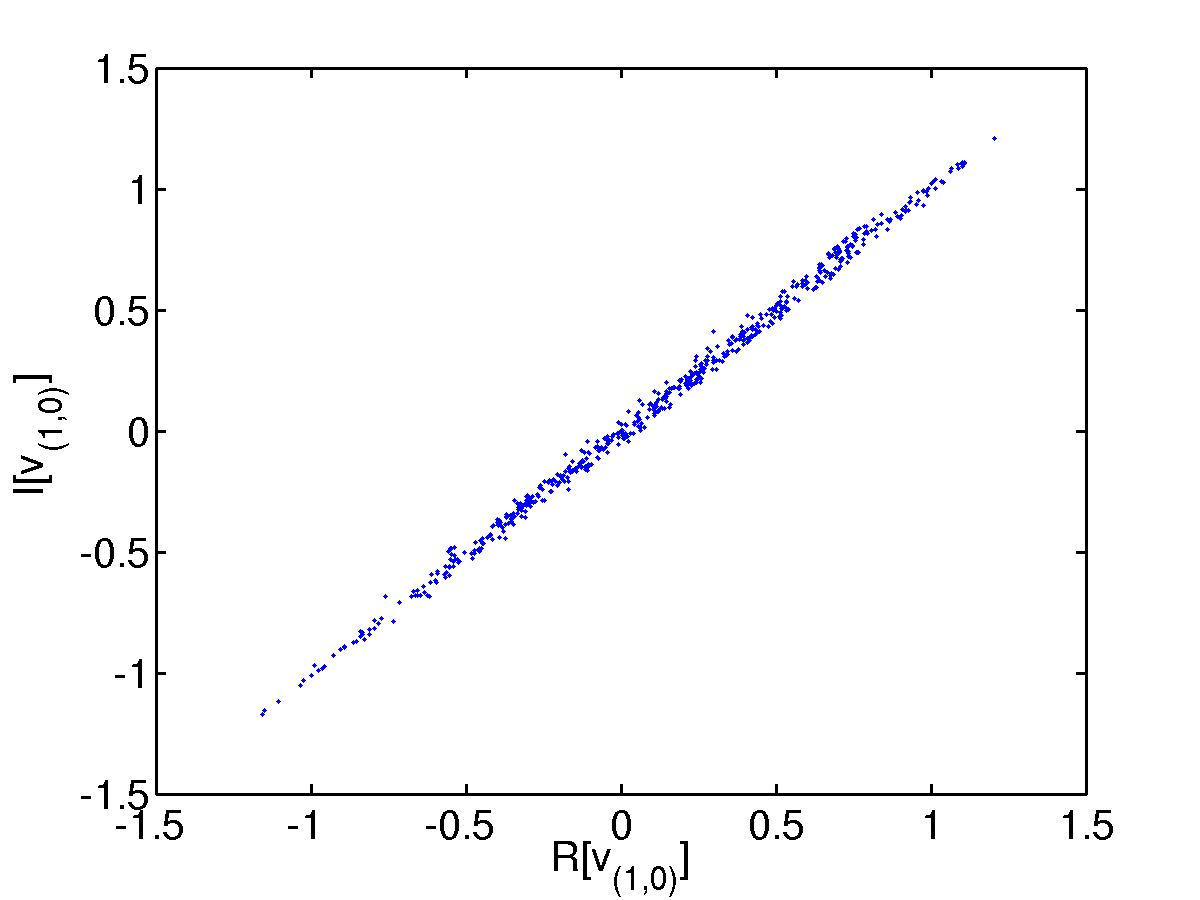}
  \put(-53,95){\color{black}{ {\bf (c)} } }\\
  \includegraphics[width=0.323\linewidth]{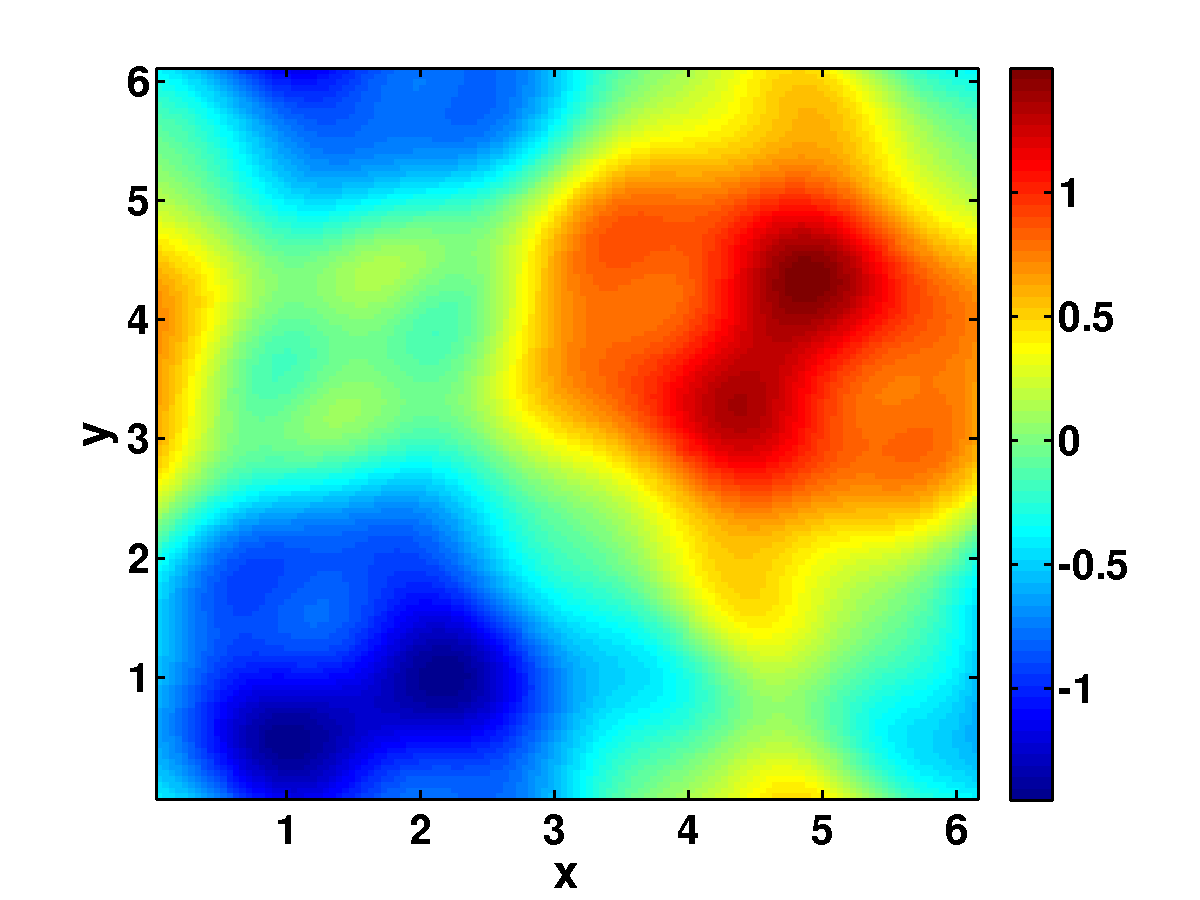}
  \put(-53,95){\color{black}{ {\bf (d)} } }
  \includegraphics[width=0.323\linewidth]{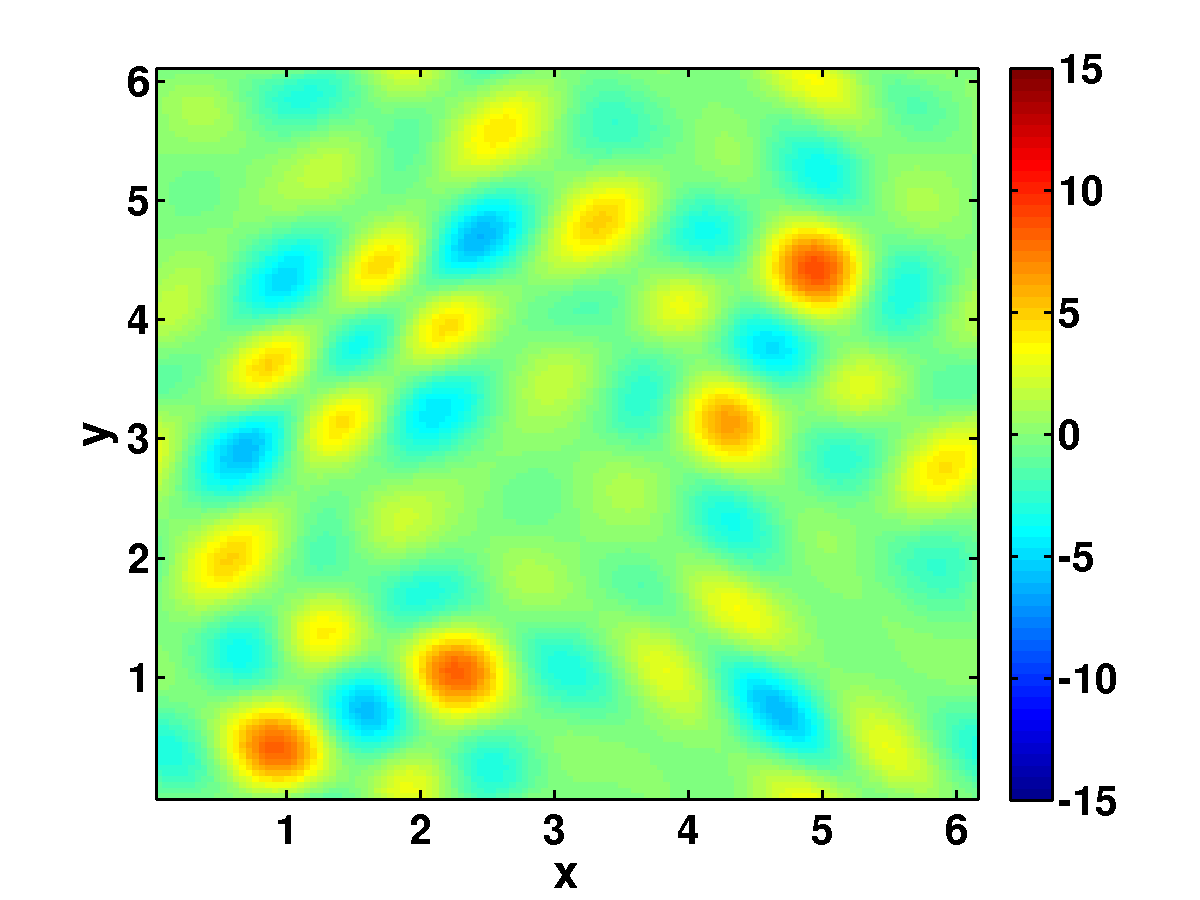}
  \put(-53,95){\color{black}{ {\bf (e)} } }
  \includegraphics[width=0.323\linewidth]{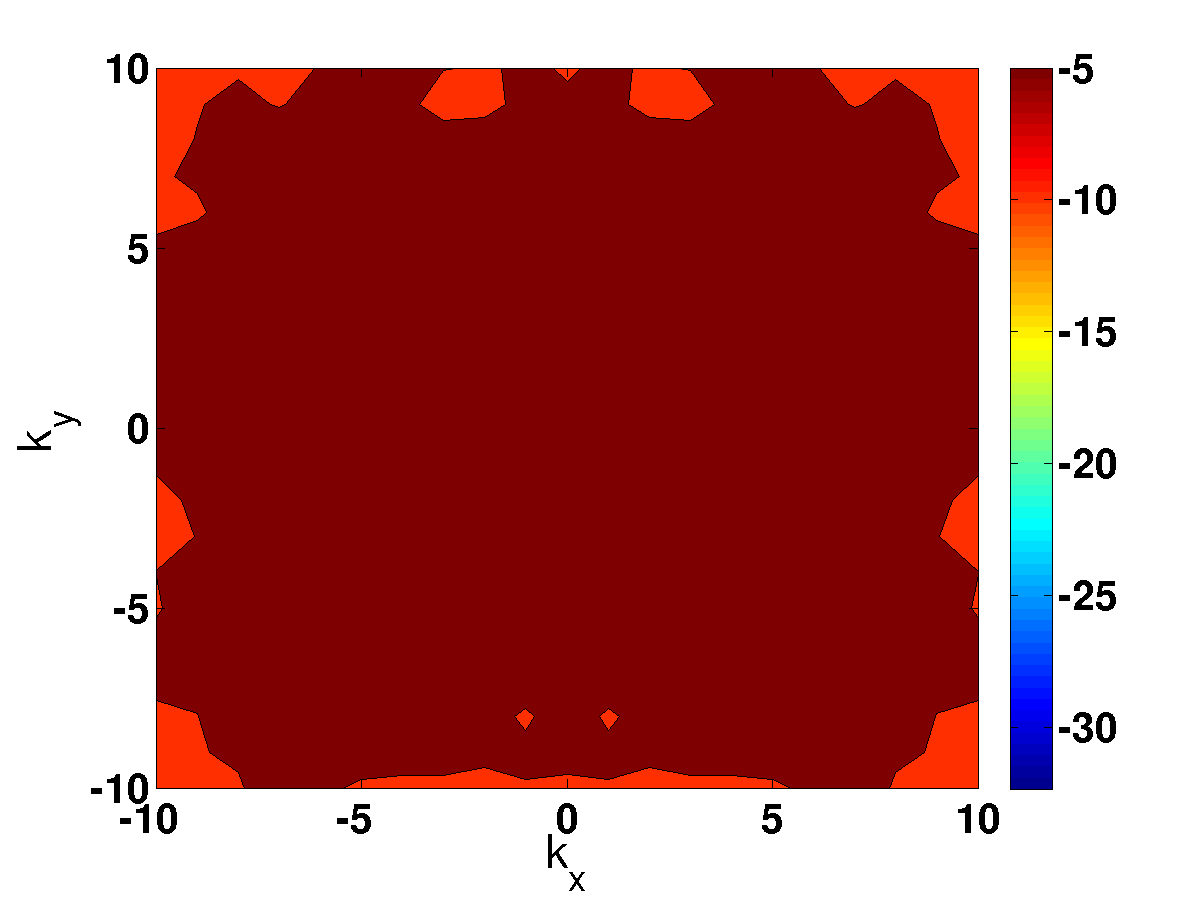}
  \put(-53,95){\color{black}{ {\bf (f)} } }
  \caption{\label{figch5:R45psilamk}(Color online) Plots for $\Omega=22$ and $\Wi=0$ : 
  (a) the time evolution of the energy $E(t)$, (b) its power spectrum 
  $|E(f)|$ versus the frequency $f$, and (c) the \Poincare-type section in
  the plane $(\Re[\hat{v}(1,0)],\Im[\hat{v}(1,0)])$. Pseudocolor plots of 
  (d) the streamfunction $\psi$ and (e) the Okubo-Weiss parameter 
  $\Lambda$; and (f) a filled contour 
  plot of the reciprocal-space spectrum $E_{\Lambda}$, which shows
  a large number of Fourier modes in addition to the ones at the
  forcing wave vectors.}
\end{figure*}

If we do not have any polymers and if $\Omega = 22$, which is greater than
$\Omega_{s,n=4}$, the steady-state solution $\omega_{s,n=4}$ is unstable, and
the temporal evolution of our system is chaotic as we can see from the time
series of the energy in Fig.~\ref{figch5:R45psilamk}(a). Its power spectrum
$|E(f)|$, which is plotted in Fig.~\ref{figch5:R45psilamk}(b), has several
peaks in a broad background, and the \Poincare-type section in the
$(\Re[\hat{v}(1,0)],\Im[\hat{v}(1,0)])$ (Fig.~\ref{figch5:R45psilamk} (c))
displays points that cover a two-dimensional area. The disordered spatial
structure of this state, shown by the pseudocolor plots of $\psi$ and $\Lambda$
in Figs.~\ref{figch5:R45psilamk} (d) and (e), respectively, indicates that this
state is spatiotemporally chaotic and turbulent.
Figure~\ref{figch5:R45psilamk} (f) shows the reciprocal-space energy spectrum
$E_{\Lambda}$, which exhibits a large number of Fourier modes in addition to
the ones at the forcing wave vectors.

\begin{figure*}[]
\includegraphics[width=0.32\linewidth]{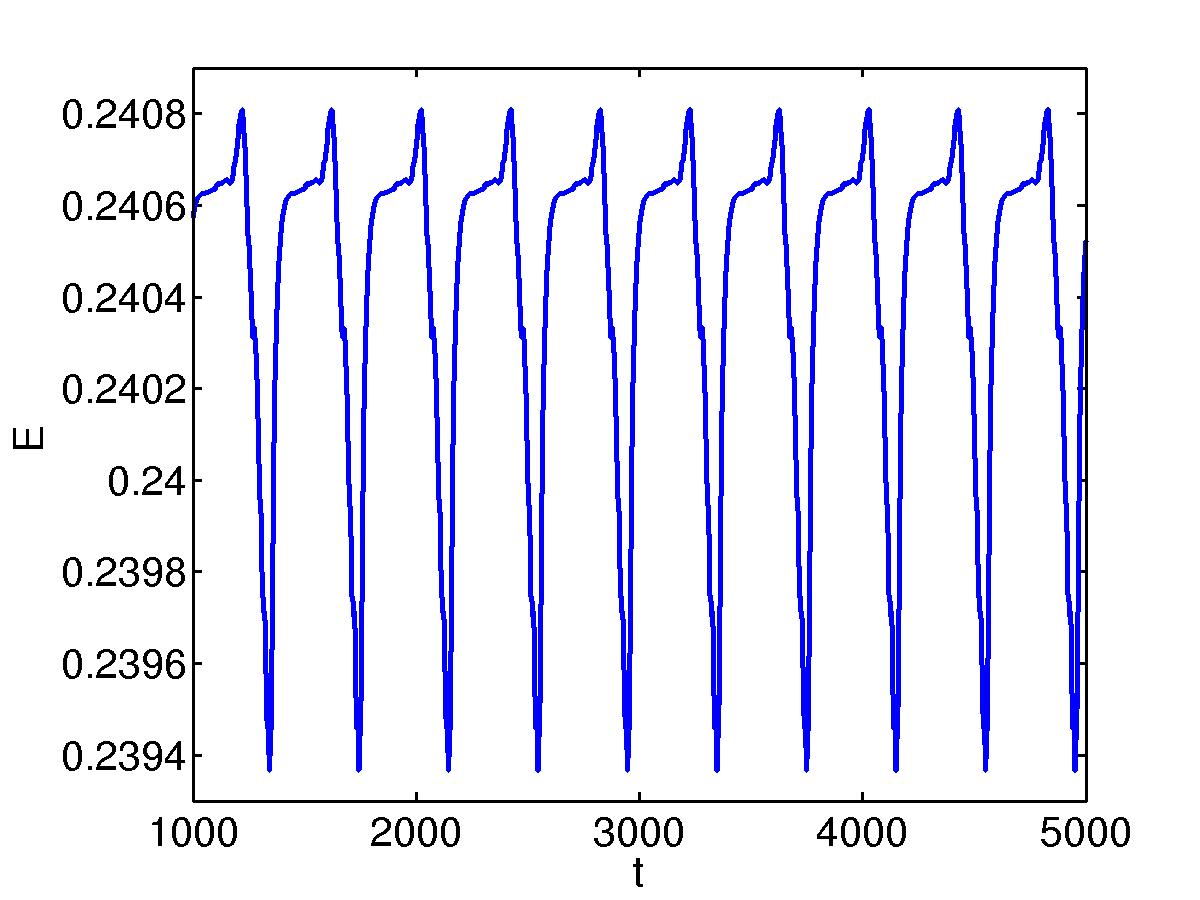}
  \put(-53,95){\color{black}{ {\bf (a)} } }
\includegraphics[width=0.32\linewidth]{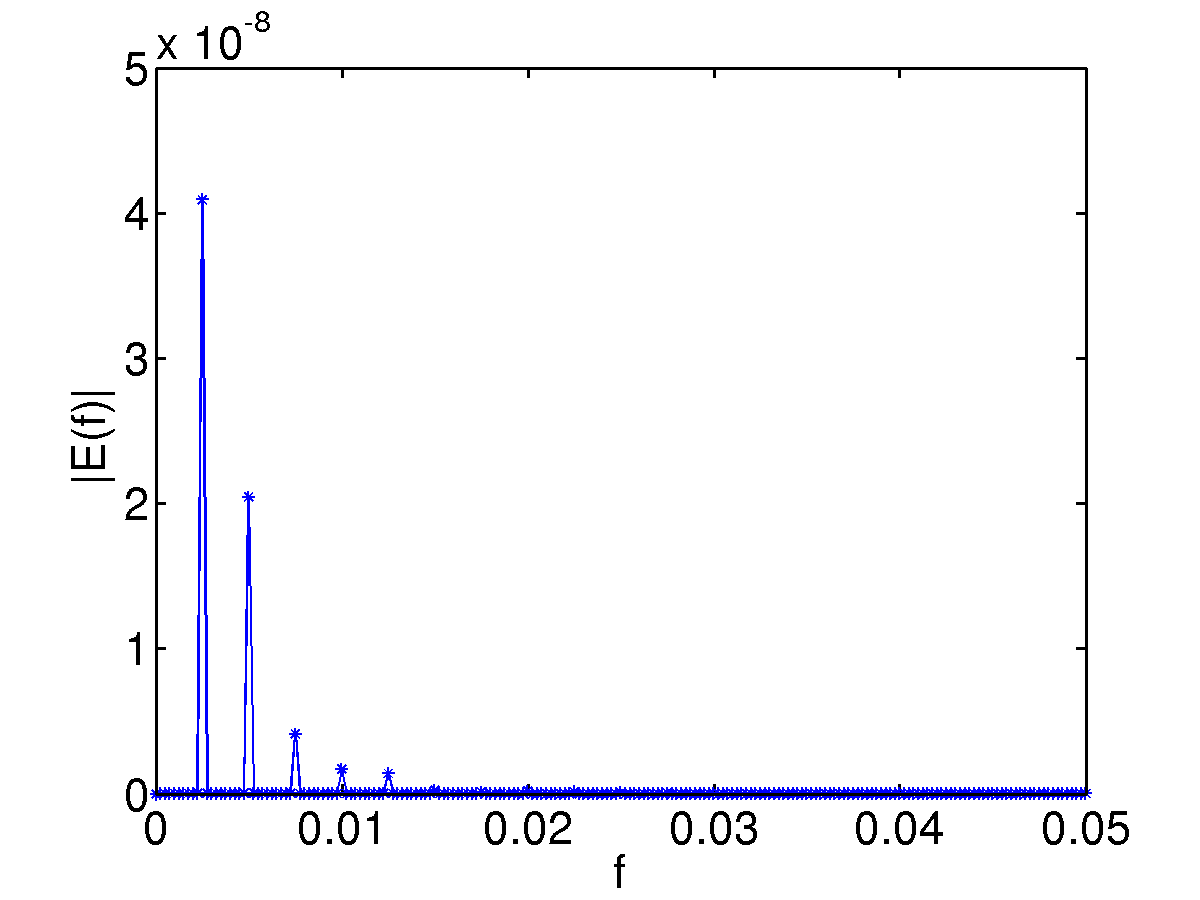}
  \put(-53,95){\color{black}{ {\bf (b)} } }
\includegraphics[width=0.32\linewidth]{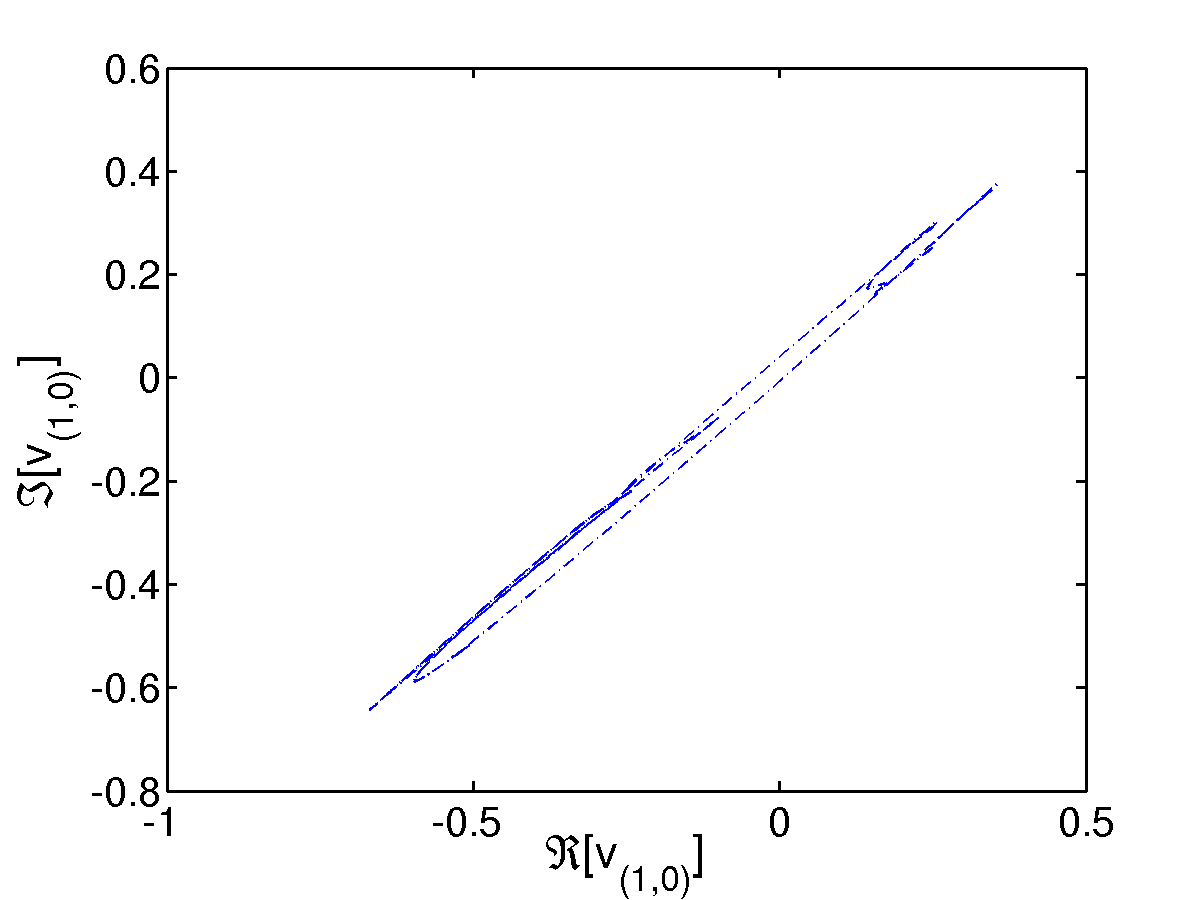}
  \put(-53,95){\color{black}{ {\bf (c)} } }\\
\includegraphics[width=0.323\linewidth]{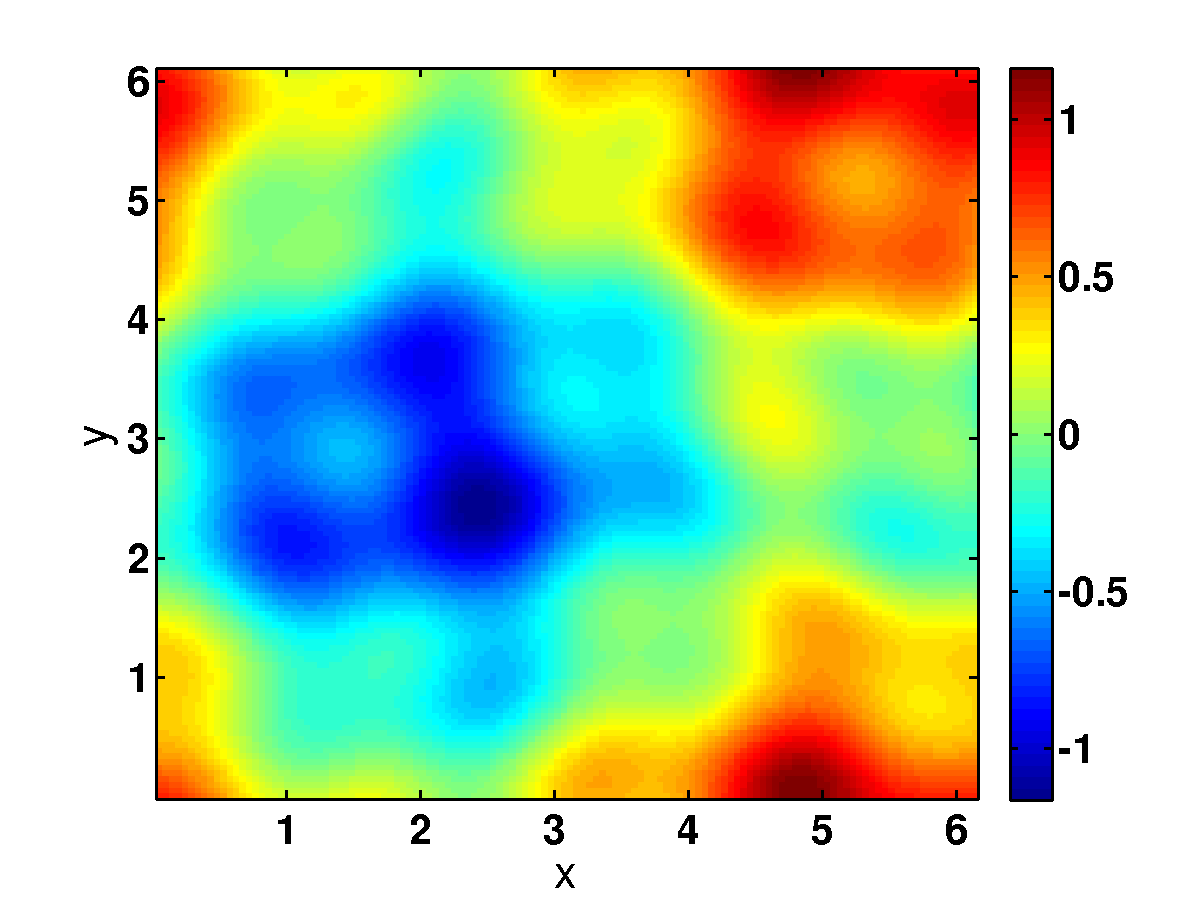}
  \put(-53,95){\color{black}{ {\bf (d)} } }
\includegraphics[width=0.323\linewidth]{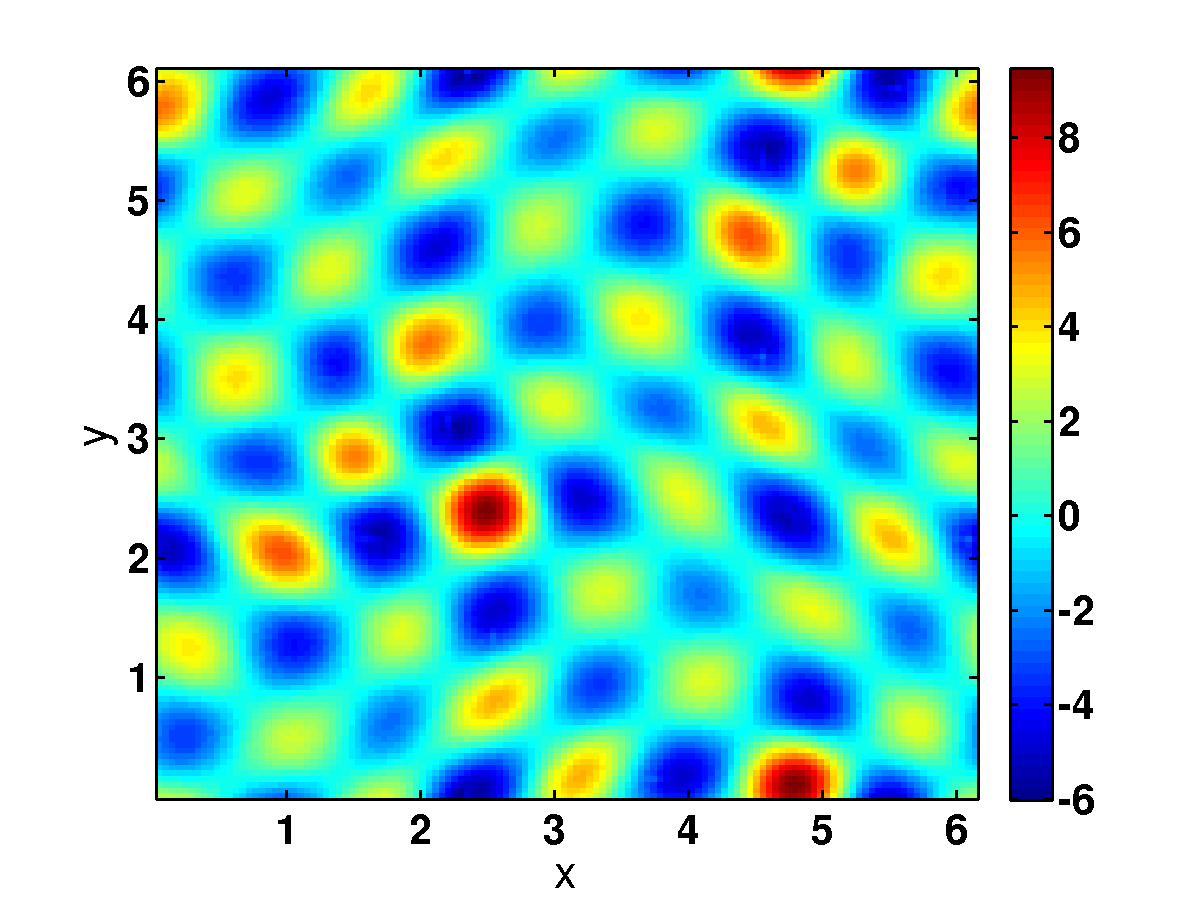}
  \put(-53,95){\color{black}{ {\bf (e)} } }
\includegraphics[width=0.323\linewidth]{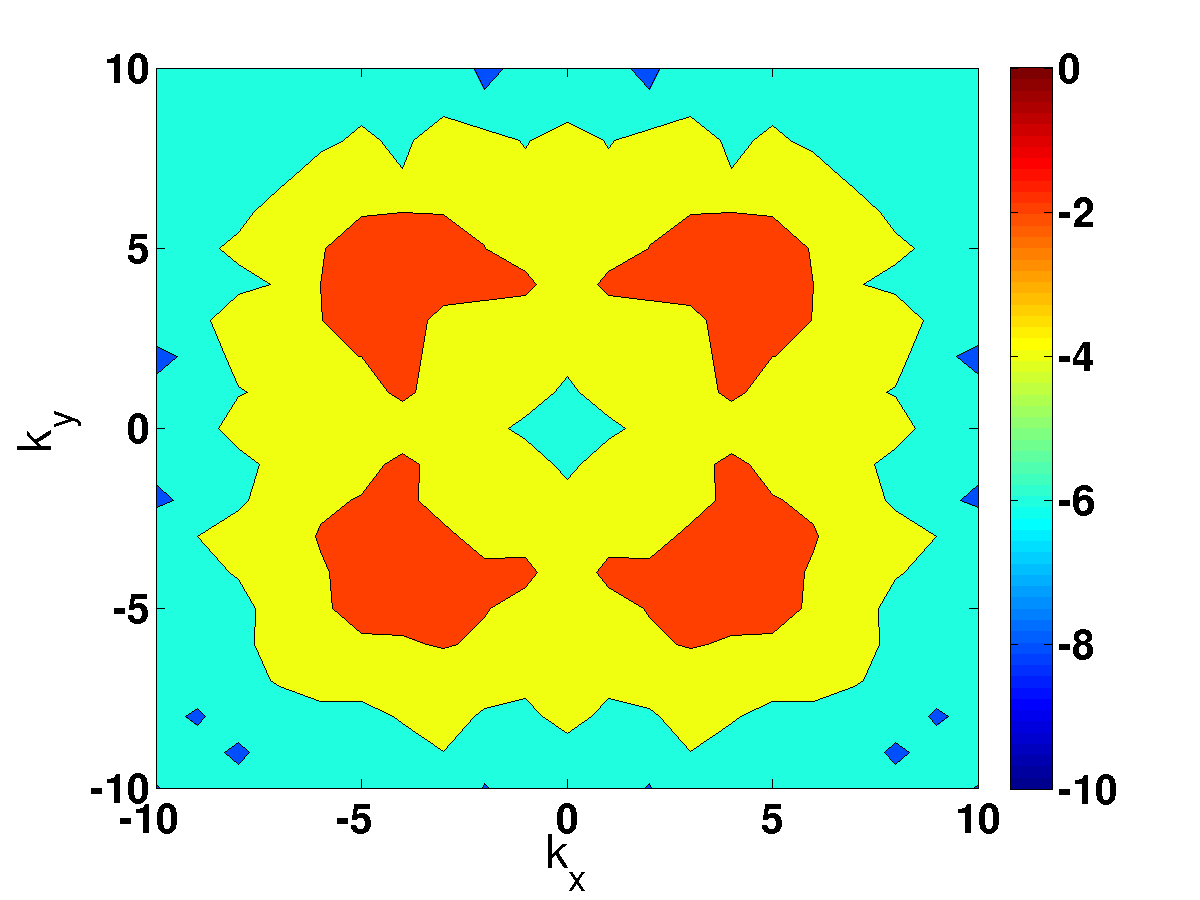}
  \put(-53,95){\color{black}{ {\bf (f)} } }
\caption{\label{figch5:R46psilamk}(Color online) Plots for
$\Omega=22$ and $\Wi=0.3$ : (a) the time evolution of the energy $E(t)$, (b)
the power spectrum $|E(f)|$ versus the frequency $f$, and (c) the 
\Poincare-type section in the plane
$(\Re[\hat{v}(1,0)],\Im[\hat{v}(1,0)])$. Pseudocolor plots of (d) the
streamfunction $\psi$ and (e) the Okubo-Weiss parameter $\Lambda$; (f) a filled contour plot of the
reciprocal-space spectrum $E_{\Lambda}$.} 
\end{figure*}

When we add polymers to our 2D Navier-Stokes system at $\Omega =
22$, these polymers lead first to a reappearance of states that
are periodic in time; this occurs, e.g., for the range of
$\Wi$ covered in runs $\tt R22-2$. For $\Wi = 0.3$,
Fig.~\ref{figch5:R46psilamk}(a) shows that time series of the
energy is periodic in time and its power spectrum $|E(f)|$
(Fig.~\ref{figch5:R46psilamk}(b)) has a fundamental peak at
$f_0 = 2.5 \times 10^{-3}$ and harmonics at $f_1
= 2 f_0$, $f_2 = 3 f_0$, $f_4 = 4 f_0$, and $f_5 = 5 f_0$. The
\Poincare-type section in the
$(\Re[\hat{v}(1,0)],\Im[\hat{v}(1,0)])$ plane in
Fig.~\ref{figch5:R46psilamk} (c) exhibits a simple attractor,
which is consistent with the periodic behaviors that emerge from
Figs.~\ref{figch5:R46psilamk}(a) and (b). The spatial
organization of this state, shown by pseudocolor plots of $\psi$
and $\Lambda$ in Figs.~\ref{figch5:R46psilamk} (d) and (e),
respectively, is disordered, although some vestiges of the unit
cells of the original crystal are visible in the pseudocolor plot
of $\Lambda$; and Fig.~\ref{figch5:R46psilamk} (f) shows the
reciprocal-space spectrum $E_{\Lambda}$. 

\begin{figure*}[]
\includegraphics[width=0.323\linewidth]{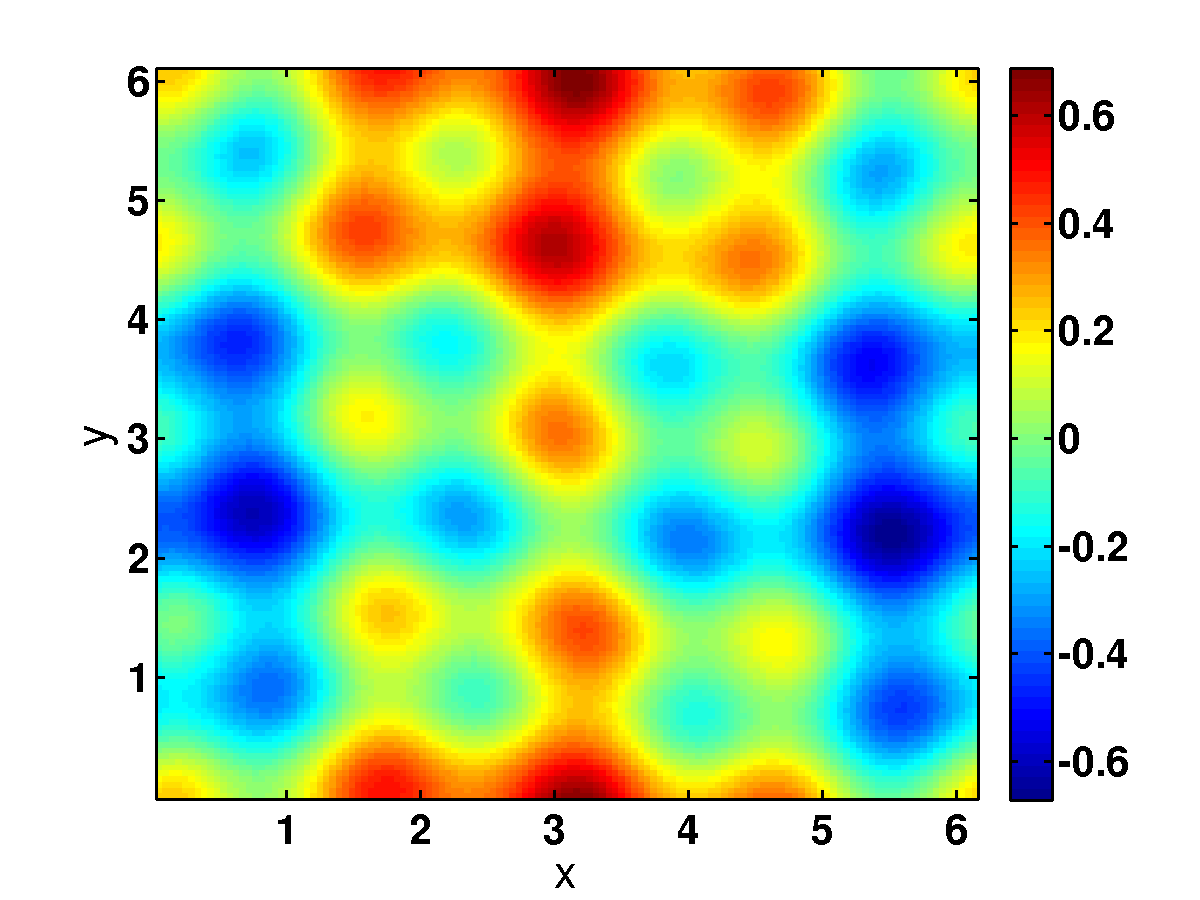}
  \put(-53,95){\color{black}{ {\bf (a)} } }
\includegraphics[width=0.323\linewidth]{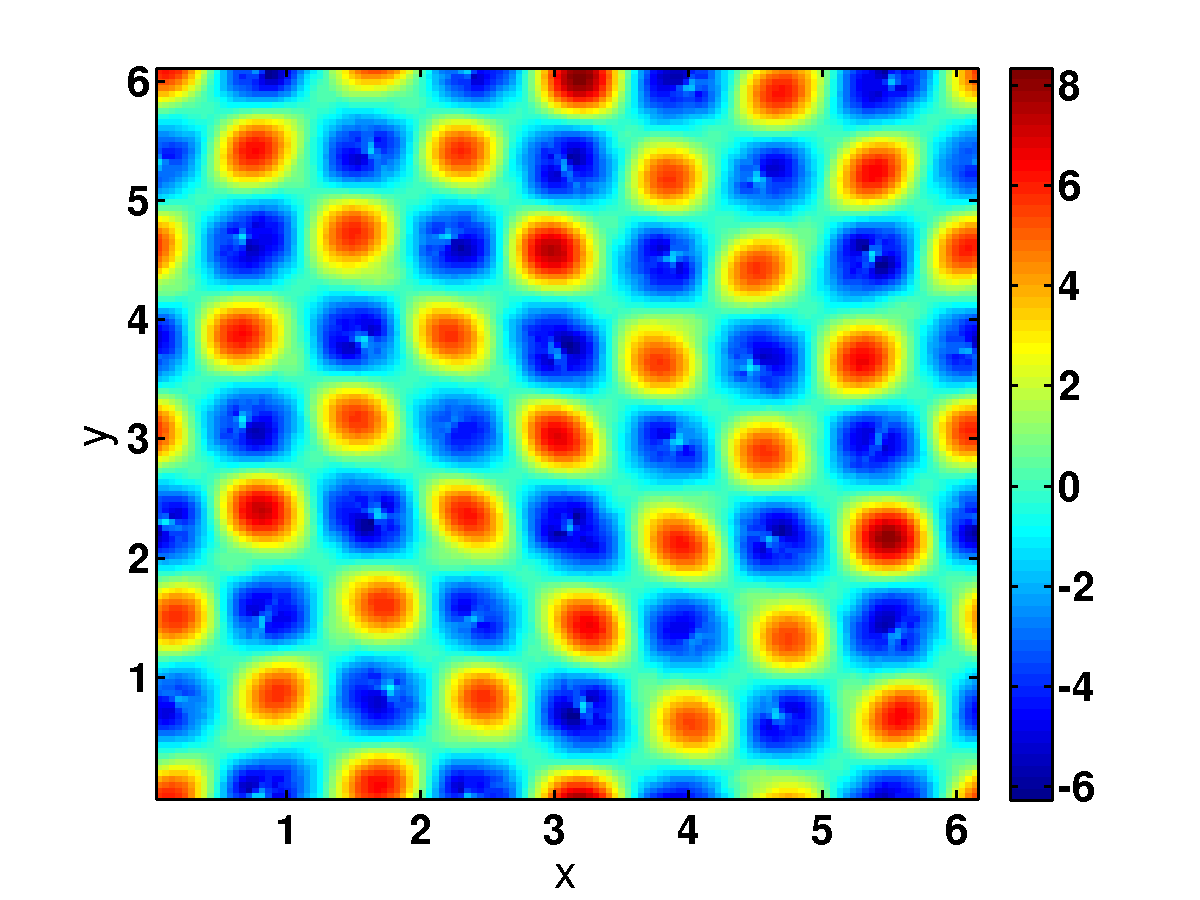}
  \put(-53,95){\color{black}{ {\bf (b)} } }
\includegraphics[width=0.323\linewidth]{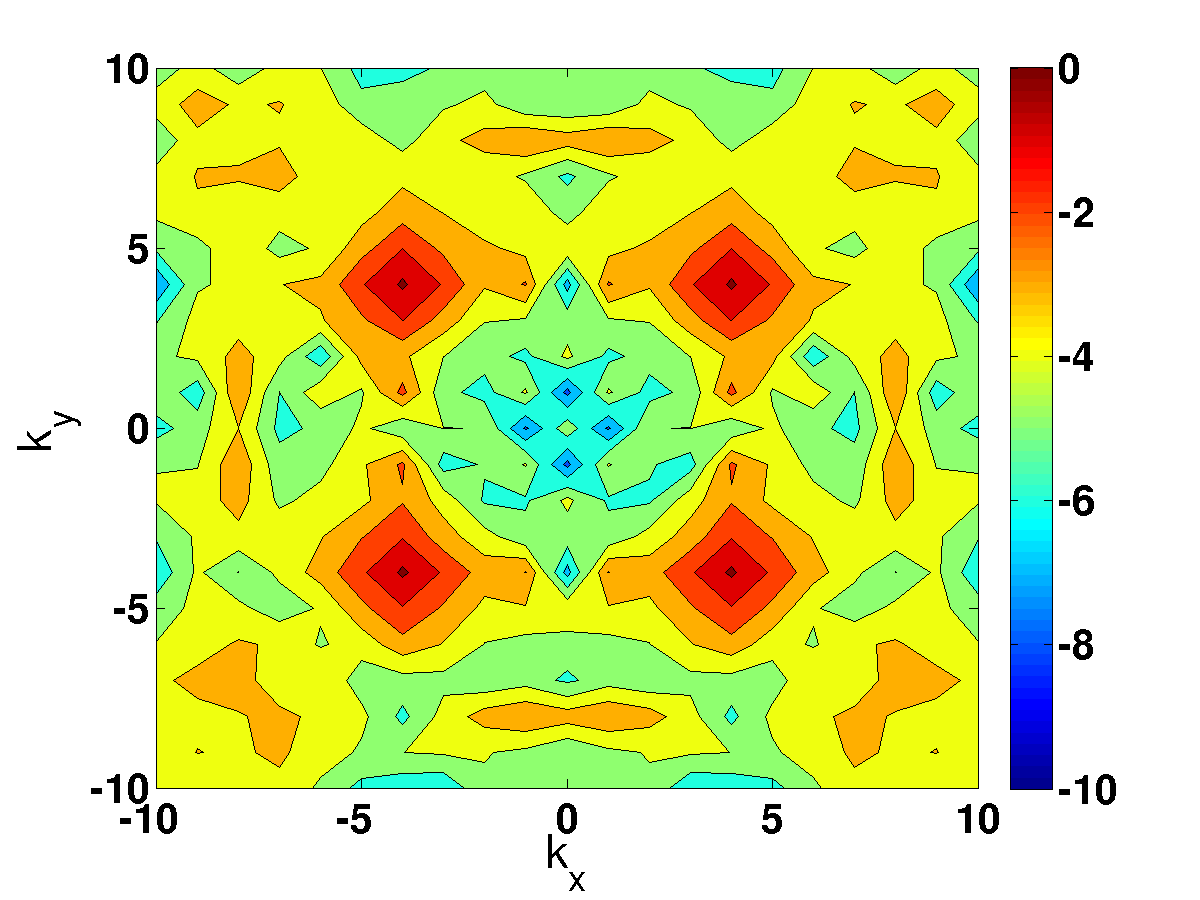}
  \put(-53,95){\color{black}{ {\bf (c)} } }
\caption{\label{figch5:R47psilamk}(Color online) Pseudocolor plot for $\Omega=22$ and
$\Wi=0.5$ : (a) the streamfunction $\psi$ and (b) the Okubo-Weiss parameter
$\Lambda$; and (c) a filled contour plot of the reciprocal-space spectrum
$E_{\Lambda}$, which shows peaks at the forcing wave vectors, but with
amplitudes that are lower than for their counterparts in
Fig.\ref{figch5:R42psilamk}(c).}
\end{figure*}

If we increase the Weissenberg number to $\Wi = 0.5$ in run $\tt
R22-3$, we obtain a new steady state that is a vortex crystal; however, it is
not exactly the same as the original vortex crystal as can be seen clearly by
comparing the pseudocolor plots of $\Lambda$ in Figs.~\ref{figch5:inlam}(a) and
~\ref{figch5:R47psilamk}(b). This difference is even more striking if we
compare the pseudocolor plots of $\psi$ shown in
Figs.~\ref{figch5:R47psilamk}(a) and \ref{figch5:inpsi}(a). Not
surprisingly, then, the reciprocal-space energy spectrum $E_{\Lambda}$ shows
new peaks in addition to the ones at the forcing wave vectors, which now appear
with a reduced amplitude. 

\begin{figure*}[]
  \includegraphics[width=0.32\linewidth]{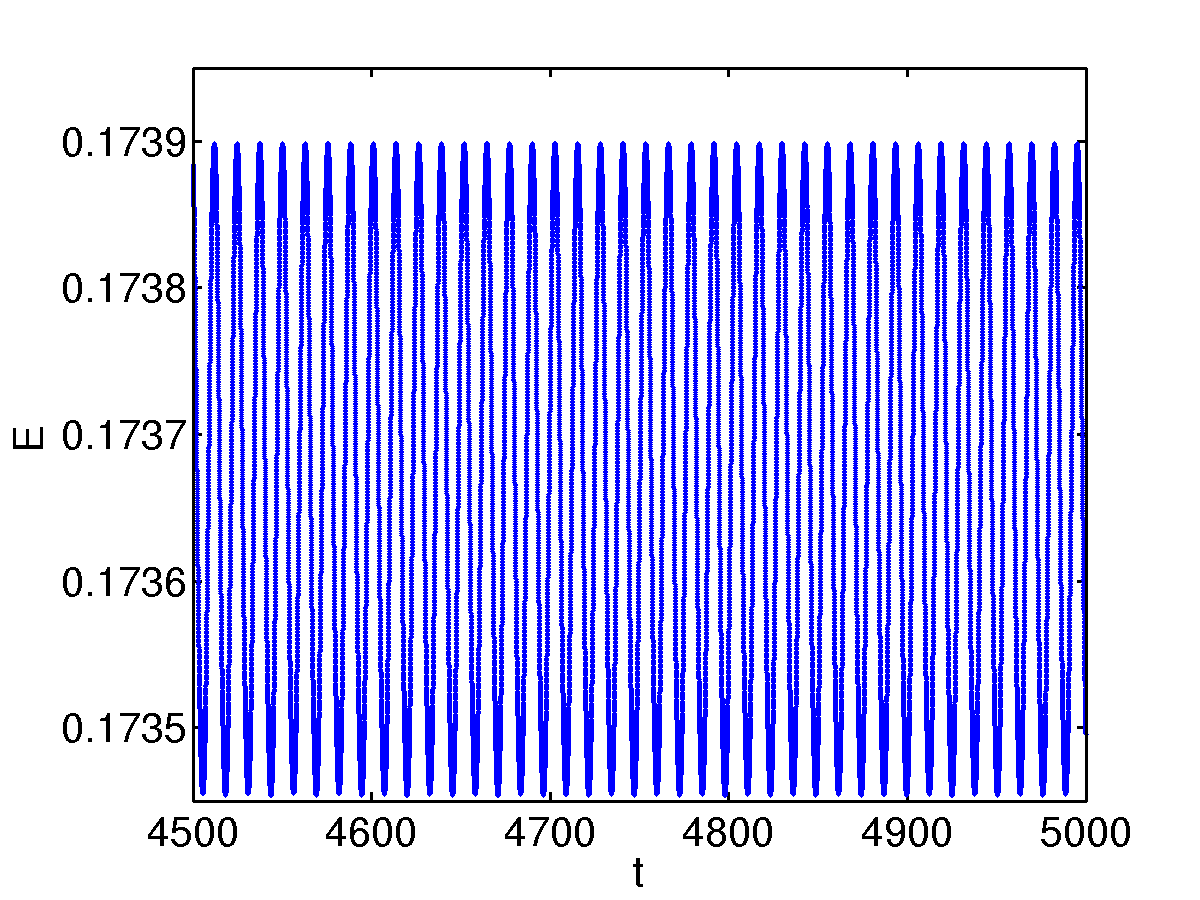}
  \put(-53,95){\color{black}{ {\bf (a)} } }
  \includegraphics[width=0.32\linewidth]{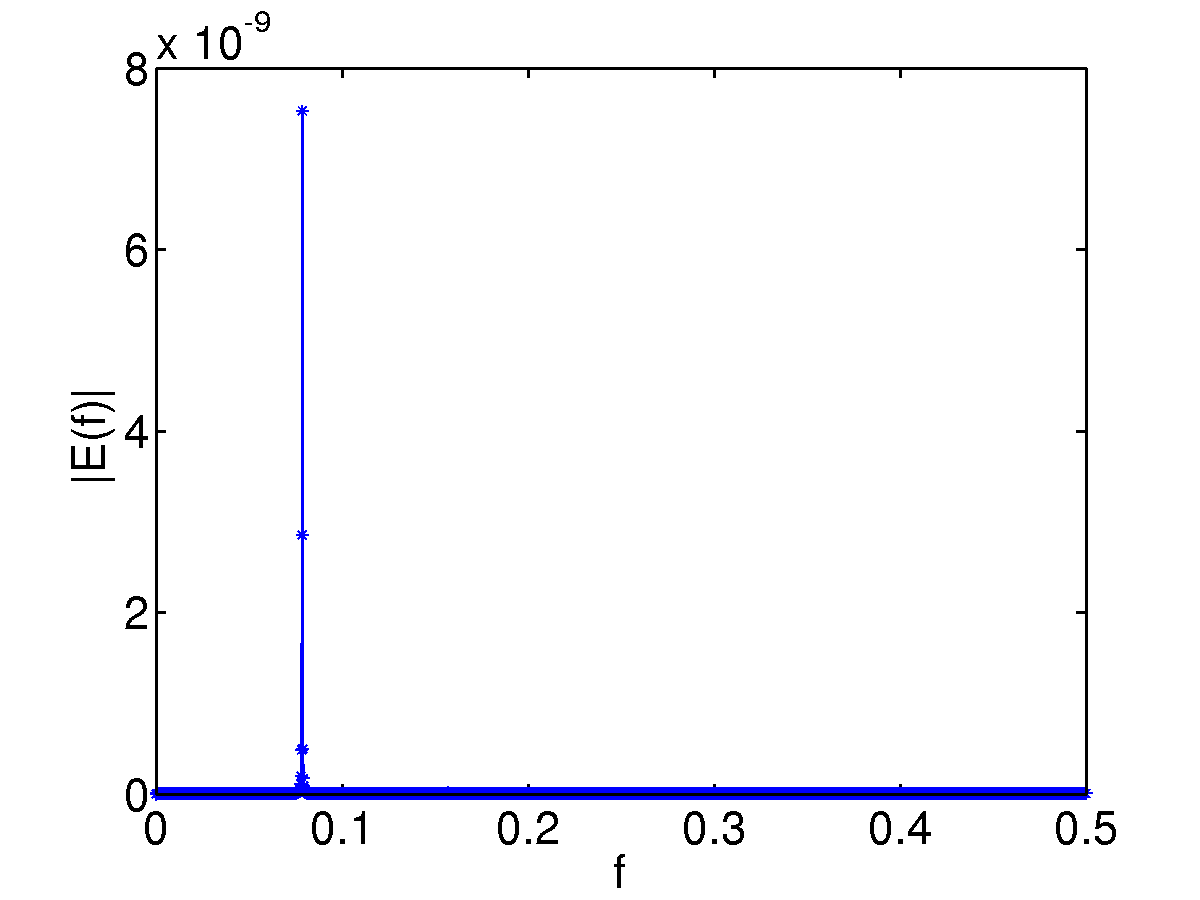}
  \put(-53,95){\color{black}{ {\bf (b)} } }
  \includegraphics[width=0.32\linewidth]{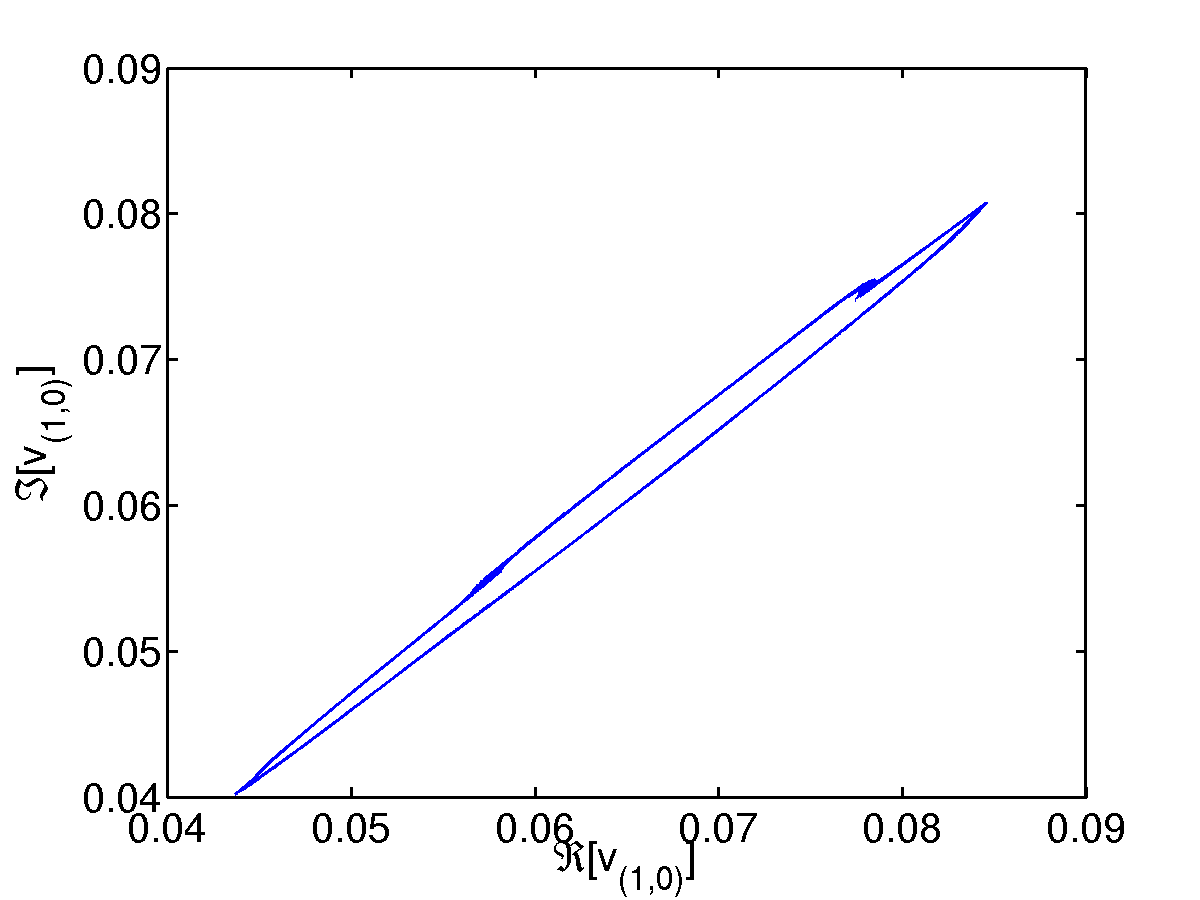}
  \put(-53,95){\color{black}{ {\bf (c)} } }\\
  \includegraphics[width=0.323\linewidth]{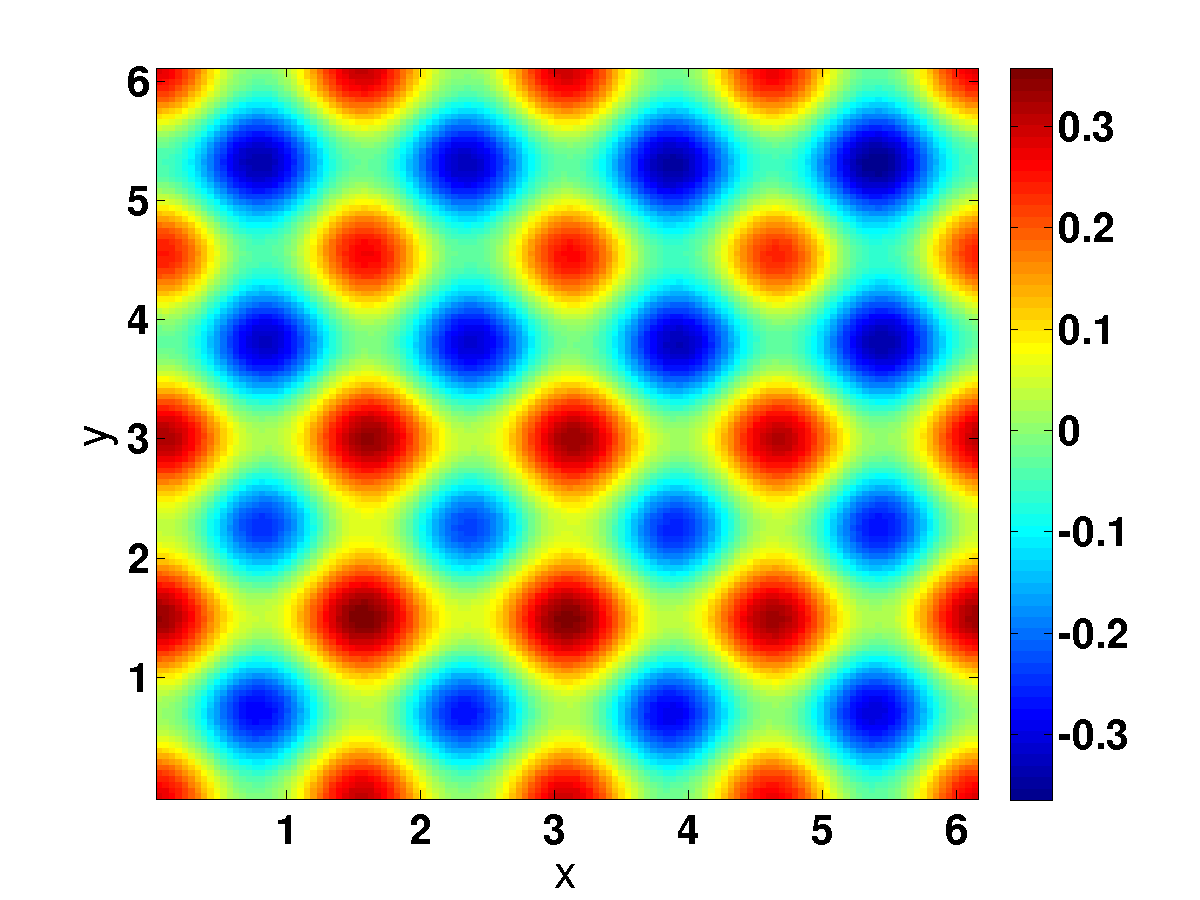}
  \put(-53,95){\color{black}{ {\bf (d)} } }
  \includegraphics[width=0.323\linewidth]{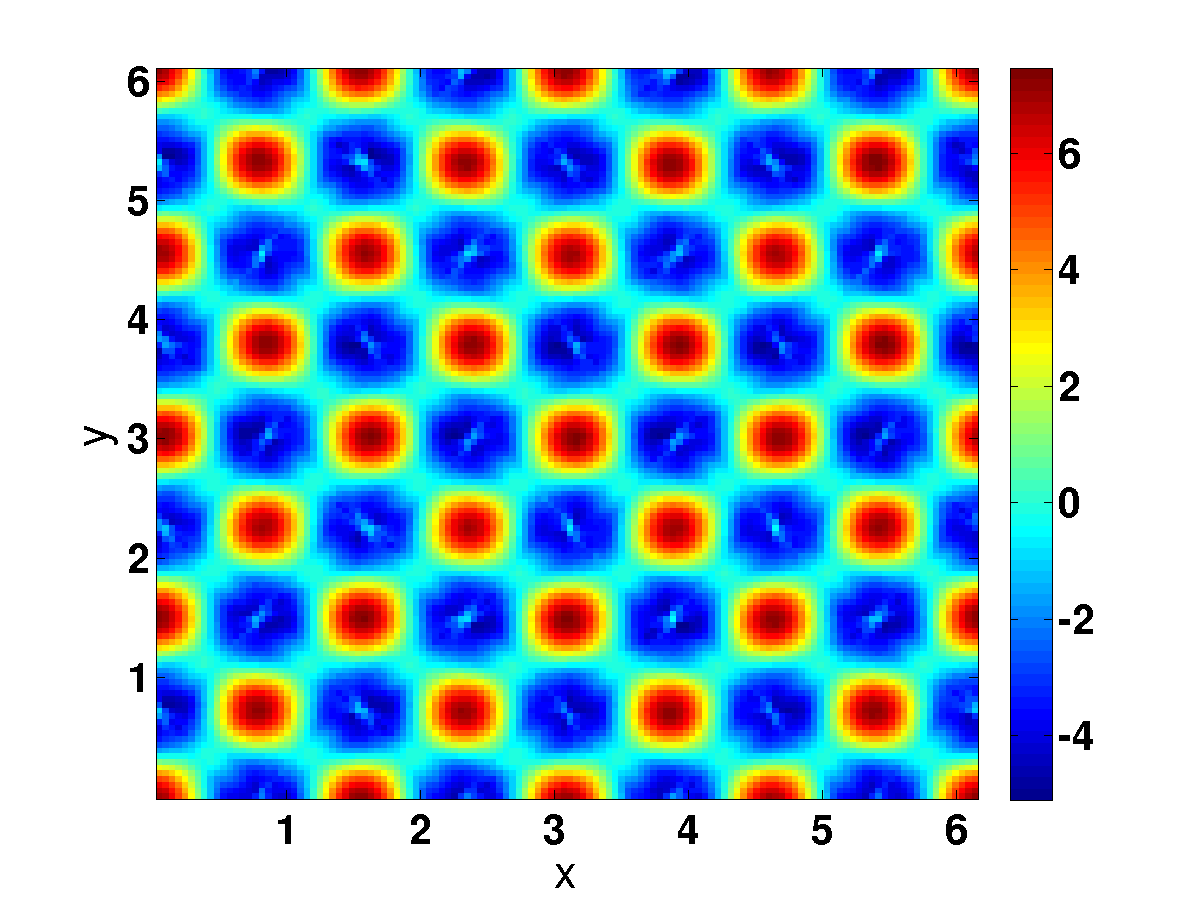}
  \put(-53,95){\color{black}{ {\bf (e)} } }
  \includegraphics[width=0.323\linewidth]{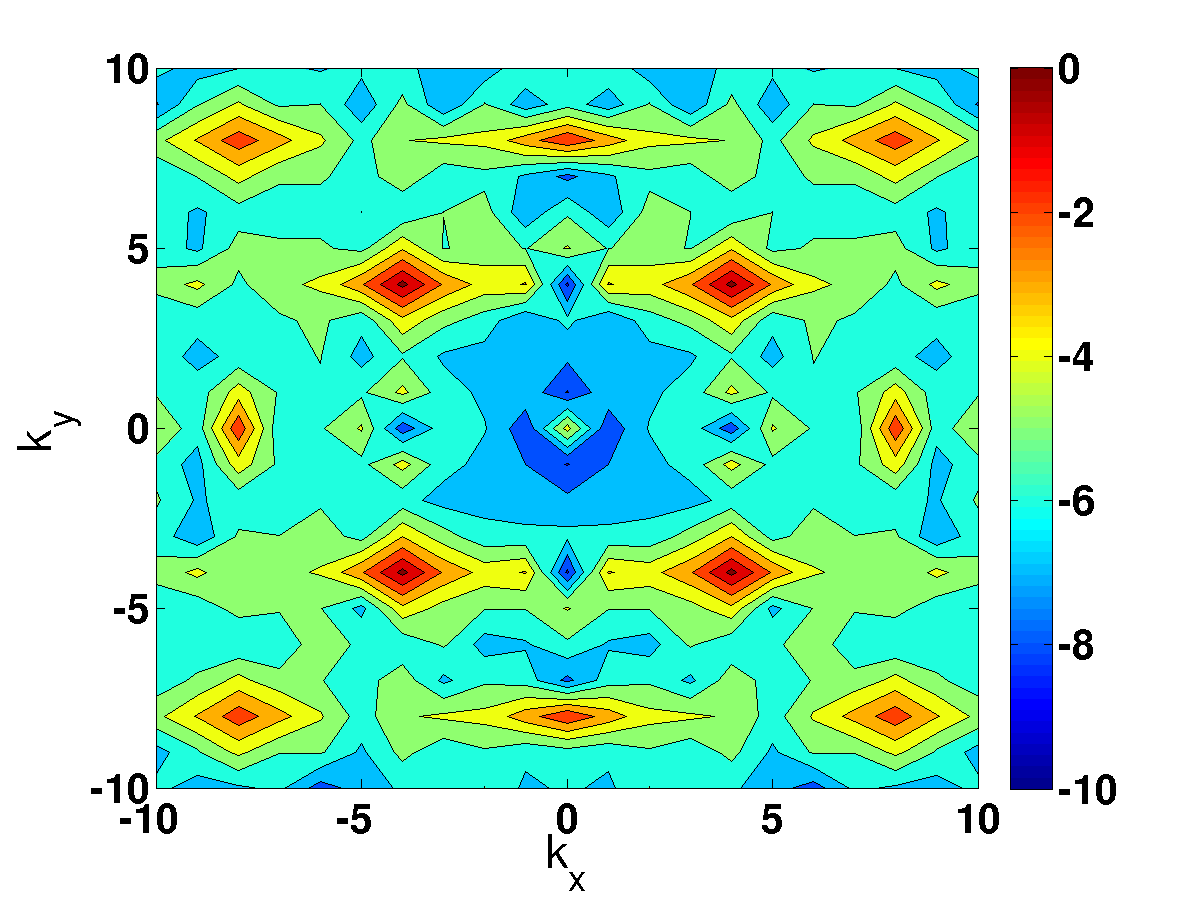}
  \put(-53,95){\color{black}{ {\bf (f)} } }
  \caption{\label{figch5:R48psilamk}(Color online) Plots for
    $\Omega=22$ and $\Wi=0.7$ : (a) the time evolution of the 
    energy $E(t)$, (b) the power spectrum $|E(f)|$ versus the frequency $f$, 
    and (c) the \Poincare-type section in the plane 
    $(\Re[\hat{v}(1,0)],\Im[\hat{v}(1,0)])$.
    Pseudocolor plots of (d) the streamfunction
    $\psi$ and (e) the Okubo-Weiss parameter $\Lambda$; and 
    (f) a filled contour plot of the
    reciprocal-space spectrum $E_{\Lambda}$.}
\end{figure*}

If we increase $\Wi$ some more (runs $\tt R22-4$ with $0.55 \leq \Wi \leq
0.8$), we obtain a spatiotemporal crystal, i.e., one that is spatially periodic
and which also oscillates in time~\cite{perlekar2010turbulence}.  For instance,
if $\Wi = 0.7$, the time series of the energy is periodic in time, as shown
in Fig.~\ref{figch5:R48psilamk}(a); and Fig.~\ref{figch5:R48psilamk}(b) shows
that its power spectrum $|E(f)|$ has only one dominant peak, which is a clear
indication that the temporal evolution is periodic.
Figure~\ref{figch5:R48psilamk}(c) shows the \Poincare-type section in the
$(\Re[\hat{v}(1,0)],\Im[\hat{v}(1,0)])$ plane, which displays that the
projection of the attractor on this plane is a closed loop.
Figures~\ref{figch5:R48psilamk}(d) and (e) show pseudocolor plots of $\psi$ and
$\Lambda$, respectively; and the associated reciprocal-space spectrum
$E_{\Lambda}$ (Fig.~\ref{figch5:R48psilamk}(f)) shows dominant peaks
principally at the forcing wave vectors.

\begin{figure*}[]
  \includegraphics[width=0.32\linewidth]{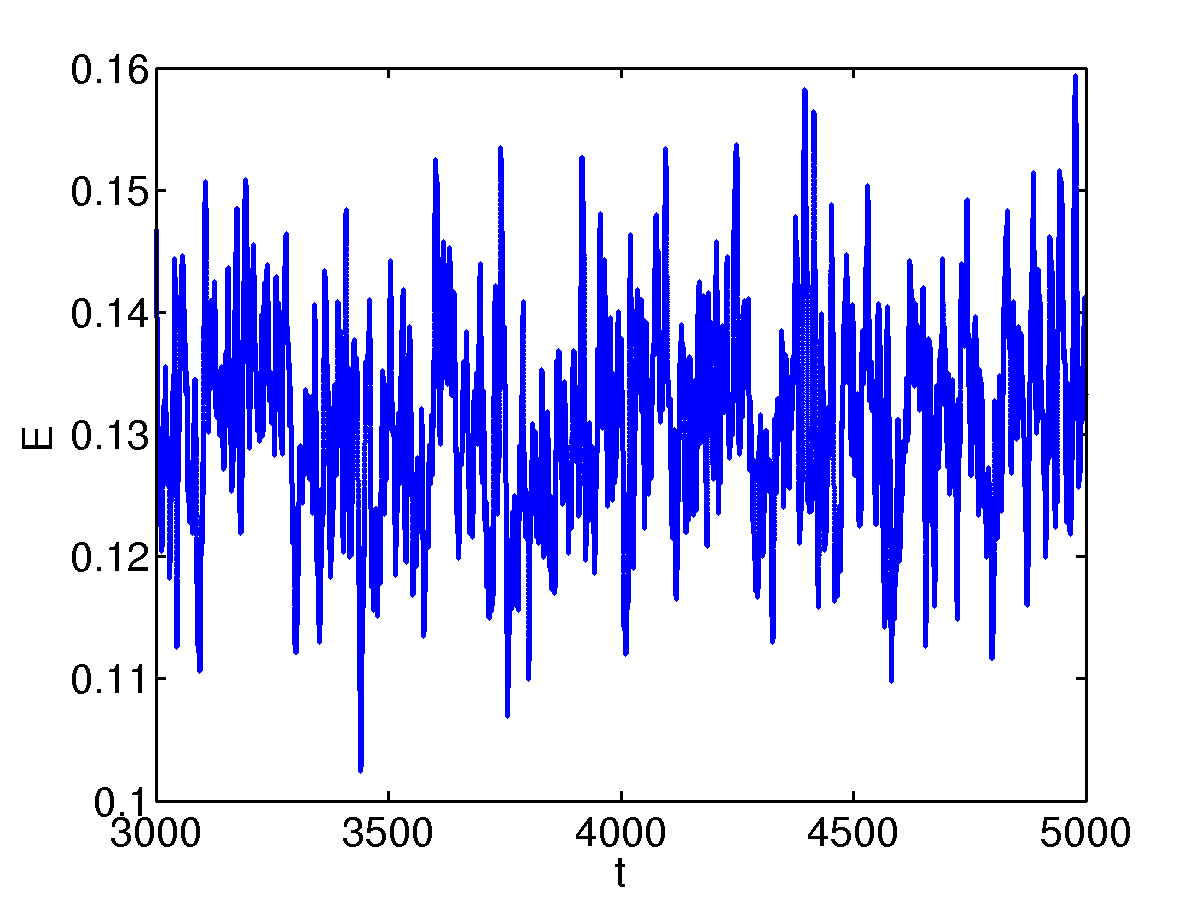}
  \put(-53,95){\color{black}{ {\bf (a)} } }
  \includegraphics[width=0.32\linewidth]{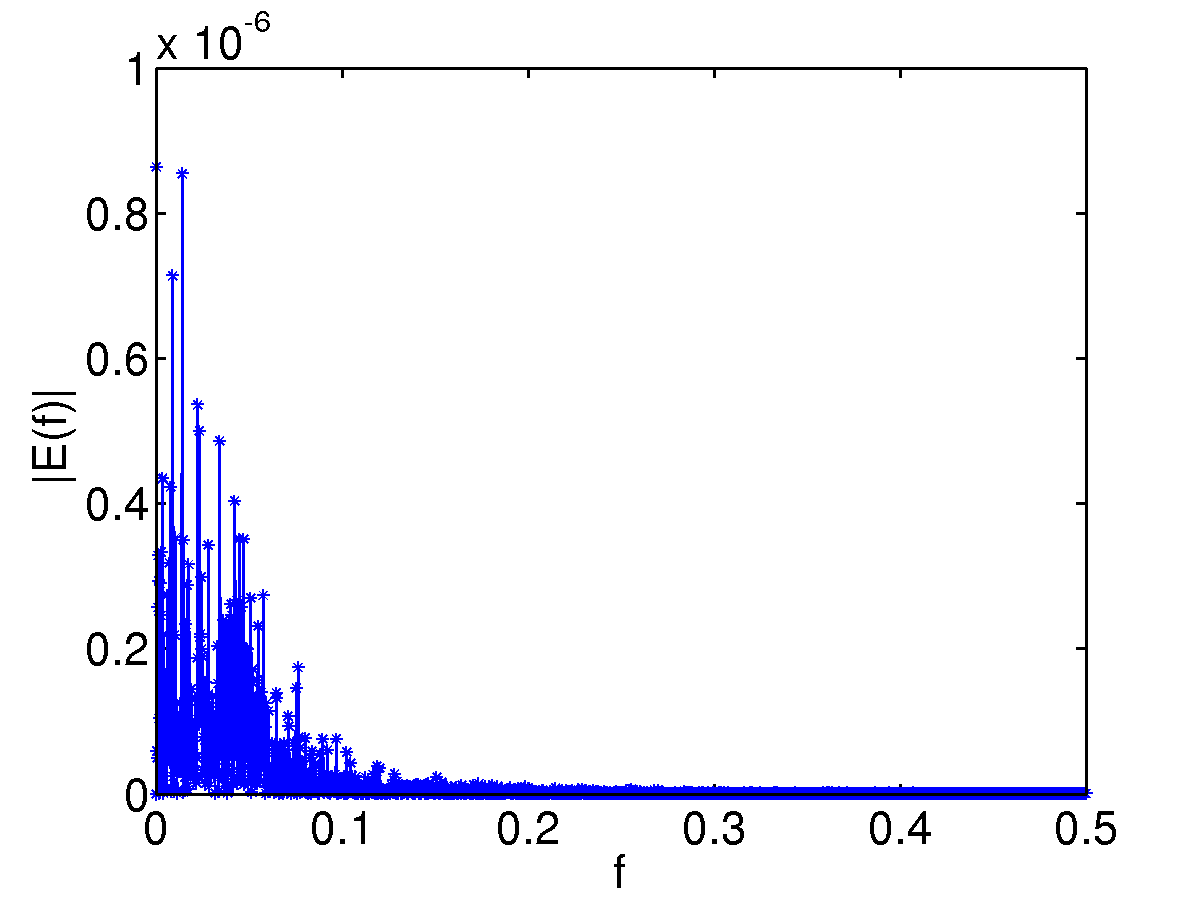}
  \put(-53,95){\color{black}{ {\bf (b)} } }
  \includegraphics[width=0.32\linewidth]{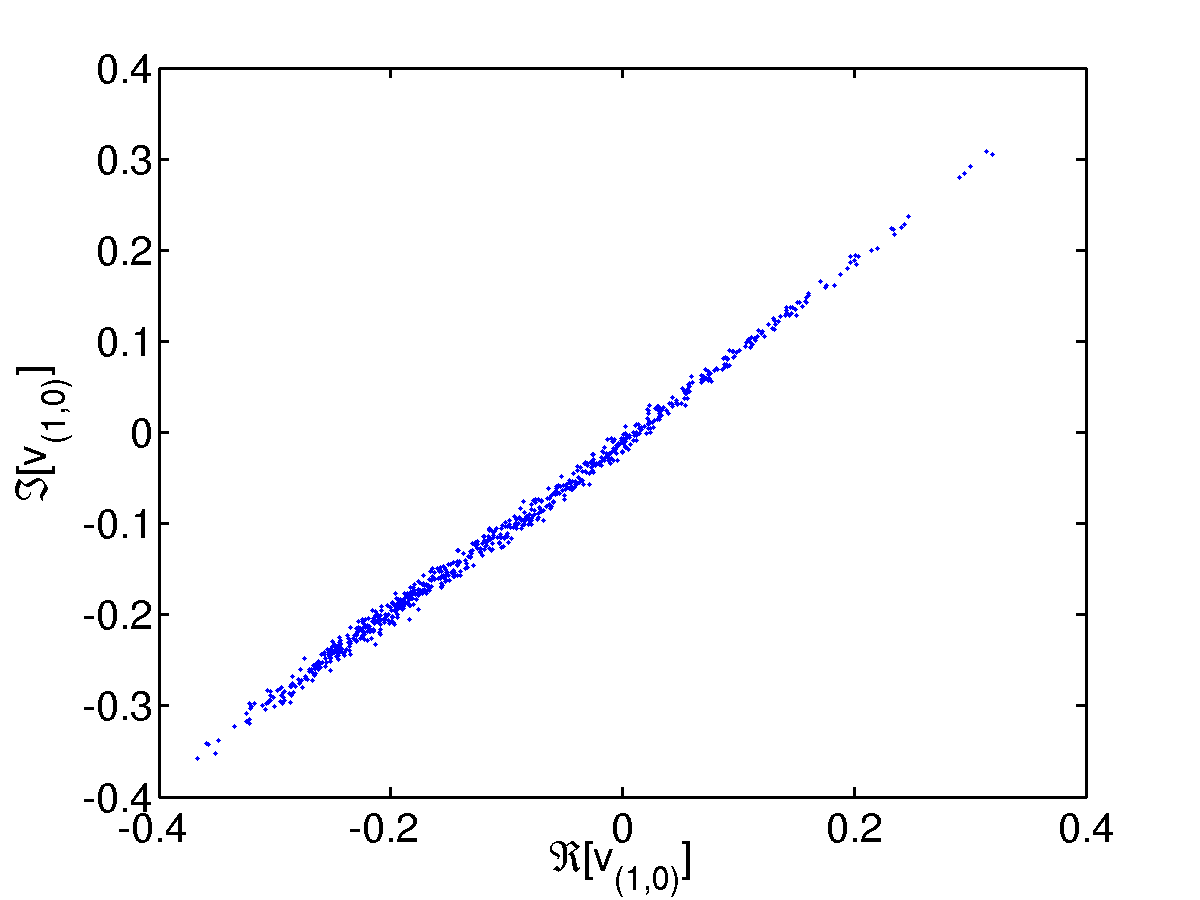}
  \put(-53,95){\color{black}{ {\bf (c)} } }\\
  \includegraphics[width=0.323\linewidth]{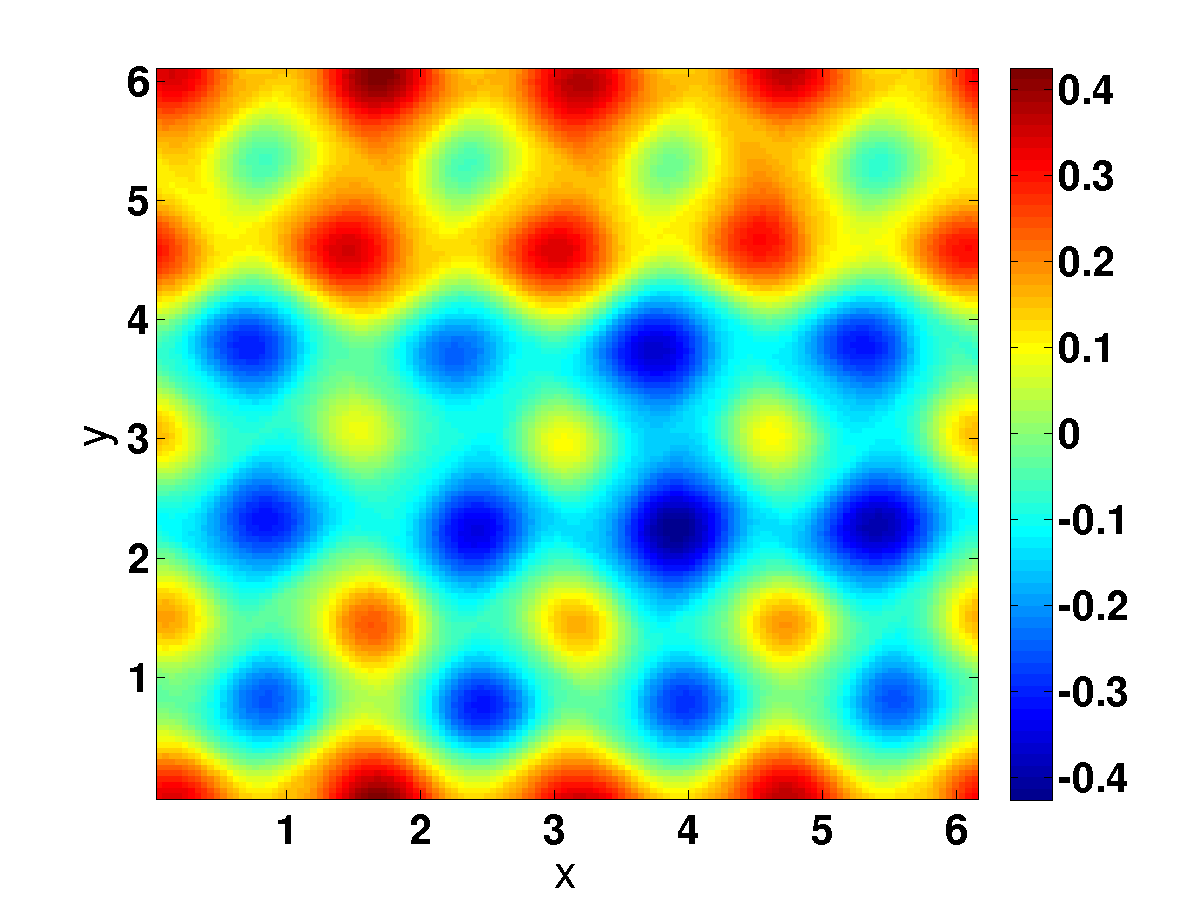}
  \put(-53,95){\color{black}{ {\bf (d)} } }
  \includegraphics[width=0.323\linewidth]{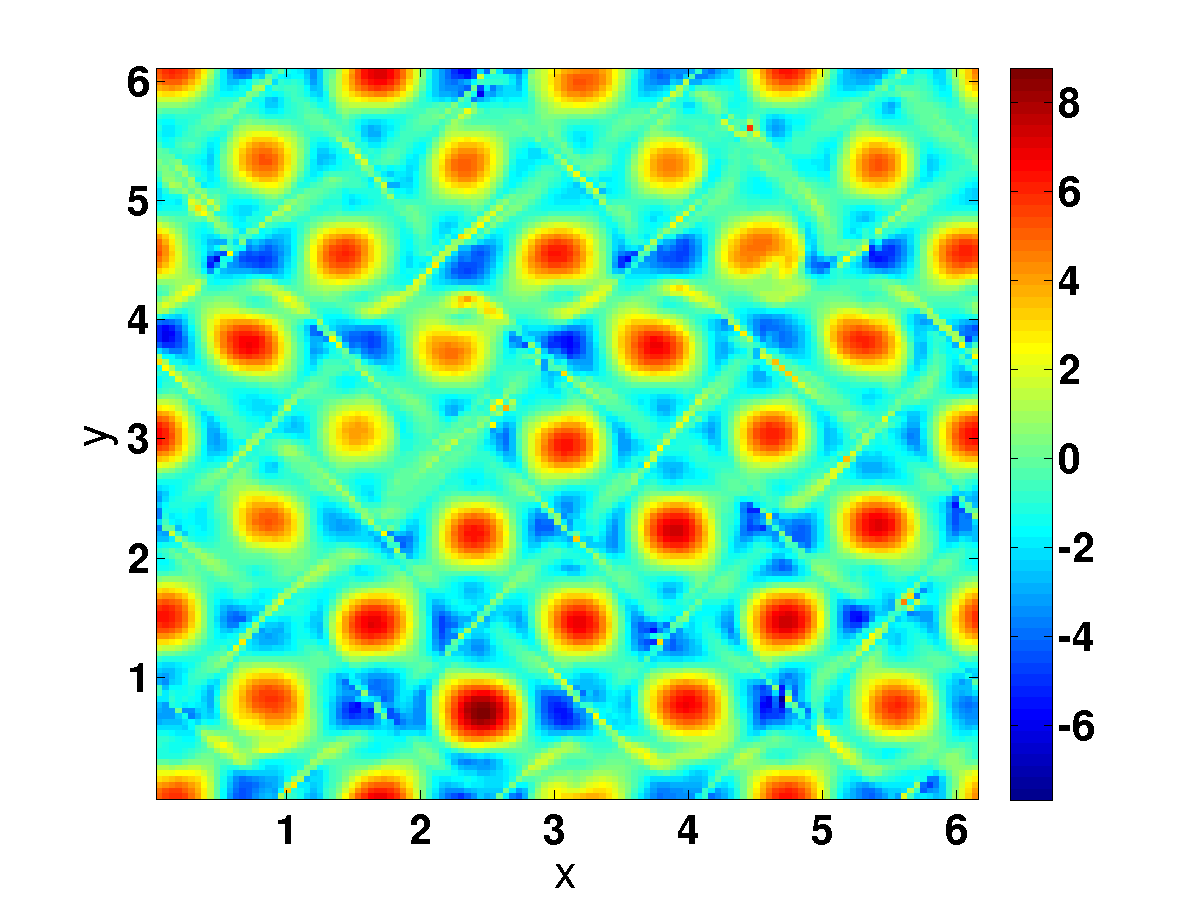}
  \put(-53,95){\color{black}{ {\bf (e)} } }
  \includegraphics[width=0.323\linewidth]{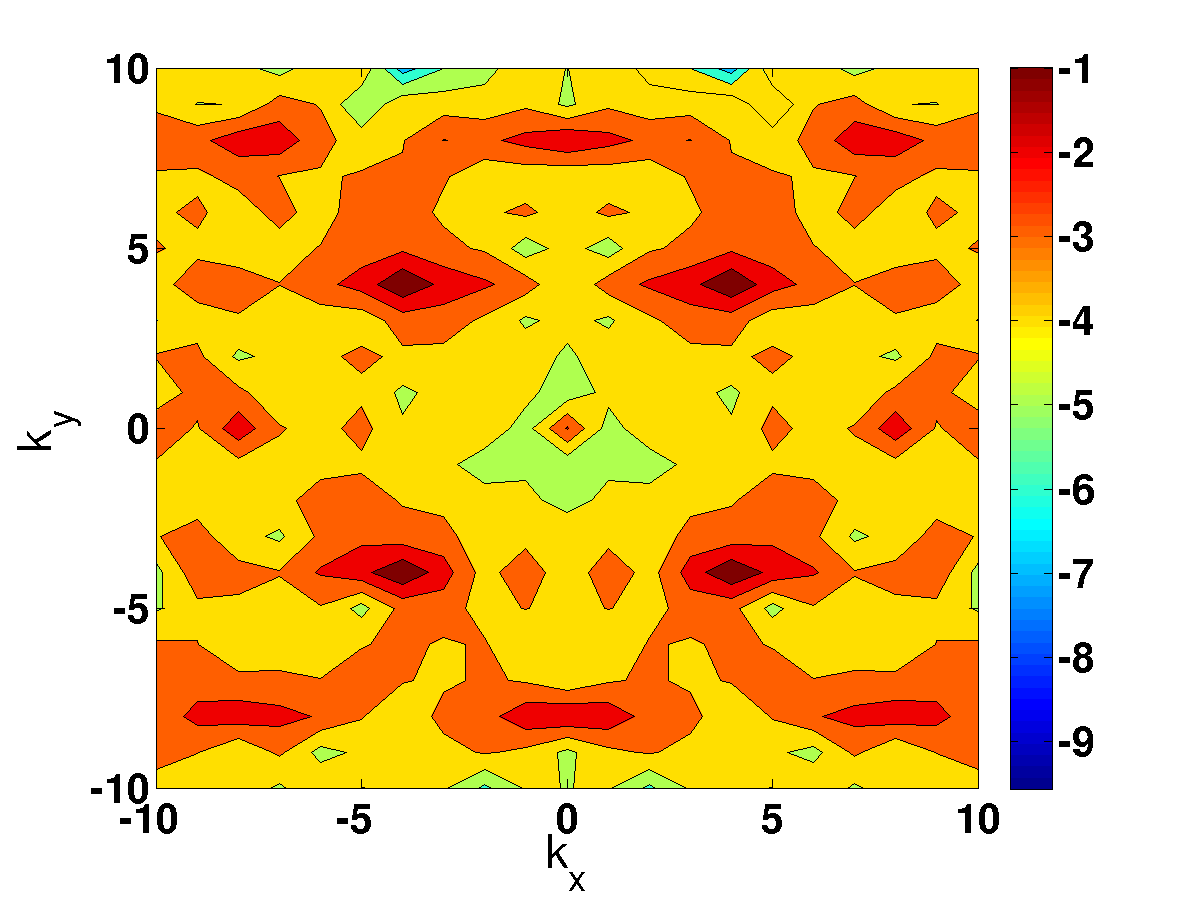}
  \put(-53,95){\color{black}{ {\bf (f)} } }
  \caption{\label{figch5:R49psilamk}(Color online) Plots for
    $\Omega=22$ and $\Wi=20$ : (a) the time evolution of the 
    energy $E(t)$, (b) the power spectrum $|E(f)|$ versus the frequency
    $f$, and (c) the \Poincare-type section in the plane 
    $(\Re[\hat{v}(1,0)],\Im[\hat{v}(1,0)])$.
    Pseudocolor plots of (d) the streamfunction
    $\psi$ and (e) the Okubo-Weiss parameter $\Lambda$; and (f) a filled contour plot of the
    reciprocal-space spectrum $E_{\Lambda}$.}
\end{figure*}

We now increase $\Wi$ again (run $\tt R22-5$); the temporal evolution of the
system becomes chaotic and the spatial organization of the vortex crystal is
distorted.  Thus, we finally enter a state with spatiotemporal chaos and
turbulence (SCT); this is our analog of the disordered, liquid state, which
appears on the melting of an equilibrium
crystal~\cite{ramakrishnan1979first,chaikin2000principles,oxtoby1991liquids,singh1991density}.
However, some vestiges of the incipient crystalline ordering can still be found
in reciprocal-space spectra, as we illustrate for $\Wi = 20$ in
Fig.~\ref{figch5:R49psilamk}. From the time series $E(t)$ of the energy
(Fig.~\ref{figch5:R49psilamk}(a)), it is clear that system is chaotic; this is
confirmed by the power spectrum $|E(f)|$ (Fig.~\ref{figch5:R49psilamk}(b)),
which displays several peaks, and  the \Poincare-type section in the
$(\Re[\hat{v}(1,0)],\Im[\hat{v}(1,0)])$ plane
(Fig.~\ref{figch5:R49psilamk}(c)), in which the points fill a two-dimensional
area. The disordered spatial organization of vortical structures is shown in
Figs.~\ref{figch5:R49psilamk}(d) and (e) via pseudocolor plots of $\psi$ and
$\Lambda$, respectively.  The reciprocal-space spectrum $E_{\Lambda}(k)$
(Fig.~\ref{figch5:R49psilamk}(f)) shows residual peaks at the forcing wave
vectors, which indicate the incipient vortex crystal, but the other Fourier
modes show that long-range spatial periodicity has been lost in this SCT state.

Thus, our careful study of the case $\Omega = 22$, for various values of
$\Wi$, shows how the turbulent SCT state, at $\Wi = 0$, gives way to the
states OPXA, SXA, and OPXA states, as we increase $\Wi$
(Table~\ref{tablech5:para}); finally these states show reentrant melting into
the SCT state. 

\subsection{Order Parameters, Spatial Correlation Functions, and 
Kinetic-Energy Spectra}

We now use statistical measures that are employed in the density-functional
theory~\cite{ramakrishnan1979first,chaikin2000principles,oxtoby1991liquids,singh1991density}
of freezing. In particular, we examine the dependence of the order parameters
$\langle{\hat{\Lambda}}_{\bf k}\rangle$ and the spatial autocorrelation
function $G({\bf r})$, defined in Eq. (\ref{eq:auto_cf}), on $\Wi$ and
$\Omega$. Recall that equilibrium melting is a first-order transition at which
$\rho_{\bf G}$ jumps discontinuously from a nonzero value in the crystal to
zero in the liquid. The melting of our nonequilibrium vortex crystal is far
more complicated, in all parts of the parameter space of the phase diagram of
Fig.~\ref{figch5:phase}, because there are many transitions.

\begin{figure*}[]
  \includegraphics[width=0.47\linewidth]{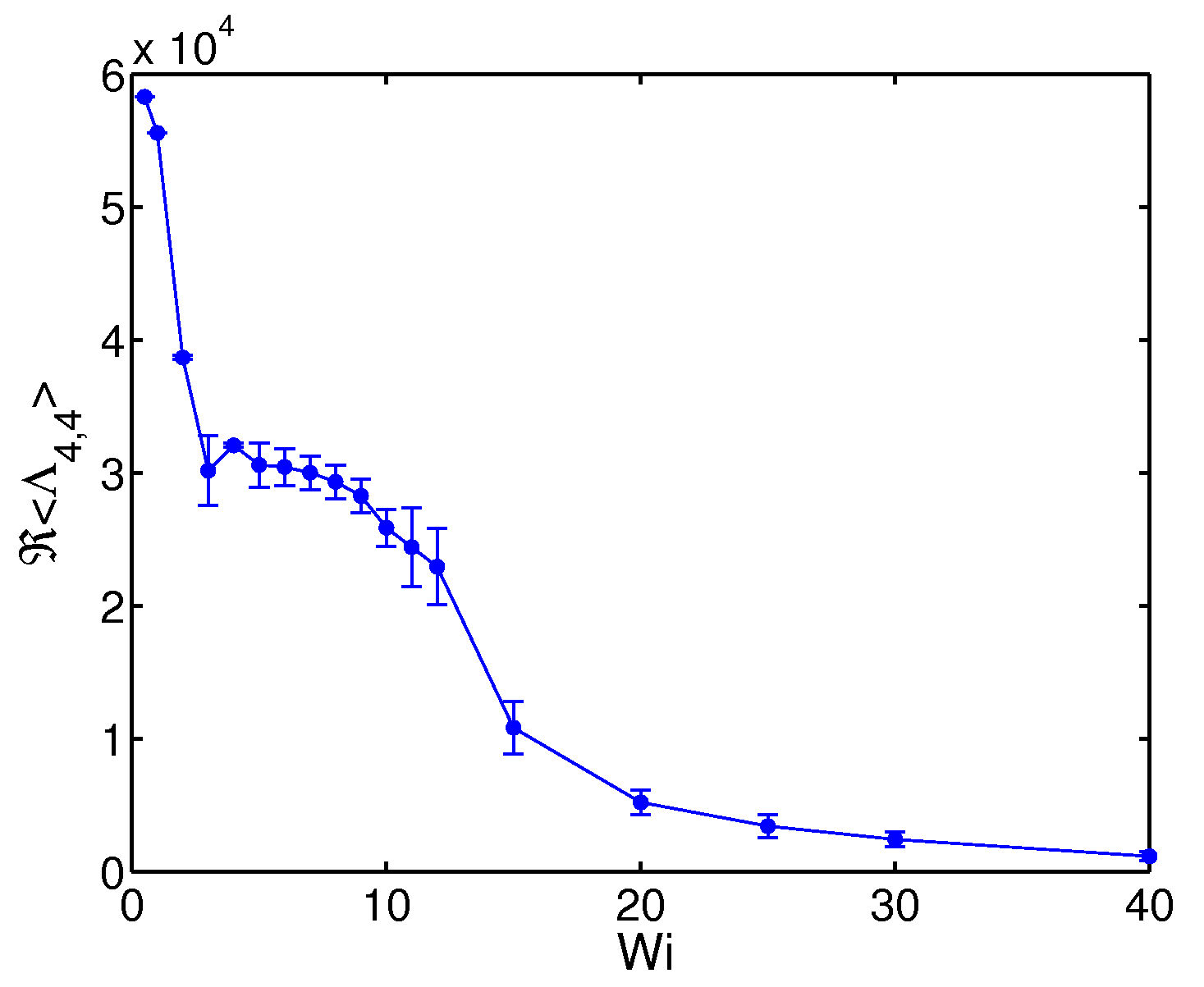}
  \put(-53,130){\color{black}{ {\bf (a)} } }
  \includegraphics[width=0.47\linewidth]{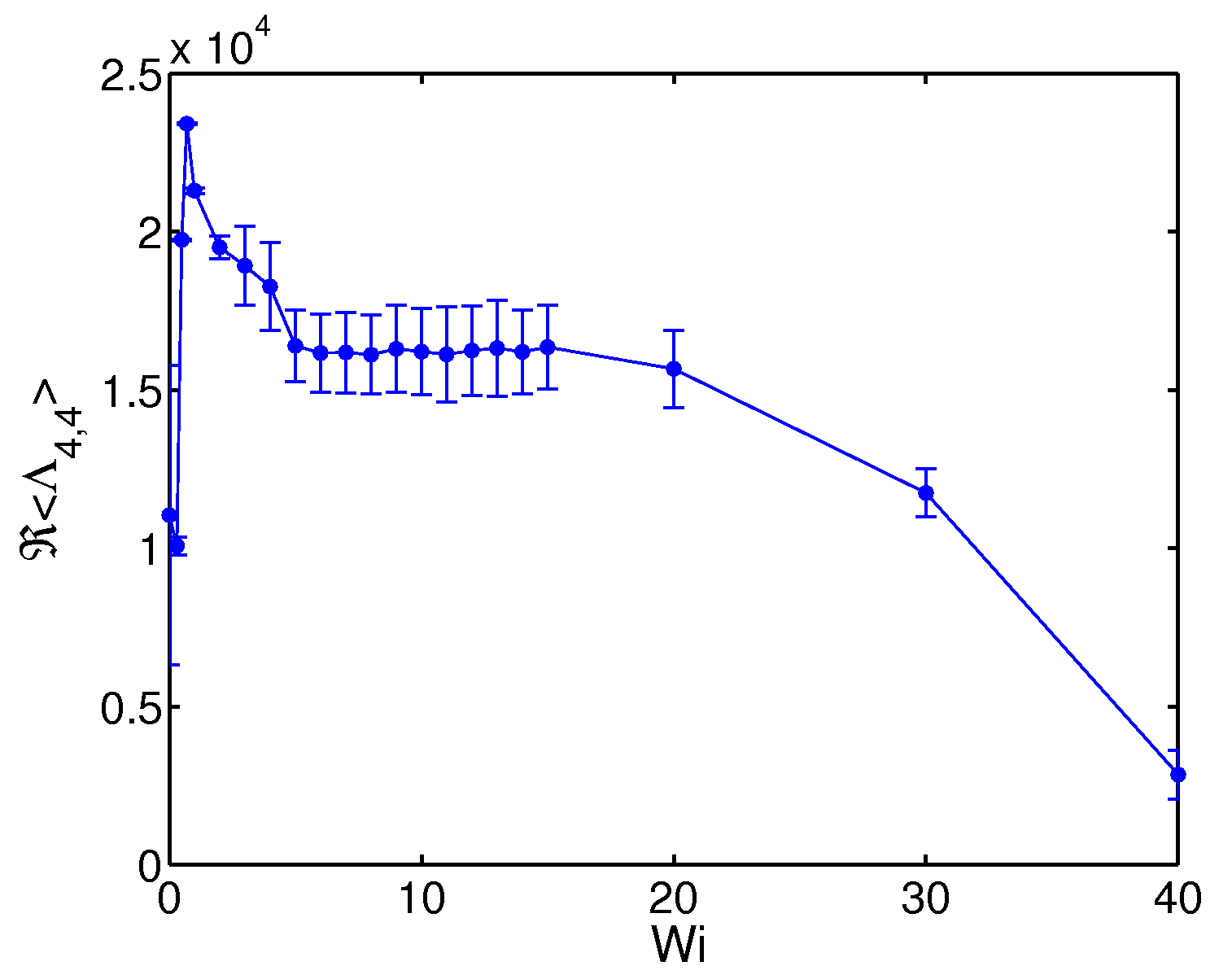}
  \put(-53,130){\color{black}{ {\bf (b)} } }
  \caption{\label{figch5:rhog}(Color online) Plots of 
    $\langle{\hat{\Lambda}}_{{\bf k} = (4,4)}\rangle$ versus $\Wi$
    for (a) $\Omega = 1$ and for (b) $\Omega=22$; in (b) 
    $\langle{\hat{\Lambda}}_{4,4}\rangle$ first increases with
    $\Wi$, goes through a maxima, which corresponds to an OPXA state, and
    then decreases after that.}
\end{figure*}

For our nonequilibrium vortex crystal $\Re\langle{\hat{\Lambda}}_{\bf
k}\rangle$, which is obtained by summing ${\hat{\Lambda}}_{\bf k}$ over the
four forcing wave vectors, is the analog of the order parameters
$\Re\langle\rho_{\bf G}\rangle$ in a conventional crystal.
$\Re\langle{\hat{\Lambda}}_{\bf k}\rangle$ changes with $\Wi$ as shown,
respectively, for (a) $\Omega=1$ and ${\bf k} = (4,4)$ and (b) $\Omega=22$ and
${\bf k} = (4,4)$ in Figs.~\ref{figch5:rhog} (a) and (b). For $\Omega = 1$ and
small values of $\Wi$, the steady state of our system is SX, so the Fourier
modes at the forcing wave vector are the most significant ones. As we increase
$\Wi$, our nonequilibrium system undergoes a series of transitions from  SX
to the turbulent state SCT (see above); therefore,
$\Re\langle{\hat{\Lambda}}_{\bf k}\rangle$ decreases with $\Wi$ as we show
in Fig.~\ref{figch5:rhog}(a).  For $\Omega = 22$ and $\Wi=0$, we begin with
a system that is disordered and turbulent; if we then increase $\Wi$, our
system first goes to the state SX, so $\Re\langle{\hat{\Lambda}}_{\bf
k}\rangle$ first increases as we show in Fig.~\ref{figch5:rhog}(b); as we
increase $\Wi$, the system becomes turbulent and disordered again, so
$\Re\langle{\hat{\Lambda}}_{\bf k}\rangle$ decreases at large $\Wi$
(Fig.~\ref{figch5:rhog}(b)).

\begin{figure*}[]
  \includegraphics[width=0.47\linewidth]{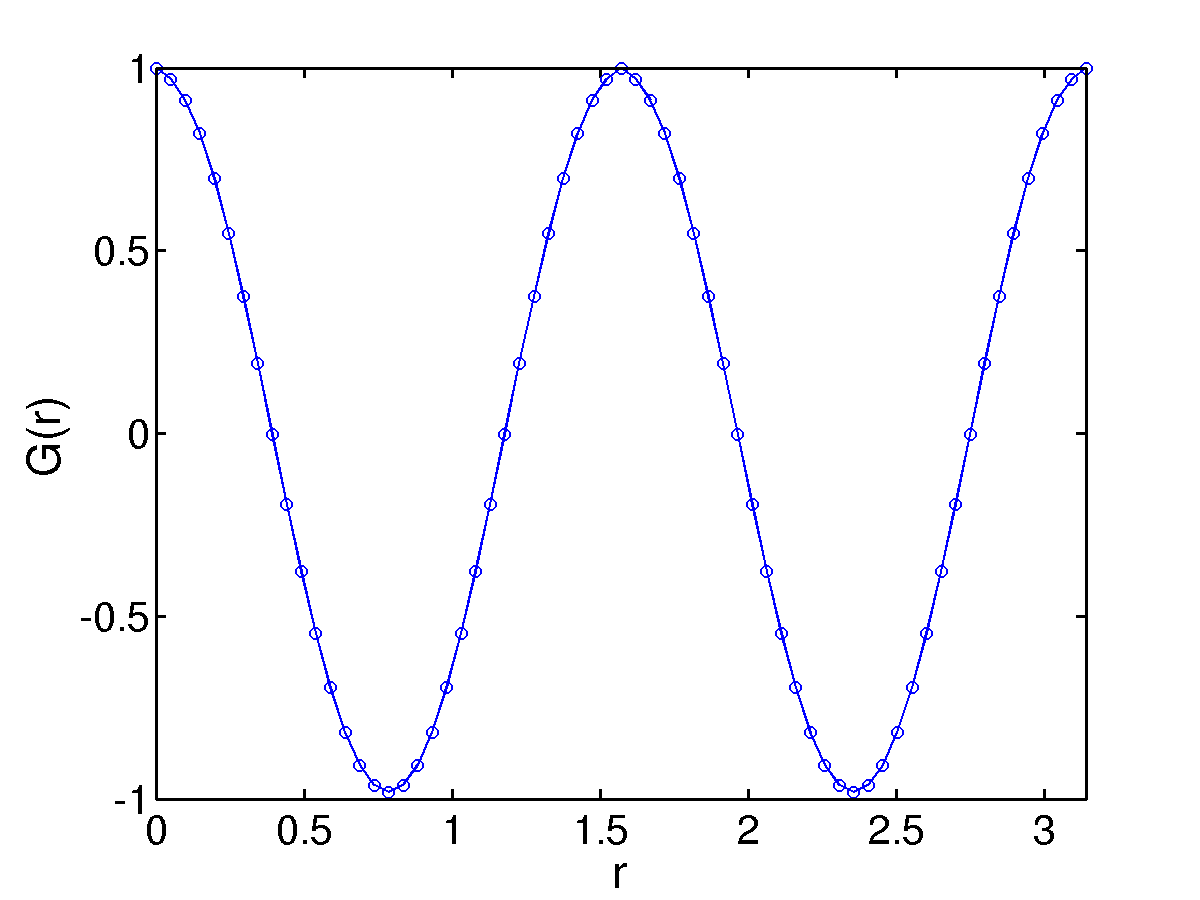}
  \put(-53,140){\color{black}{ {\bf (a)} } }
  \includegraphics[width=0.47\linewidth]{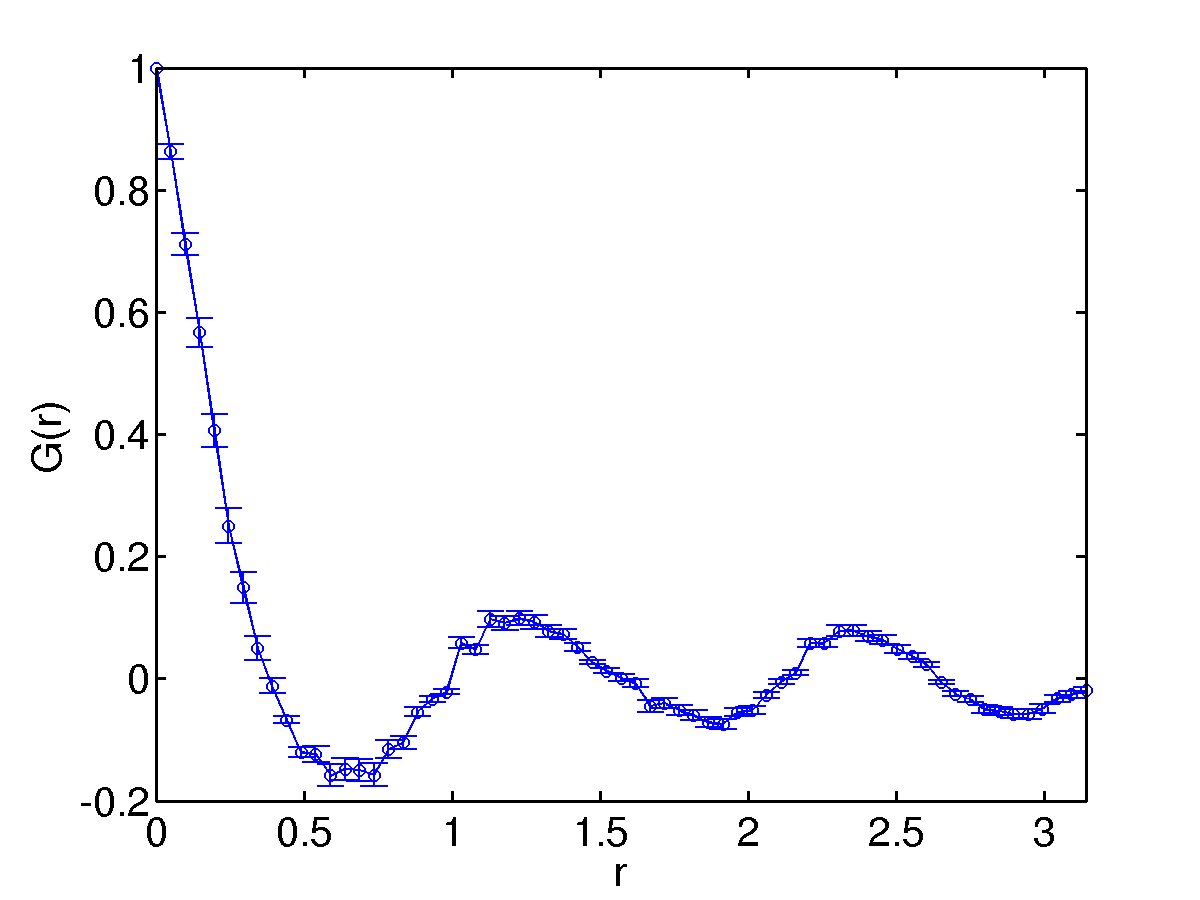}
  \put(-53,140){\color{black}{ {\bf (b)} } }\\
  \includegraphics[width=0.47\linewidth]{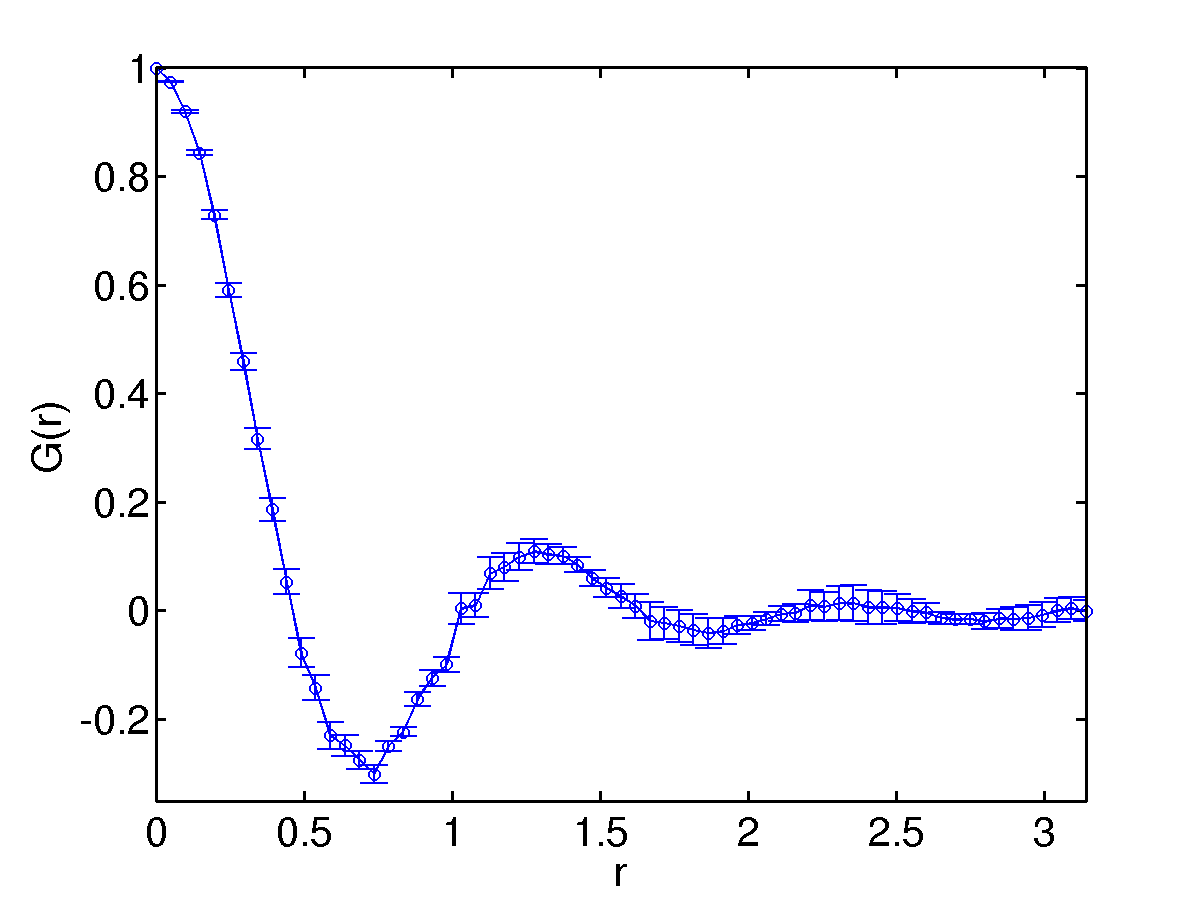}
  \put(-53,140){\color{black}{ {\bf (c)} } }
  \includegraphics[width=0.47\linewidth]{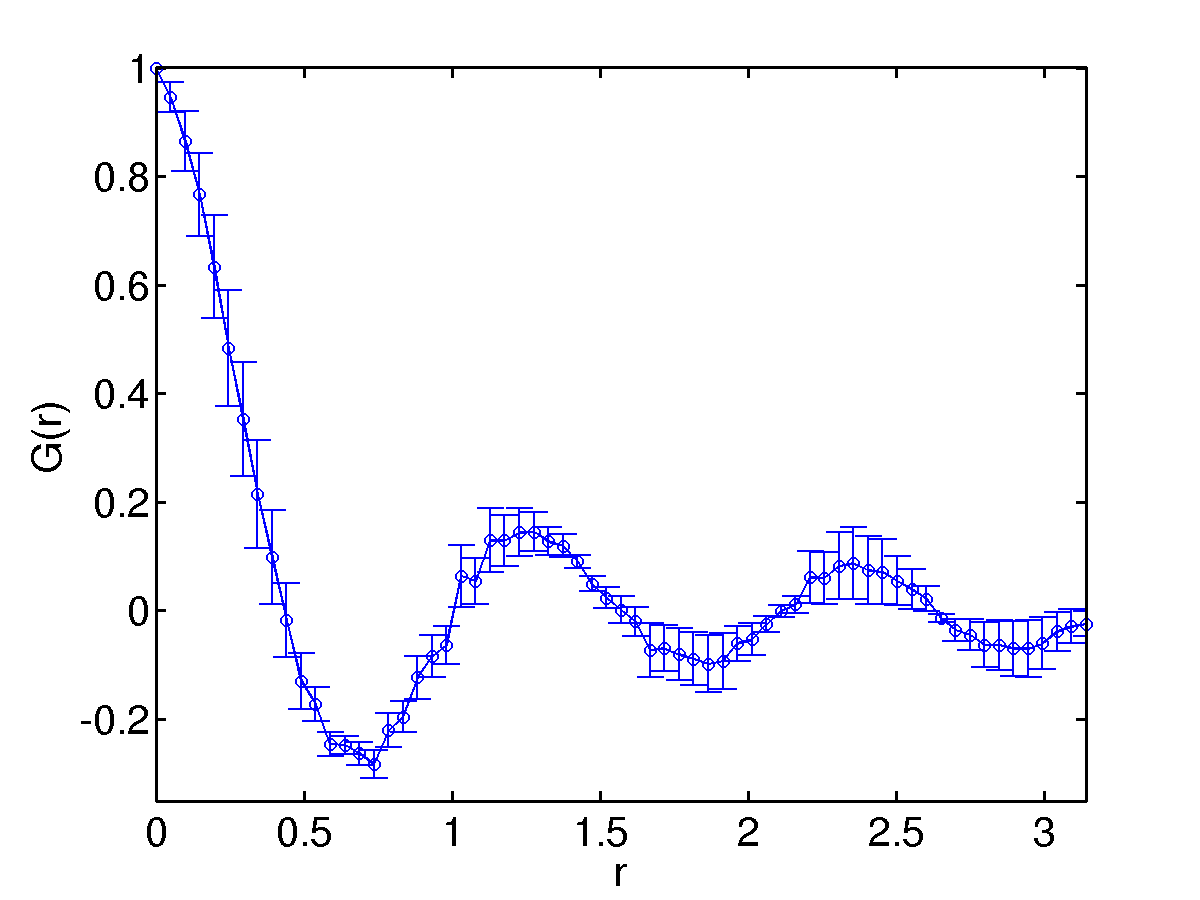}
  \put(-53,140){\color{black}{ {\bf (d)} } }
  \caption{\label{figch5:Gr}(Color online) (a) Plot of $G({\bf r})$ for the crystalline 
  state with $\Omega = 1$ and $\Wi=1$ along the line connecting 
  ${\bf r} = (\pi/2,\pi/2)$ and ${\bf r} = (\pi/2,\pi)$; and plots of the 
  circularly averaged $G({\bf r})$ in the turbulent state with (b) 
  $\Omega = 1$ and $\Wi=20$, (c) $\Omega = 22$ and $\Wi=0$, and
  (d) $\Omega = 22$ and $\Wi=20$.}
\end{figure*}

We give representative plots of the correlation function  $G({\bf r})$ (see
Eq.(\ref{eq:auto_cf})) in Figs.~\ref{figch5:Gr} (a)-(d).  For the crystal, we
calculate $G({\bf r})$ along the line from ${\bf r} = (\pi/2,\pi/2)$ and ${\bf
r} = (\pi/2,\pi)$; this shows periodic peaks [Figs.~\ref{figch5:Gr} (a) for
$\Omega=1, \Wi=1$] with widths that are related to those of vortical or
strain-dominated regions.  In the turbulent phase, we use a circular average of
$G$ [Figs.~\ref{figch5:Gr} (b), (c), and (d) for $\Omega=1$ and $\Wi=20$,
$\Omega=22$ and $\Wi=0$, and $\Omega=22$ and $\Wi=20$, respectively];
these peaks decay over a length scale that yields the degree of short-range
order.  This decay is similar to the decay of spatial correlation functions in
a disordered liquid phase in an equilibrium system.
For similar, but not
the same, correlation functions in a viscoelastic Taylor-Couette flow see
Refs.~\cite{latrache2012transition,latrache2016defect}.

\begin{figure*}[]
\begin{center}
  \includegraphics[width=0.47\linewidth]{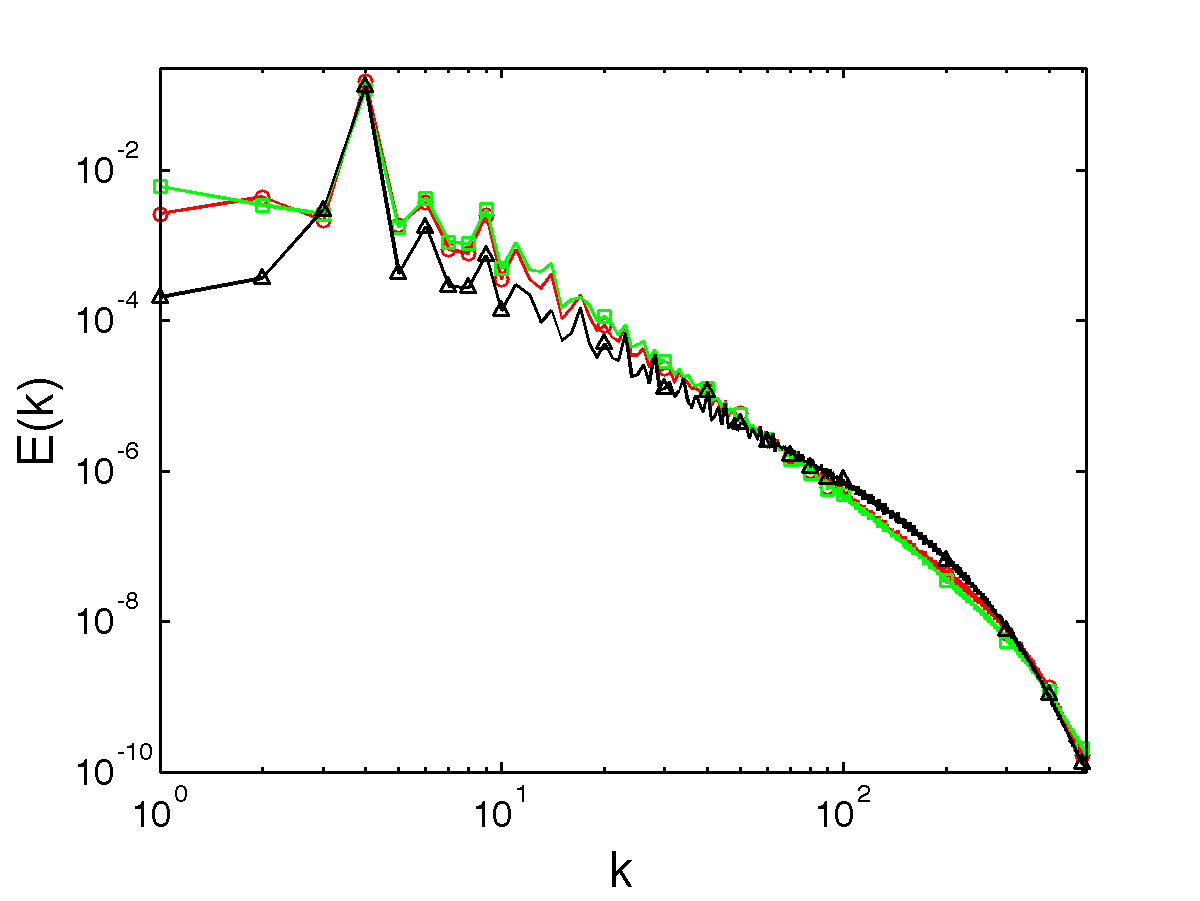}
  \put(-53,140){\color{black}{ {\bf (a)} } }
  \includegraphics[width=0.47\linewidth]{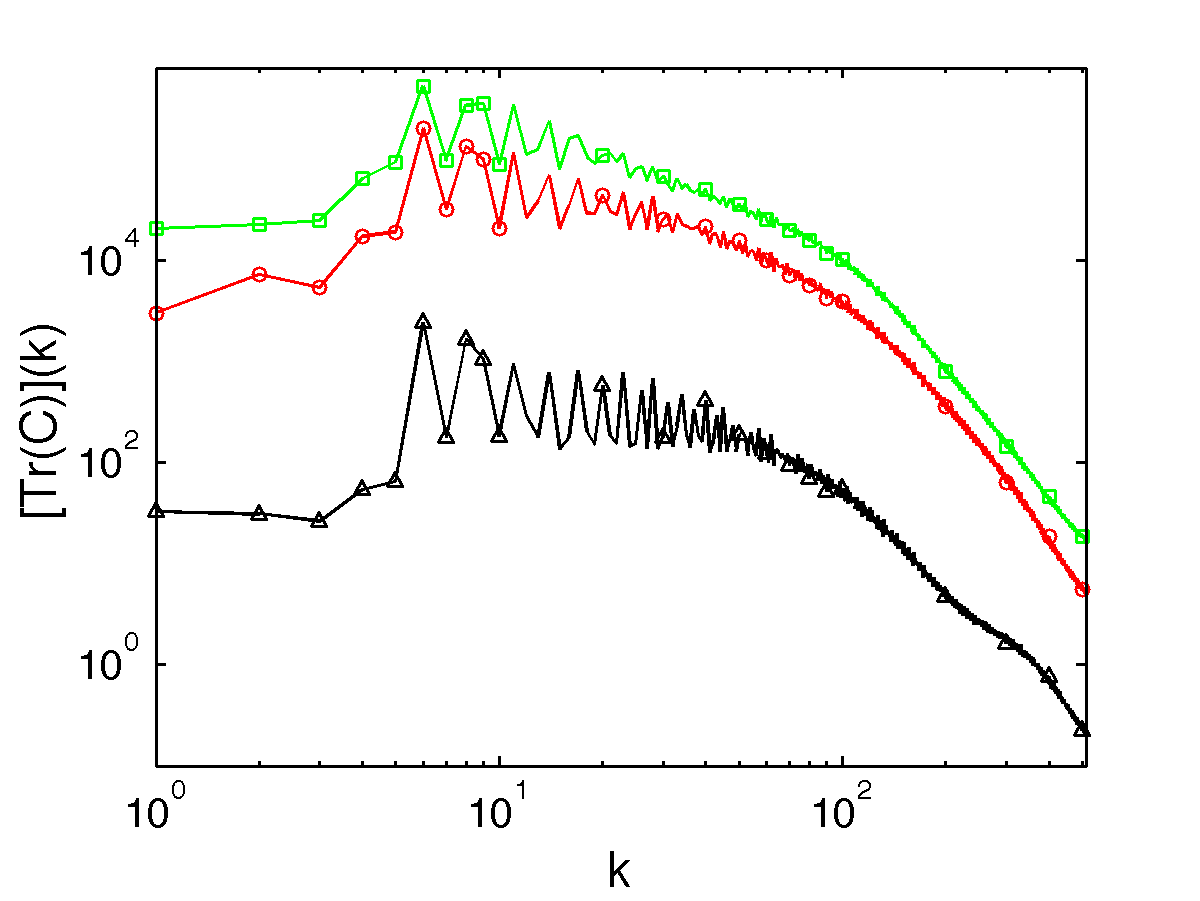}
  \put(-53,140){\color{black}{ {\bf (b)} } }
  \caption{\label{figch5:spectra}(Color online) Plots of (a) energy spectra 
    $E(k)$ versus $k$ and (b) polymer stretching spectra  
    $[{\rm Tr} {\cal C}](k)$ versus $k$, 
    for $\Omega = 4$ and $\Wi = 10$ (red line with circles), $\Omega = 4$
    and $\Wi = 40$ (green line with squares), and $\Omega = 50$ and $\Wi = 5$ 
    (black line with triangles).}
\end{center}
\end{figure*}

We have also obtained the energy spectra $E(k)$ in two parts of the turbulent
state of our system; the first part is dominated by elastic turbulence and the
other one lies in the region of fluid turbulence, where polymers lead to
dissipation reduction.  For a given resolution, there is one qualitative
difference between the energy spectra in these two regimes. In the
forward-cascade part of $E(k)$, the slope of the energy spectrum is higher in
the elastic-turbulence regime than in the dissipation-reduction regime as we
show in Fig.~\ref{figch5:spectra} (a), for $\Omega = 4$ and $\Wi = 10$ (red
line with circles), $\Omega = 4$ and $\Wi = 40$ (green line with squares), and $\Omega = 50$ and
$\Wi = 5$ (black line with triangles). 
The exponent of the energy spectra is $\simeq -3.2$,
which is in reasonable agreement with the exponent in laboratory experiments of
Groisman and Steinberg~\cite{groisman2004elastic}.
Note that, in both elastic-turbulence and dissipation-reduction regimes, the energy spectra show that the energy of the polymeric fluid is spread over many decades of $k$ (or, equivalently, over many length scales); this is a well-known signature of turbulent flows, whether they are engendered by large $\Rey$ or large ${\Wi}$. 
We have also obtained the spectrum of polymer
stretching, which is defined as $[Tr(C)](k) \equiv\sum_{k-1/2< k' \le k+1/2}
k'^2 \langle |\widehat { C_{11}}({\bf k'},t)|^2 + |\widehat {C_{22}}({\bf
k'},t)|^2 \rangle _t$, where $\langle \rangle_t$ denotes the time average over
the statistically steady state.  In Fig.~\ref{figch5:spectra} (b) we show the
polymer-stretching spectra, for $\Omega = 4$ and $\Wi = 10$ (red line with circles),
$\Omega = 4$ and $\Wi = 40$ (green line with squares), and $\Omega = 50$ and $\Wi = 5$
(black line with triangles).

%


\subsection{Spatial Patterns of Polymers and Inertial Particles in the
Nonequilibrium States of our System}

\begin{figure*}[]
  \includegraphics[width=0.47\linewidth]{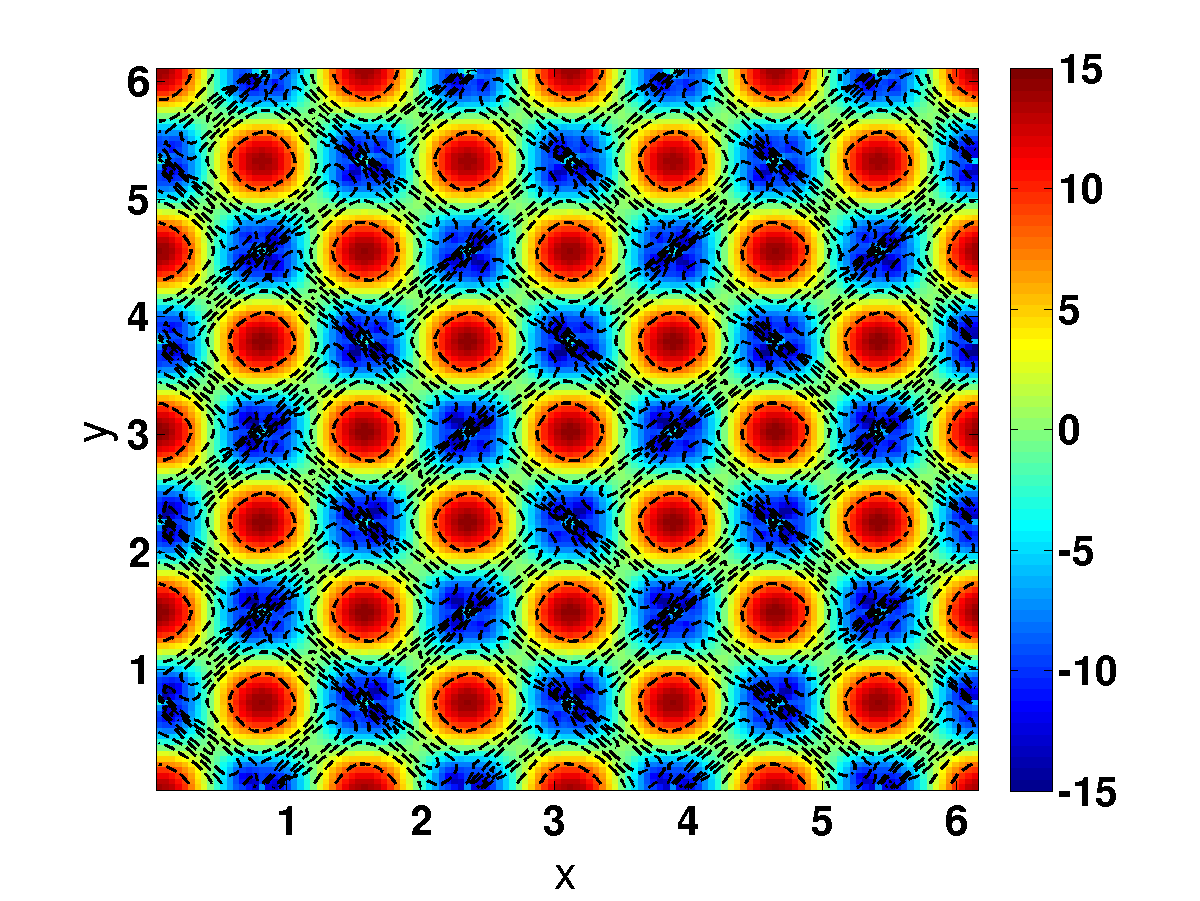}
  \put(-63,140){\color{white}{ {\bf (a)} } }
  \includegraphics[width=0.47\linewidth]{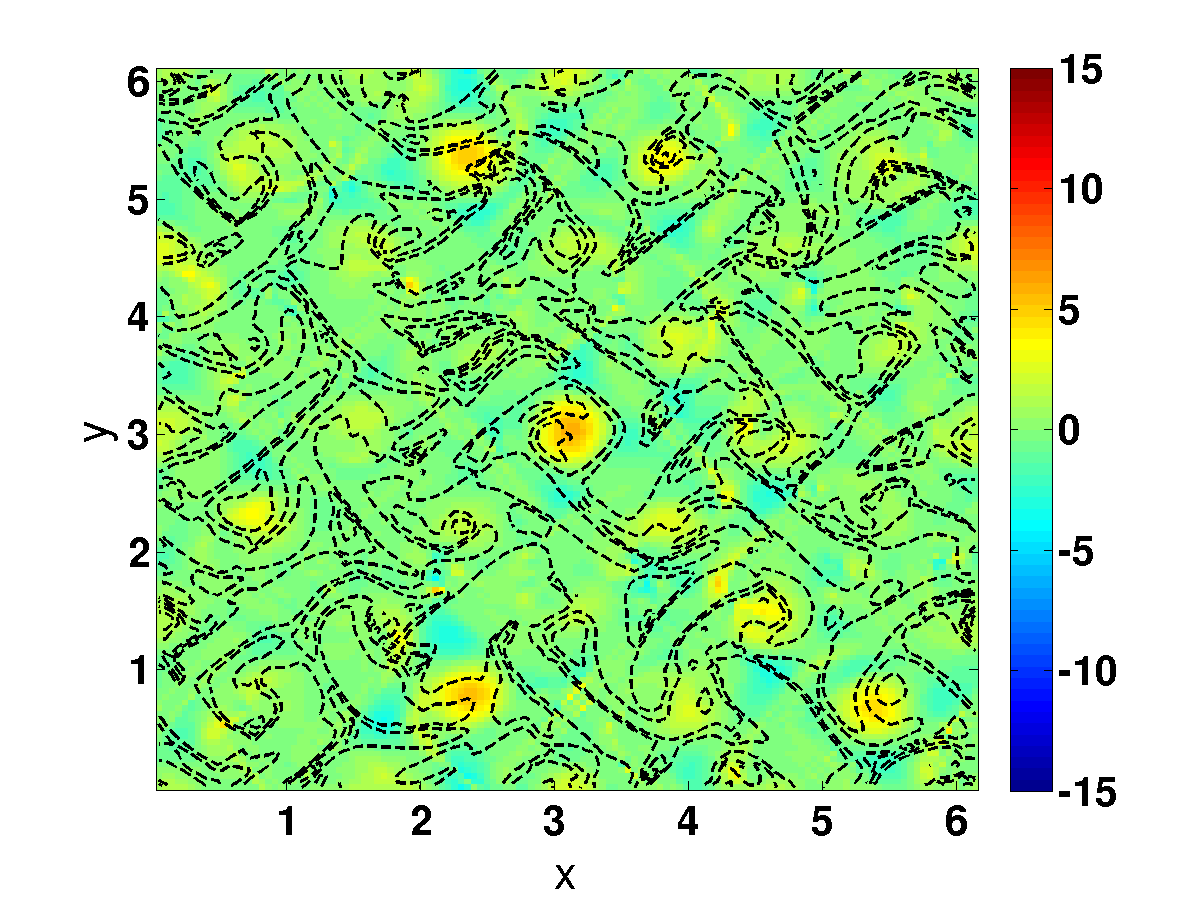}
  \put(-63,140){\color{magenta}{ {\bf (b)} } }\\
  \includegraphics[width=0.47\linewidth]{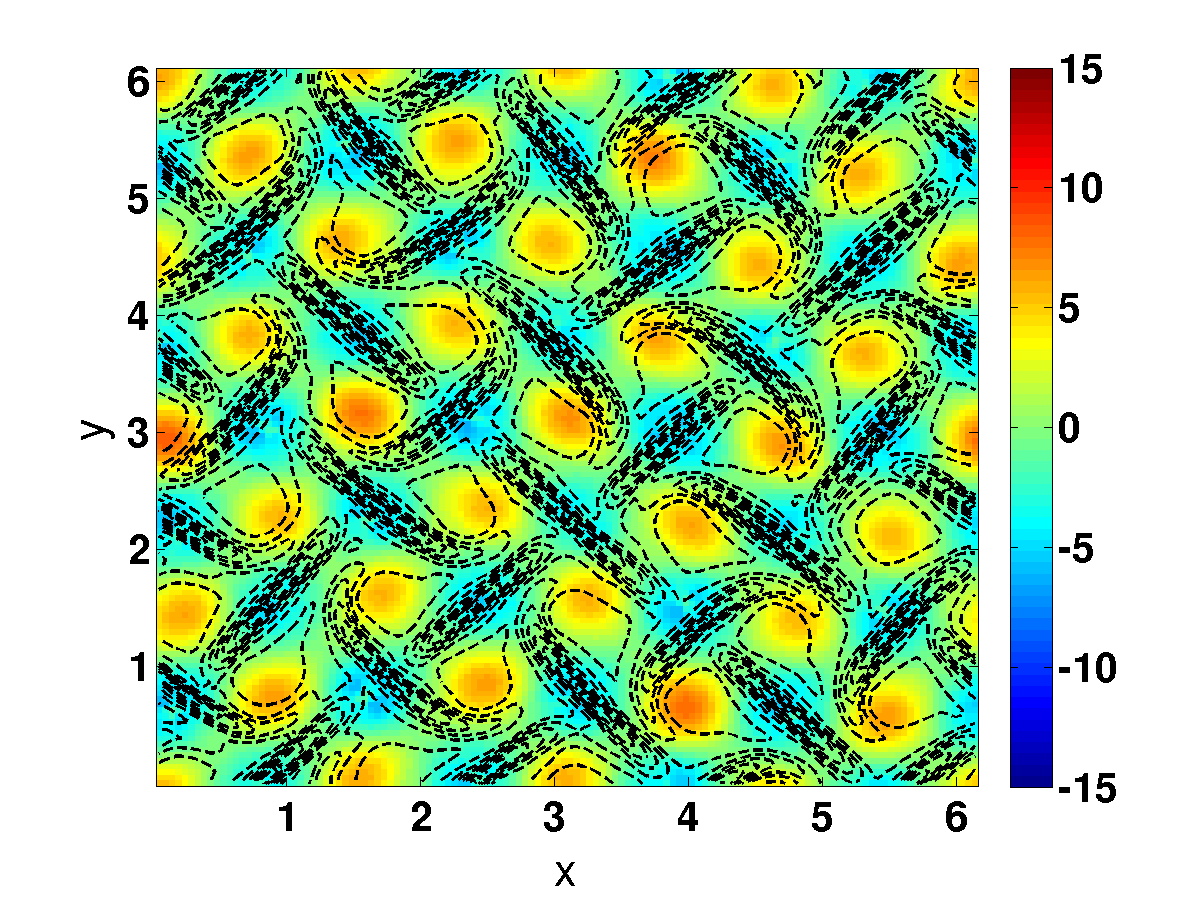}
  \put(-63,140){\color{magenta}{ {\bf (c)} } }
  \includegraphics[width=0.47\linewidth]{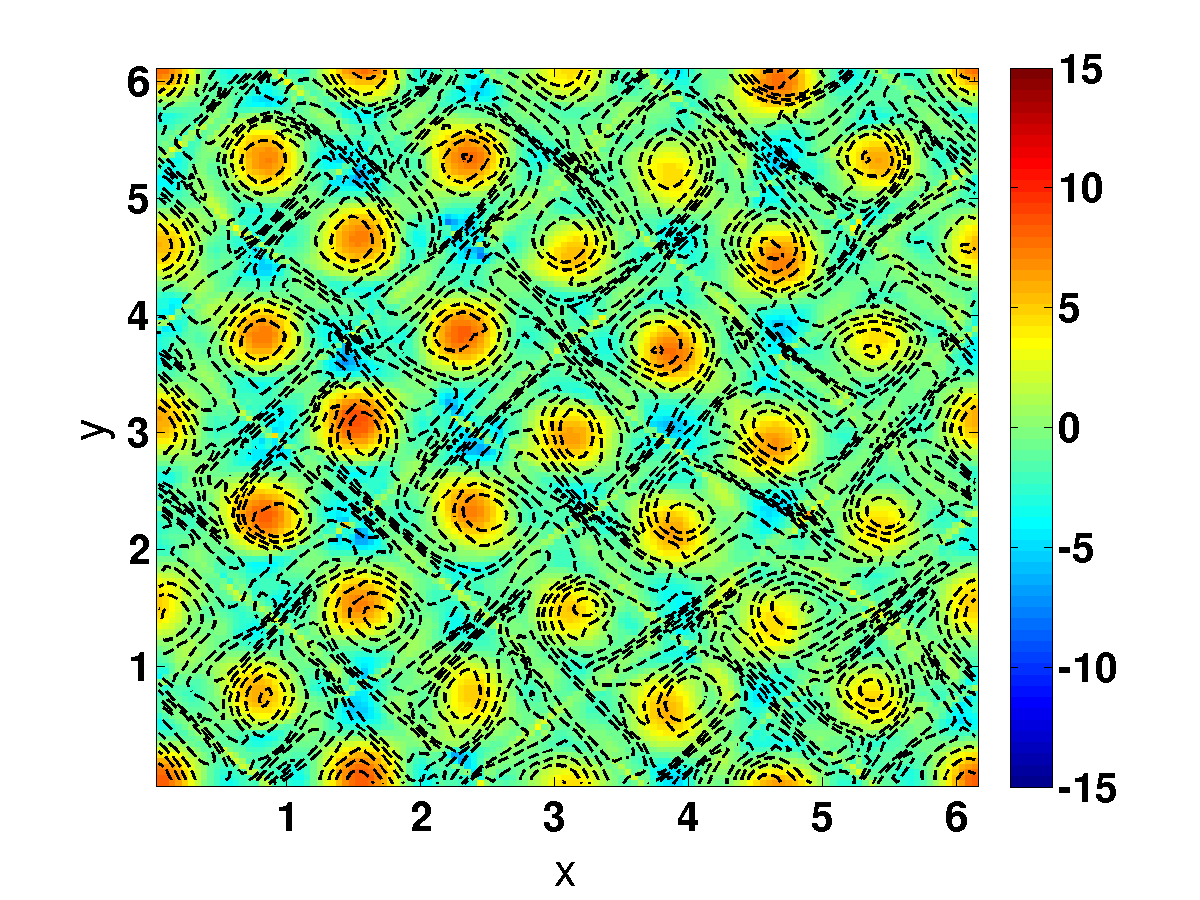}
  \put(-63,140){\color{magenta}{ {\bf (d)} } }
  \caption{\label{figch5:Lambda_rp}(Color online) Plots of contours of 
    the polymer stretching $r_P^2$ superimposed on a pseudocolor plot of the Okubo-Weiss
    field  $\Lambda$ for (a) $\Omega = 1$ and $\Wi=1$ (crystalline phase), 
    (b) $\Omega = 1$ and $\Wi=20$ (melted crystal in the
    elastic-turbulence regime), (c) $\Omega = 22$ and  $\Wi=0.5$ 
    (frozen crystal), and (d) $\Omega = 22$ and $\Wi=20$ (melted crystal 
    in the dissipation-reduction regime).}
\end{figure*}

It is instructive to study the spatial patterns of polymers and inertial
particles in the nonequilibrium states of our system; we show below that such
patterns also provide clear signatures of these states. In
Figs.~\ref{figch5:Lambda_rp} (a), (b), (c), and (d) we show contours of the
polymer stretching $r_P^2$ superimposed on pseudocolor plots of the Okubo-Weiss
field  $\Lambda$ for $\Omega = 1$ and $\Wi=1$ (crystalline state), $\Omega
= 1$ and $\Wi=20$ (melted crystal in the elastic-turbulence regime),
$\Omega = 22$ and $\Wi=0.5$ (frozen crystal), and $\Omega = 22$ and
$\Wi=20$ (melted crystal in the dissipation-reduction regime),
respectively. As we have seen in Ref~\cite{gupta2015two}, the polymers stretch
preferentially in extensional regions, where $\Lambda < 0$; thus, the patterns
of their extension mirror the spatial periodicity or lack thereof in the
crystalline and turbulent phases, respectively. The spatiotemporal evolution of
the plots in Fig.~\ref{figch5:Lambda_rp} is given in Video crystal-rp-a, b, c,
and d at http://www.physics.iisc.ernet.in/$\sim$rahul/movies.html .

\begin{figure*}[]
  \includegraphics[width=0.47\linewidth]{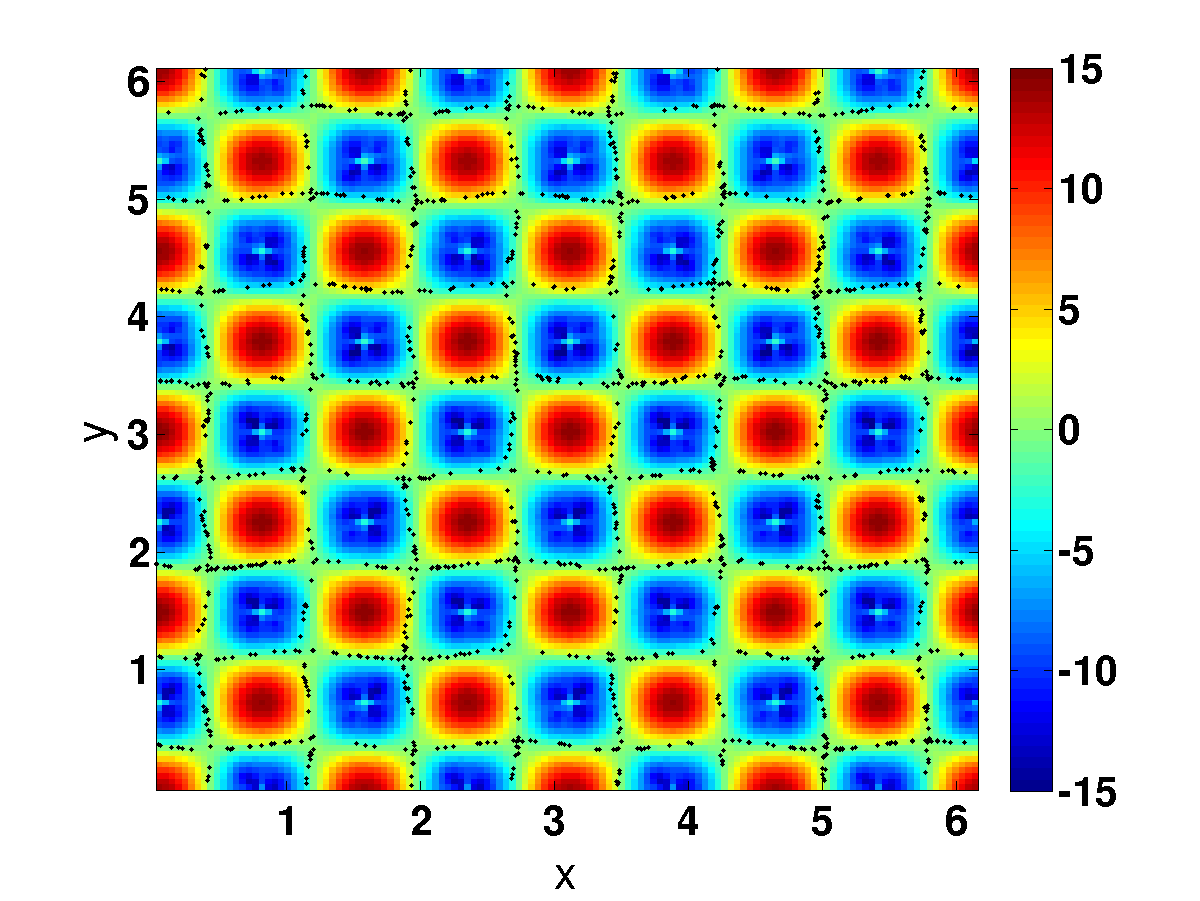}
  \put(-63,140){\color{white}{ {\bf (a)} } }
  \includegraphics[width=0.47\linewidth]{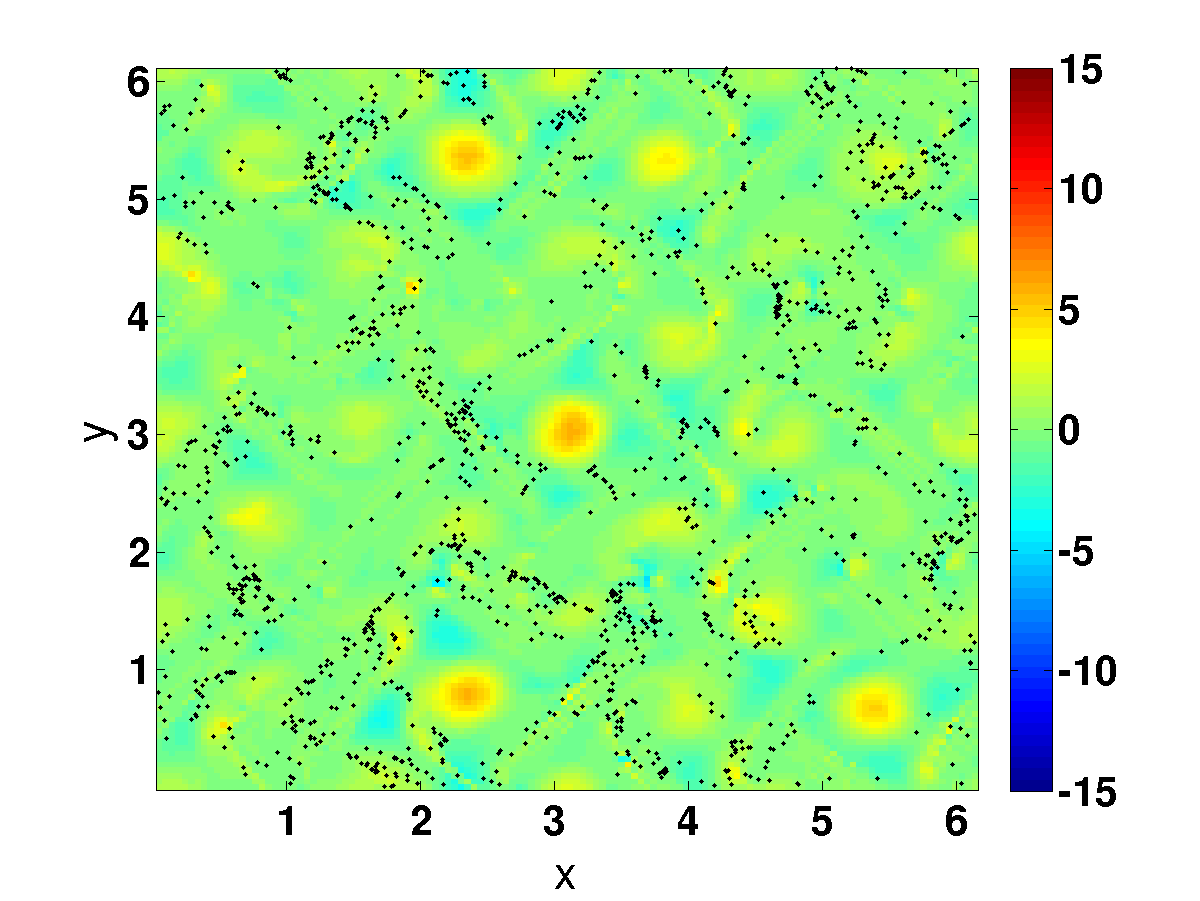}
  \put(-63,140){\color{black}{ {\bf (b)} } }\\
  \includegraphics[width=0.47\linewidth]{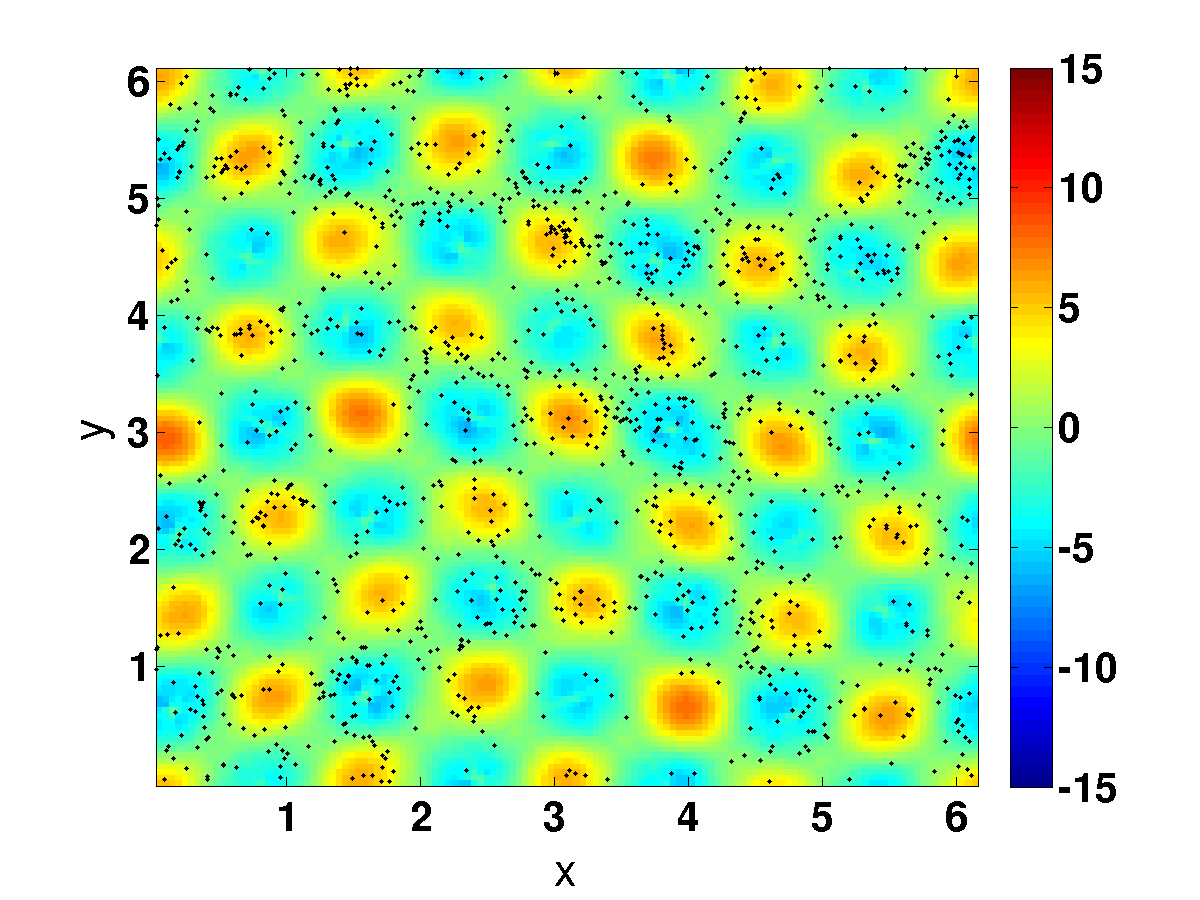}
  \put(-63,140){\color{black}{ {\bf (c)} } }
  \includegraphics[width=0.47\linewidth]{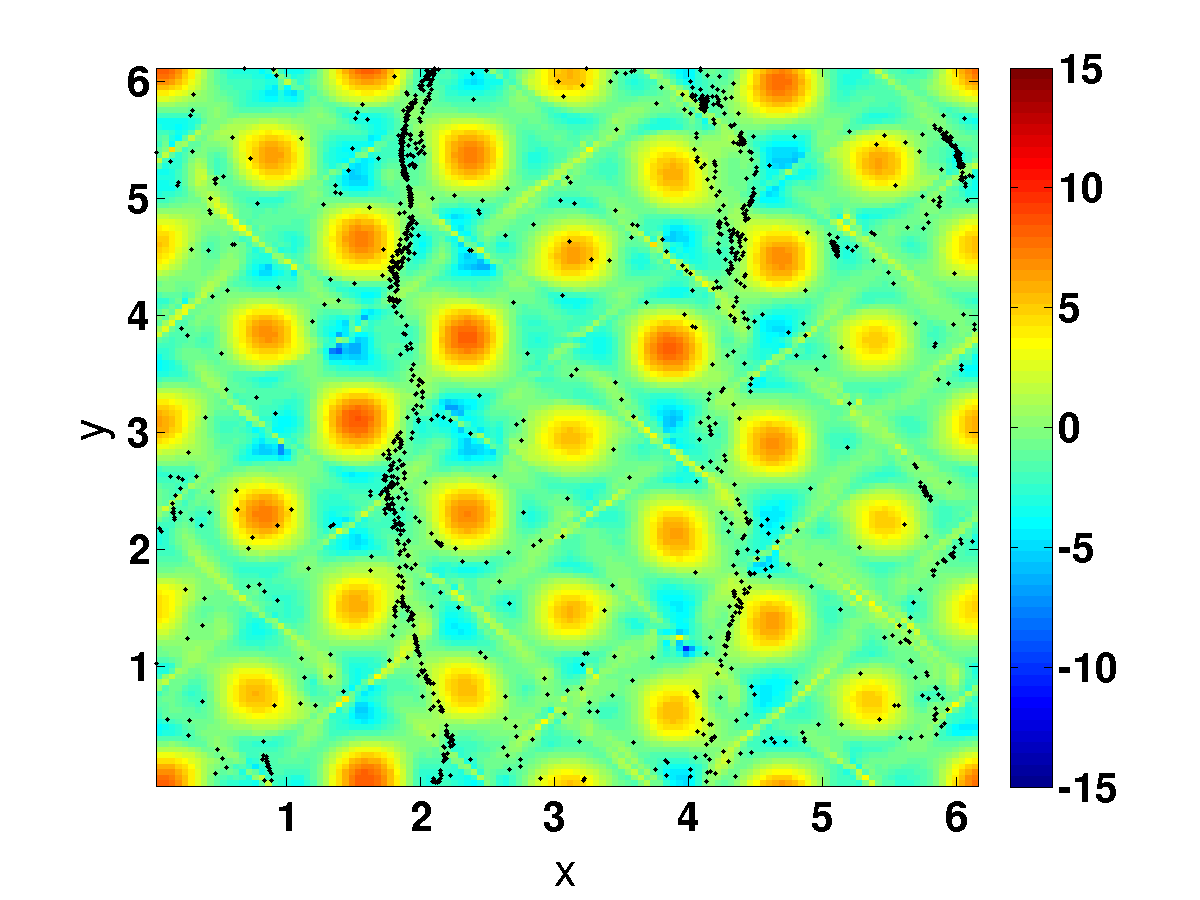}
  \put(-63,140){\color{black}{ {\bf (d)} } }
  \caption{\label{figch5:Lambda_part}(Color online) Particle positions (black dots)
    superimposed on pseudocolor plots of the Okubo-Weiss
    field  $\Lambda$ for (a) $\Omega = 1$ and $\Wi=1$ (crystalline phase), 
    (b) $\Omega = 1$ and $\Wi=20$ (melted crystal in the
    elastic-turbulence regime), (c) $\Omega = 22$ and  $\Wi=0.5$ 
    (frozen crystal), and (d) $\Omega = 22$ and $\Wi=20$ (melted crystal 
    in the dissipation-reduction regime).}
\end{figure*}

The analogs of Figs.~\ref{figch5:Lambda_rp} (a), (b), (c), and (d), with
polymers replaced by inertial particles (black dots), are given in
Figs.~\ref{figch5:Lambda_part} (a), (b), (c), and (d), respectively.  We see
from these plots that inertial particles organize themselves in regions where
$\Lambda \simeq 0$; thus, their spatial organization can also be used to
surmise the underlying periodicity of a nonequilibrium vortex crystal. This
method of visualization has been used in experiments on fluid films without
polymers~\cite{ouellette2007curvature,ouellette2008dynamic}. All these plots
are for a Stokes number $\St \simeq 1$. Note the wavy patterns of particles across the flow field, which are strongly influenced by the forcing field; these patterns are not permanent; and, as the elastic turbulence changes the particle positions, these wavy patterns of particles disappear; similar patterns appear later in time, but at  different places in the simulation domain.

\begin{figure*}[]
  \includegraphics[width=0.47\linewidth]{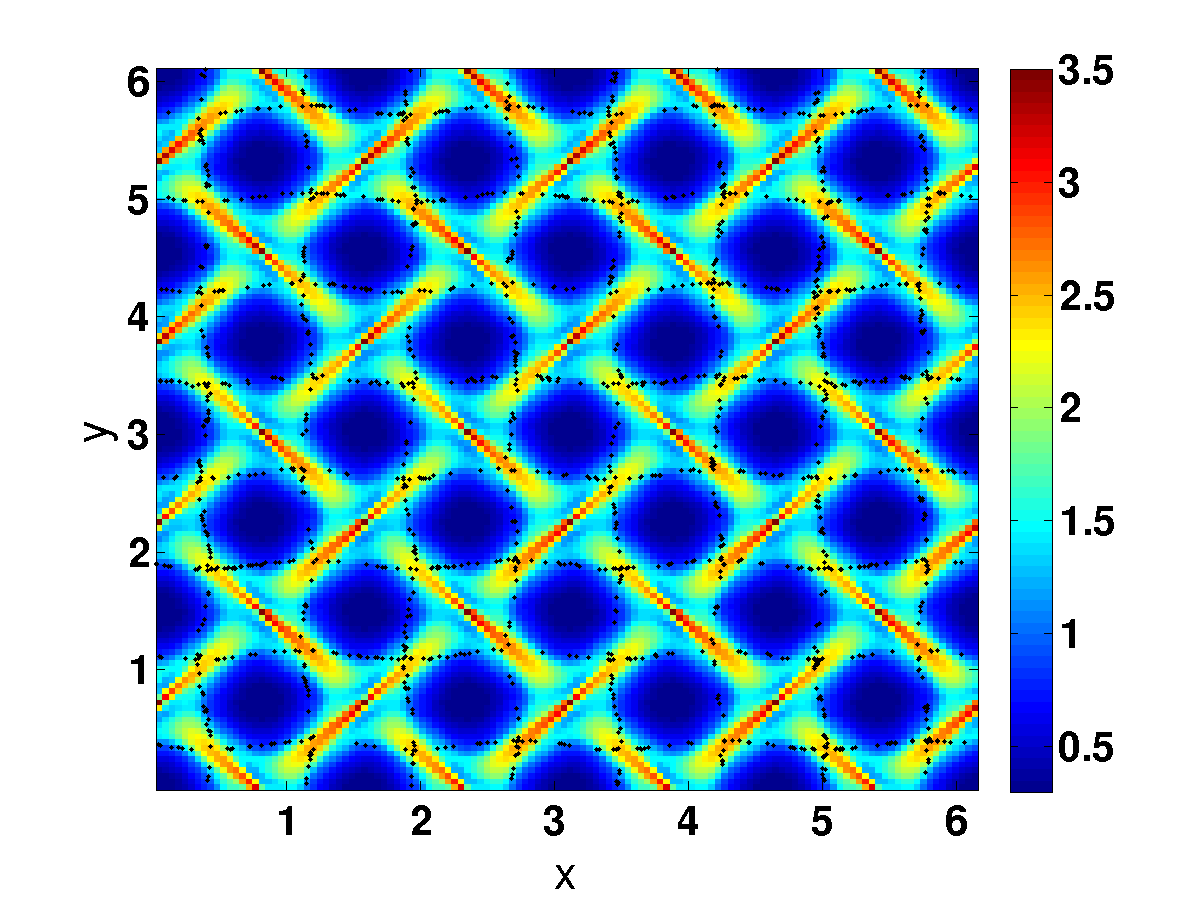}
  \put(-63,140){\color{white}{ {\bf (a)} } }
  \includegraphics[width=0.47\linewidth]{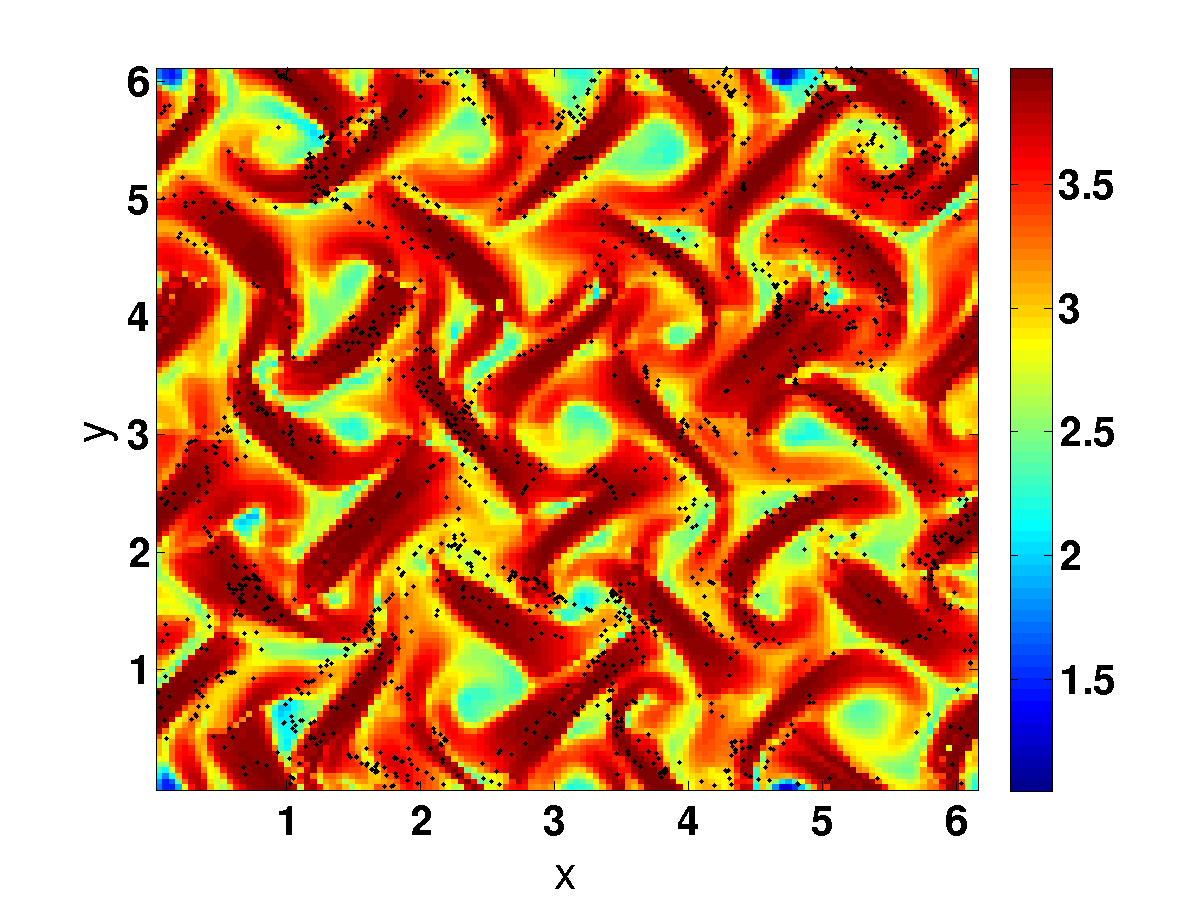}
  \put(-63,140){\color{white}{ {\bf (b)} } }\\
  \includegraphics[width=0.47\linewidth]{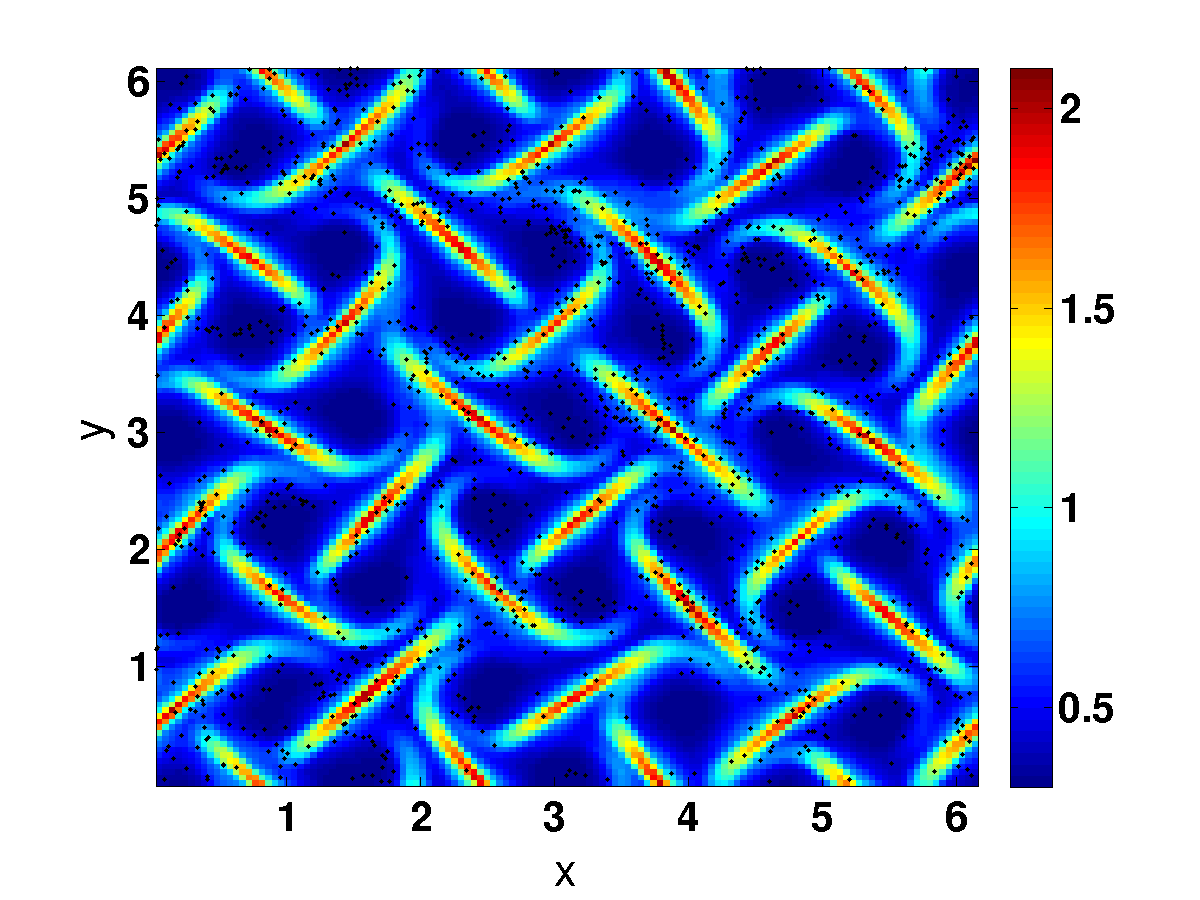}
  \put(-63,140){\color{white}{ {\bf (c)} } }
  \includegraphics[width=0.47\linewidth]{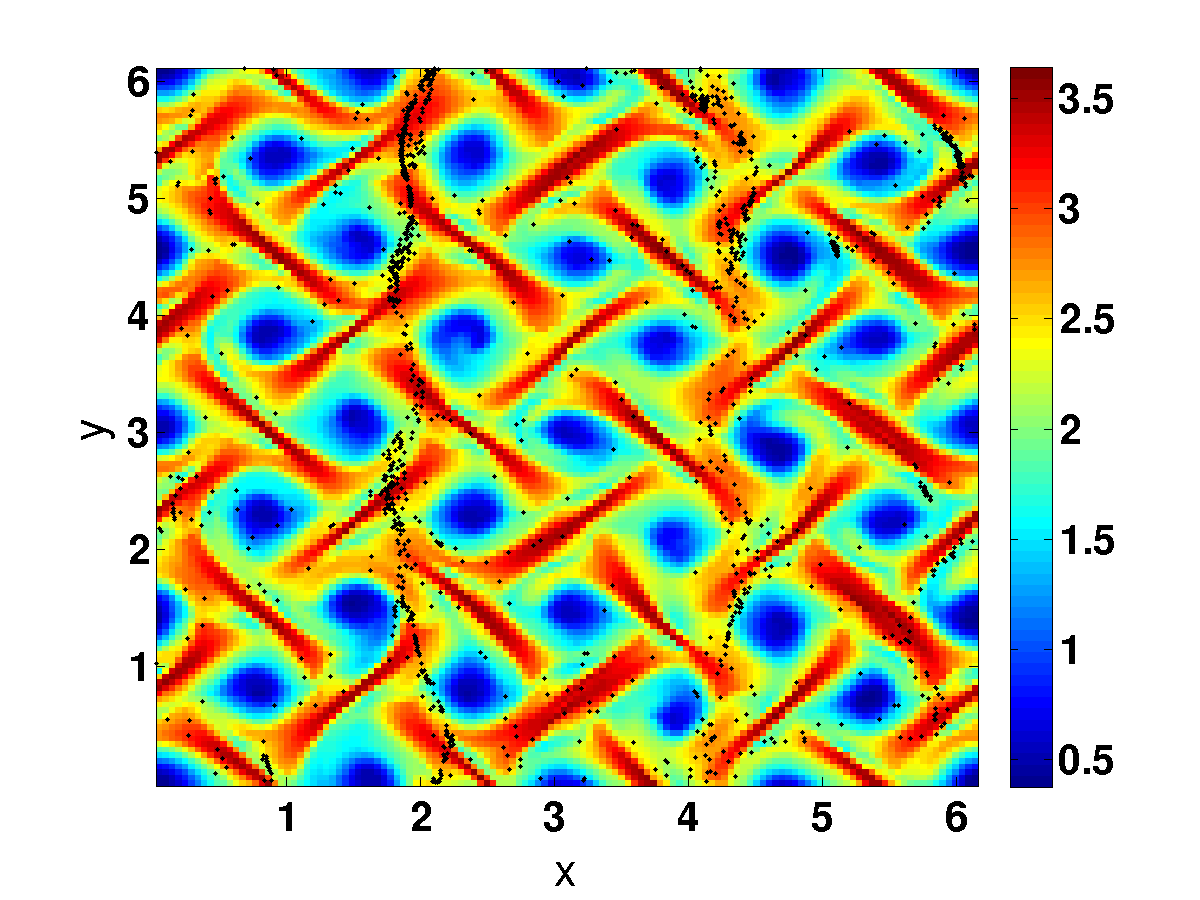}
  \put(-63,140){\color{white}{ {\bf (d)} } }
  \caption{\label{figch5:rp_part}(Color online) Particle positions (black dots)
    superimposed on a pseudocolor plot of the polymer stretching $r_P^2$
    for (a) $\Omega = 1$ and $\Wi=1$ (crystalline phase), 
    (b) $\Omega = 1$ and $\Wi=20$ (melted crystal in the
    elastic-turbulence regime), (c) $\Omega = 22$ and $\Wi=0.5$ 
    (frozen crystal), and (d) $\Omega = 22$ and $\Wi=20$ (melted crystal 
    in the dissipation-reduction regime).}
\end{figure*}

The analogs of Figs.~\ref{figch5:Lambda_part} (a), (b), (c), and (d), with the
$\Lambda$ field replaced by polymers, are given in Figs.~\ref{figch5:rp_part}
(a), (b), (c), and (d), respectively.  We see from these plots that the
correlation between the spatial organizations of polymers and inertial
particles is not dramatic. This is not surprising insofar as the polymers
stretch preferentially where $\Lambda < 0$, whereas inertial particles tend to
cluster in regions where $\Lambda \simeq 0$.


\section{Conclusions}\label{sec:conclusions}

Our detailed DNS elucidates the nature of melting of a vortex crystal in a
nonequilibrium, forced, thin fluid film with polymer additives.  This melting
can be induced by (a) elastic turbulence at low Reynolds numbers $\Rey$ (or
$\Omega$) but large Weissenberg numbers ${\Wi}$ (or $\tau_P$), (b)
fluid turbulence at low $\Wi$ but large $\Omega$, or (c) a combination of
these two types of turbulence (see Figs.~\ref{figch5:phase} (a) and (b)). Our
work leads to the rich, nonequilibrium phase diagram of
Fig.~\ref{figch5:phase}. The topology of this phase diagram and the intricate
natures of the boundaries between the different phase in it have neither been
anticipated nor been studied hitherto. 

Our work builds on the DNS study of Ref.~\cite{perlekar2010turbulence}, which
uses ideas from the density-functional theory of
freezing~\cite{ramakrishnan1979first,chaikin2000principles,oxtoby1991liquids,singh1991density},
nonlinear dynamics, and turbulence to characterize the phases and transitions
in the melting, by turbulence, of a nonequilibrium vortex crystal in a 2D fluid
film. Most studies of the turbulence-induced melting of a vortex crytal in a
fluid film do not include polymer additives; for an overview of such studies we
refer the reader to Ref.~\cite{perlekar2010turbulence}. The DNS study of 2D
fluid films with polymer
additives~\cite{gupta2015two,berti2008two,berti2010elastic}, which uses the
Oldroyd-B model, is an exception; it has studied the elastic-turbulence-induced
melting of spatially periodic Kolmogorov-flow pattern, which is, in our
terminology, a one-dimensional crystal. To the best of our knowledge, there is
no study of vortex crystals in fluid films with polymers, which brings together
the variety of methods we use to analyze the melting of such crystals. 

We follow Ref.~\cite{perlekar2010turbulence}, which does not include polymers,
in (a) identifying the order parameters for the vortex crystal, (b)
characterizing the series of transitions in terms of $\psi$ and $\Lambda$, the
energy time series $E(t)$, and their Fourier transforms, and (c) using the
spatial correlation functions $G$ in crystalline and turbulent phases.
However, we go considerably beyond the study of
Ref.~\cite{perlekar2010turbulence} by including polymers and, therefore, (a) a
new dimensionless control parameter, the Weissenberg number ${\Wi}$ (or
$\tau_P$), (b) the polymer-conformation tensor field ${\mathcal C}$, and (c)
inertial particles.  Therefore, we can examine the nonequilibrium phase diagram
of this system in a two-dimensional parameter space (see
Figs.~\ref{figch5:phase} (a) and (b)) and elucidate the organization of
polymers and inertial particles in the flow, both in crystalline and turbulent
phases. We show also that our system provides a natural laboratory in which we
can study the crossover from elastic-turbulence to dissipation-reduction
regimes in the turbulent phase.

We refer the reader to Ref.~\cite{perlekar2010turbulence} for a discussion of
the thermodynamic limit and finite-size effects in such systems of
nonequilibrium, vortex crystals. No study (including ours) has addressed these
issues because, for large system sizes, we need very large DNSs to characterize
the temporal evolution of the system. If we extract a correlation length from
the correlation function $G$, it is much smaller than the linear size of our
simulation domain in the turbulent phase (SCT), so our results should not
change if we increase the system size.  However, subtle size dependence might
occur in the ordered phase because of the inverse cascade, well known in 2D
fluid turbulence, which might lead to undulations that give rise to crystals
with ever larger unit cells, whose size can be controlled only by the inclusion
of friction. A systematic study of such subtle finite-size effects lies beyond
the scope of our study.

The sequence of transitions that take us from a vortex crystal to the turbulent
state is far richer than conventional equilibrium melting.  The former differ
from the latter in another important way: To maintain the steady states,
statistically or strictly steady, we must impose the force $F_{\omega}$; in the
language of phase transitions, this force is a symmetry-breaking field, both in
the ordered and disordered phases. Therefore, there is no symmetry difference
between the disordered, turbulent state and the ordered vortex crystal, as we
see directly from the vestiges of the dominant peaks in the reciprocal-space
spectra $E_{\Lambda}({\bf k})$; and the order parameters
$\langle{\hat{\Lambda}}_{\bf k}\rangle$, with ${\bf k} = (4,4)$ the forcing
wave vectors, do not become exactly zero in the turbulent phase;
however, they do become very small.  Moreover, our nonequilibrium vortex
crystal undergoes a sequence of transitions from an ordered state to an
undulating crystal and then to a fully turbulent state; there is no external or
thermal noise here and hence no fluctuations in the perfect vortex crystal;
this has no equilibrium analog.

We hope that our study will encourage experimental groups to try to obtain the
rich phase diagram of the melting of a nonequilibrium vortex crystal in a fluid
film with polymer additives. In particular, experiments on such a system could look for spatiotemporal crystals, which are periodic both in space and time and which cannot be obtained easily in nonequilibrium, hard-matter settings. Furthermore, the reentrant crystallization, i.e., the recrystallization, with an increase in ${\Wi}$, of the melted crystal, which emerges from our detailed DNS, is certainly worth exploring in experiments. We have recently learnt of a shell-model study of 
elastic turbulence that is just published~\cite{ray2016elastic}.

\section{Acknowledgments}

We thank the Council for Scientific and Industrial Research, University Grants
Commission, and Department of Science and Technology  (India) for support, SERC
(IISc) for computational resources, and P.Perlekar, D. Mitra, S.S. Ray, and D.
Vincenzi for discussions.  AG acknowledges gratefully funding from the European
Research Council under the European Community's Seventh Framework Programme
(FP7/2007-2013)/ERC Grant Agreement N. 279004.

\section*{References}

\bibliography{sample}

\end{document}